\documentclass[oldversion]{aa} 
\usepackage{graphicx}
\usepackage{txfonts}
\usepackage{longtable} 

\begin{document}

   \title{The massive star initial mass function of the Arches cluster\thanks{Based on observations obtained with the ESO/YEPUN telescope at Paranal Observatory. Program 73.D-0815.}\fnmsep\thanks{Tables 2, 3, and 5 are only available in electronic form at the CDS via anonymous ftp to cdsarc.u-strasbg.fr (130.79.128.5)
or via http://cdsweb.u-strasbg.fr/cgi-bin/qcat?J/A+A/}}

    \author{P. Espinoza \inst{1,2} \and F. J. Selman \inst{3} \and
          J. Melnick \inst{3}}
   \offprints{P. Espinoza, \\ \email{pespinoza@as.arizona.edu}}

   \institute{Steward Observatory, The University of Arizona, 933 North Cherry Avenue, Tucson AZ 85719, USA;\\ \email{pespinoza@as.arizona.edu} \and Previous Address: Departamento de Astronom\'ia y Astrof\'isica, Pontificia Universidad Cat\'olica de Chile, Santiago, Chile \and European Southern Observatory, Alonso de C\'ordova 3107, Santiago, Chile\\
              \email{[fselman;jmelnick]@eso.org}
  	     }
 \authorrunning{Espinoza, Selman, and Melnick}
 \titlerunning{The massive star IMF of the Arches cluster}
   \date{Received day month / Accepted day month }

   \abstract
   { The massive Arches cluster near the Galactic center should be an ideal laboratory for investigating massive star formation under extreme conditions.  But it comes at a high price: the cluster is hidden behind several tens of magnitudes of visual extinction. Severe crowding requires space or AO-assisted instruments to resolve the stellar populations, and even with the best instruments interpreting the data is far from direct. Several investigations using NICMOS and the most advanced AO imagers on the ground revealed an overall top-heavy IMF for the cluster, with a very flat IMF near the center. There are several effects, however, that could potentially bias these results, in particular the strong differential extinction and the problem of transforming the observations into a standard photometric system in the presence of strong reddening.  We present new observations obtained with the NAOS-Conica (NACO) AO-imager on the VLT. The problem of photometric transformation is avoided by working in the natural photometric system of NACO, and we use a Bayesian approach to determine masses and reddenings from the broad-band IR colors. A global value of $\Gamma=-1.1 \pm 0.2$ for the high-mass end ($M>10M_{\odot}$) of the IMF is obtained, and \textit{we conclude that a power law of Salpeter slope cannot be discarded for the Arches cluster}. The flattening of the IMF towards the center is confirmed, but is less severe than previously thought. We find $\Gamma=-0.88\pm0.20$, which is incompatible with previous determinations. Within $0.4\;pc$ we derive a total mass of $\sim2.0(\pm0.6)\times10^{4}M_{\odot}$ for the cluster and a central mass density $\rho = 2(\pm0.4)\times10^{5}\;M_{\sun}\;pc^{-3}$ that confirms Arches as the densest known young massive cluster in the Milky Way.}


   \maketitle

\section{Introduction}
\label{Introduction}

The young  clusters  near the Galactic center provide ideal laboratories for testing theories of the formation and evolution of massive stars and massive clusters (e.g. Figer et al. \cite{figer99}), but they also pose formidable observational and theoretical challenges.  Even at IR wavelengths the large amount of interstellar extinction towards these clusters renders the already difficult interpretation of broad-band photometry of massive stars even more difficult.  And this difficulty is compounded by how the extinction, even at two micron, can vary by more than one magnitude from star to star. These clusters evolve in the strong tidal field of the Galactic center and, on the theoretical side, they may lose up to several 1000 $M_{\odot}/Myr$ (Portegies Zwart et al. \cite{portegies_zwart0}). This implies that the timescale for dynamical evolution is comparable to the lifetimes of the most massive stars. 

The best studied of these clusters is the massive Arches cluster located about $25\;pc$ in projection from the Galactic center. From NICMOS observations, Figer et al. (\cite{figer99}) find that the IMF of the Arches was significantly flatter than the Salpeter law (\cite{salpeter}) and concluded that the Galactic center environment favors the formation of massive stars. Stolte et al. (\cite{stolte02}; SGB02) using the adaptive optics (AO) camera on Gemini North, found a sharp flattening of the IMF towards the center of the cluster, with a global slope still flatter than the Salpeter value. SGB02 interpreted their observations as evidence of strong mass segregation, but did not reach a conclusion about whether this was a sign of dynamical evolution, or if it was inherited from the formation of the cluster. The advent of the NACO AO camera on the VLT provided Stolte et al. (\cite{stolte05};  SBG05) with the opportunity of probing the Arches cluster to deeper levels with better adaptive optics corrections.  While they basically confirmed the strong flattening of the slope in the central regions, they also found evidence of a sharp turnover in the mass function around $6-7M_{\odot}$ that they interpreted as a low-mass truncation in the cluster IMF. A new investigation of Arches by Kim et al. (\cite{kim_arches}) using the AO camera of the Keck telescope confirmed a flatter slope and revealed the $\sim6M_{\odot}$ feature as a local maximum in the cluster mass function. With appropiate parameters as input, this bump was also reproduced later by the coalescence-collapse model of Dib et al. (\cite{dib}). However, no evidence of the low-mass truncation suggested by SBG05 was found in Kim's work.

All these investigations suffer from the problems of interpreting broad-band JHK photometry, the most insidious of which  is the large extinction variations from star to star.  This effect has been recognized by SGB02 and SBG05, but was taken into account only in an approximate manner, i.e. by modeling the extinction variations as a function of radius.  We know from our experience with the 30 Doradus cluster that interpreting broad-band photometry in the presence of high and variable extinction is a rather challenging problem (e.g. Selman et al. \cite{selman99}).  To begin with, it is not possible to transform the observations to a standard photometric system using the standard calibrators (e.g. Johnson), so one is compelled to work in the natural photometric system of the instrument (Selman, \cite{selman04}). This means that the stellar models and the extinction law must be transformed or calculated for the same photometric system. In the case of 30 Doradus, the extinction varies by as much as 3-4 magnitudes in $A_{V}$ due to dust in the parental cloud of the cluster. For Arches, the observed variations of $\sim 2$ magnitudes in $A_{K_{S}}$ (nearly 20 magnitudes in $A_{V}$) are mostly foreground to the cluster along the line of sight to the center of the Galaxy.  These variations induce a magnitude-dependent incompleteness factor which may be very large for the lowest mass bins in an otherwise complete photometric catalog. Thus, applying an average extinction correction, as done by  previous authors, can lead to serious errors in the final mass determinations. These considerations prompted us to re-observe the Arches, in order to answer the question of whether a Salpeter law can be discarded for this cluster. 

Taking advantage of the newly commissioned IR wavefront sensor in the VLT AO camera NACO allowed us to reach a Strehl ratio as high as 27\% in the $K_{S}$-band.  We find a strong radial dependence of the IMF slope, going from   $\Gamma\sim-0.88$ for the central $0.2\;pc$ of the cluster, to  $\Gamma\sim-1.28$ for $0.2<r<0.4pc$, with a global slope of  $\Gamma=-1.1\pm0.2$. Thus, while we confirm the trend found by previous investigations of a significant flattening of the IMF slope in the central region of the cluster, our values are significantly steeper than previous determinations at all radii. The discrepancy is particularly important in the center, where SBG05  find a much flatter slope of $\Gamma=-0.26\pm0.07$ over a similar mass range. Our strong conclusions are: (1)  even in the extreme environment of the Galactic center, the global IMF of the Arches is consistent with that found in other starburst clusters such as 30 Doradus in the LMC, and (2), the strong radial gradient in the present day mass distribution of the cluster stars provides a clear indication of significant mass segregation in this young cluster (see also discussions in Figer et al. \cite{figer99}, SGB02).
  
The paper is organized as follows: Section 2 discusses the observations, the data reduction procedures, and the photometric calibrations. We also discuss in this Section the incompleteness corrections derived from Monte Carlo experiments. Section 3 deals with the complicated procedure of deriving stellar physical parameters in the presence of strong and variable reddening. The IMF is computed in Section 4, where we also present additional evidence of mass segregation in the cluster. Section 5 presents a discussion of our results and a comparison with previous investigations and Section 6 summarizes the conclusions.

\section{Data Reduction}
\subsection{Observations}
\label{Observations}

The observations reported here were performed in Service Mode using the NAOS-CONICA (NACO) AO system on the VLT UT4 (Lenzen et al. \cite{lenzen}; Rousset et al. \cite{rousset}). The data were obtained using the imaging mode on September 6$^{th}$, 2004, under clear weather conditions with subarcsecond seeing. To allow the quantification of the high and variable IR background, the same amount of observing time devoted to science frames was spent on \textit{sky}, a necessary procedure when the object of interest is severely affected by crowding. No control fields were observed in order to estimate the effect of background and foreground contamination. But our photometry does not reach deep enough to sample the faint end of the IMF, and we did apply a color selection to the data. Accordingly, non-cluster members are not expected to contribute significantly to our counts, especially at radii $< 10''$.

The chosen reference star, located approximately at 10$''$ from the cluster center, had $K_{S}\sim$10 mag. To optimize the AO performance we used the $N90C10$ dichroic, i.e. $10\%$ of the light was directed to CONICA while $90\%$ went to the IR wavefront sensor. An excellent AO correction in the $K_{S}$-band resulted in a uniform PSF across the field with FWHM of 0.09$''$. In $J$ and $H$ our frames exhibited larger PSF variations (as the isoplanatic angle is smaller for shorter wavelengths) with mean FWHM of 0.17$''$ and 0.11$''$ respectively. We measured the Strehl ratio of our observations with the SCISOFT's STREHL \footnote{http://www.eso.org/scisoft} task, exceeding $27\%$ in $K_{S}$ and reaching more modest values of $5\%$ in $J$ and $11\%$ in $H$. An observations log summarizing this information is presented in Table~\ref{obs_log}, along with exposure times, $V$-band seeing, and airmasses.

\begin{table*}
 \caption{Arches ($\alpha = 17^{h}45^{m}50^{s}, \delta = -28^{\circ}49{'}28{''}$, J2000) Observing Log}             
 \label{obs_log}      
 \centering          
 \begin{tabular}{c c c c c c c c}     
 \hline\hline       
UT & Filter & $N$ $^{\mathrm{a}}$ & $t^{\mathrm{b}}$~[s]  & Airmass & Seeing$^{\mathrm{c}}$~('') & FWHM ~('') & Strehl ratio\\ 
 \hline                    
01:24:15~-~01:54:20  & $J$	&  5	& 200 $s$& 1.13~-~1.22	& 0.82	& 0.17 & 0.05\\
00:52:15~-~01:16:46  & $H$	&  8	& 100 $s$ & 1.07~-~1.11	& 1.1	& 0.11 & 0.11\\
00:14:40~-~00:47:02  & $K_{S}$  & 12	& 60 $s$ & 1.02~-~1.06	& 0.95	& 0.09 & 0.27\\
 \hline                  
 \end{tabular}
\begin{list}{}{}
\item[$^{\mathrm{a}}$] Number of exposures.
\item[$^{\mathrm{b}}$] Integration time of each exposure.
\item[$^{\mathrm{c}}$] Average value of the DIMM Seeing during observations (taken in the optical V band). It is stressed that in Paranal the DIMM values can be significantly larger than the image quality measured at the telescope.
\end{list}
\end{table*}

\subsection{Reduction}
\label{Reduction}

\begin{figure}
\centering
\includegraphics[width=8cm]{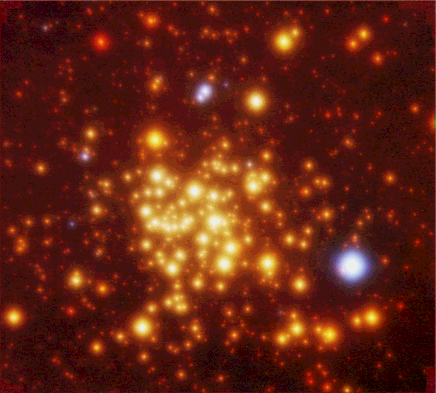}
\caption{Three-color composite image of the Arches cluster. The field of view is 28$''$. North is up, and east is to the left.}
\label{arches_color}
\end{figure}

We used ECLIPSE \footnote{\textbf{E}SO \textbf{C} \textbf{L}ibrary for an \textbf{I}mage \textbf{P}rocessing \textbf{S}oftware \textbf{E}nvironment (Devillard, \cite{devillard})} pipeline tools to reduce the data. Individual science images were flat fielded, dark subtracted, and corrected with a bad pixel mask. To obtain a reliable estimate of the IR background each pixel must see mostly sky signal during observations; this was accomplished with dithered exposures from a nearby field (with the AO loop open). All these were combined into a median average, i.e. a constant sky was assumed. To validate this procedure, we checked that during $K_{S}$ observations the variations of different sky planes compared with the median average were within $5\%$. The combined sky plane was then subtracted from all reduced science frames.

The next step was to align the individual images before co-addition. In order to shift by a sub-pixel offset it was necessary to resample frames using an interpolation kernel. In ECLIPSE, this kernel is defined in the image space and is based on the closest 16 pixel values (Devillard, \cite{devillard}). 

Finally, aligned science images were co-added by means of a simple linear average stack. In this way all the input signal is kept on the final combination, allowing a better faint object detection. The process is the same for the $J$, $H$ and $K_{S}$-band frames. The final products combined into a single three color composite image are presented in Fig.~\ref{arches_color}.

\subsection{Photometry}
\label{Photometry}

We used the DAOPHOT/IRAF\footnote{IRAF is distributed by the National Optical Astronomy Observatories, which is operated by the Association of Universities for Research in Astronomy, Inc. (AURA) under cooperative agreement with the NSF} package (Stetson, \cite{stetson87}; Davis \cite{davis}) 
to do photometry in the severely crowded environment of the Arches cluster.  
Point sources were identified using DAOFIND, with a detection threshold set to 3$\sigma$ above the local background level. Then aperture photometry was performed on the detected stars computing local sky values for each one. This is important because these values change remarkably from one star to another; especially at the cluster center, where seeing haloes dominate background variations (especially in $J$ and $H$).

The PSF was computed interactively using several bright, isolated stars located away from the frame edges, but otherwise sampling the field as uniformly as possible. We selected $9$ bright stars in the $J$, $K_{S}$ frames, and $8$ stars in $H$. The analytical component of the PSF was set to \textit{auto} to optimize the fit. This gives a Penny1 function for both $H$ and $K_{S}$ and Penny2 function for $J$. Both analytical functions are elliptical arbitrarily aligned Gaussian cores with Lorentzian wings, with the difference that the wings are also aligned arbitrarily for the Penny2 fitting function. For the empirical component, a linearly variable model is required for $J$ and $H$, while, as mentioned above, the $K_{S}$ PSF was found to be constant across the field. With these specifications we obtained the lowest residuals in the subtraction of the PSF stars and their neighbors. Once the PSF was determined, profile-fitting photometry was performed with ALLSTAR. Finally, to obtain total instrumental magnitudes for every star, we added a constant aperture correction. This correction was obtained from the aperture photometry and later curve of growth analysis of bright isolated stars in the NACO field.

\begin{figure*}
\centering
\includegraphics[width=16cm]{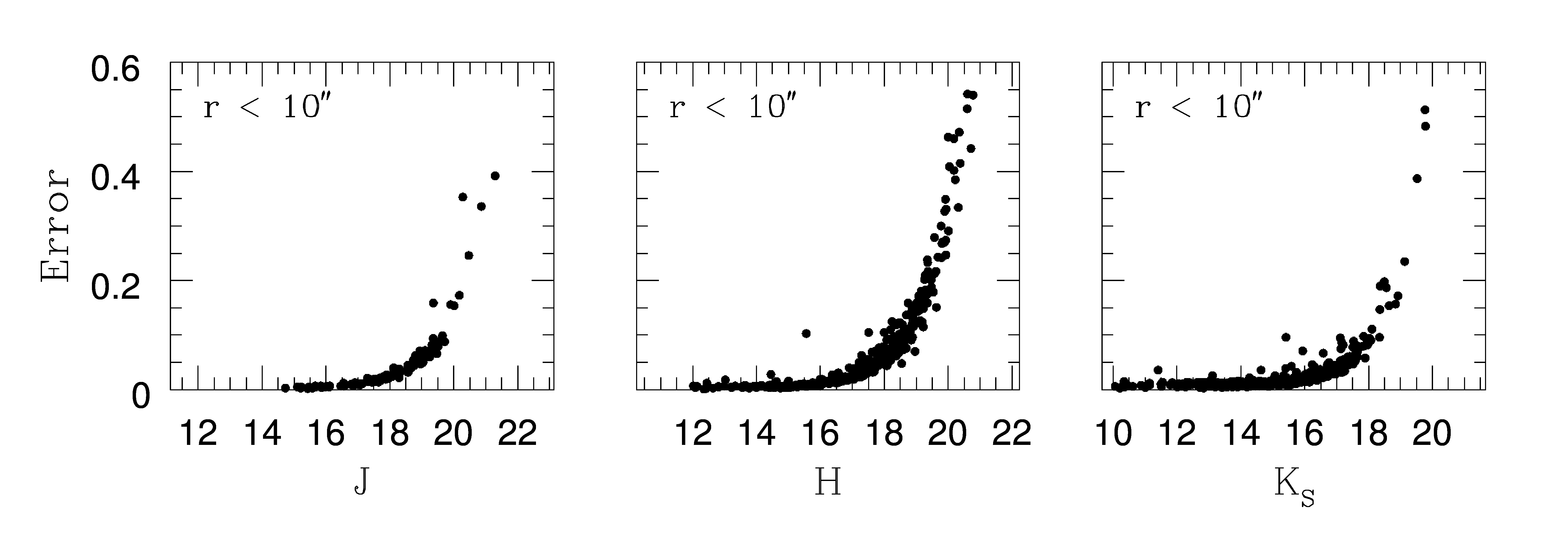}
\caption{DAOPHOT errors as a function of magnitude for the three final frames. Point source detections are obtained from a 1-pixel tolerance match between catalogs using DAOMASTER. \textbf{Left:} 143 Stars matched in the three filters. \textbf{Center} and \textbf{Right:} 603 stars matched in $H$ and $K_{S}$ only.}
\label{errores}
\end{figure*}

The photometry can now be divided into two catalogs: one for stars detected in  $HK_{S}$ only,  and another for stars detected in the three bands ($JHK_{S}$). These catalogs are obtained with DAOMATCH and DAOMASTER (Stetson \cite{stetson90}), which give a final list of stars matched between frames with a 1-pixel tolerance. Tables~\ref{ol603} and~\ref{ol143}, published electronically, present the photometry of 427 $HK_{S}$ and 126 $JHK_{S}$ stars in the innermost $10''$ of the cluster. Table~\ref{ol143} is considerably shorter due to the increasing extinction towards bluer wavelengths. Fig.~\ref{errores} shows the photometric errors as a function of magnitude as given by DAOPHOT.

\addtocounter{table}{1} 
\onllongtab{2}{
\begin{longtable}{c c c c c c c}
\caption{Photometry of $HK_{S}$ stars. The catalog contains point sources within $10''$ from the cluster center.}
\label{ol603}\\
Star ID $^{a}$  & $\Delta RA$ $^{b}$ $['']$ & $\Delta DEC$ $^{b}$ $['']$ & $K_{S}$ $^{c}$ & $\sigma_{K_{S}}$ $^{d}$ & $(H-K_{S})$ $^{e}$ & $\sigma_{(H-K_{S})}$ $^{f}$\\ 
\hline 
\endfirsthead
\caption{Continued.} \\ 
\hline 
Star ID $^{a}$  & $\Delta RA$ $^{b}$ $['']$ & $\Delta DEC$ $^{b}$ $['']$ & $K_{S}$ $^{c}$ & $\sigma_{K_{S}}$ $^{d}$ & $(H-K_{S})$ $^{e}$ & $\sigma_{(H-K_{S})}$ $^{f}$\\ 
\hline
\endhead
\hline
\endfoot    
\hline
\endlastfoot
id0001&  -1.897&   -2.321&  10.068&    0.006&    1.957&    0.009\\
id0002&   2.601&    5.739&  10.216&    0.003&    2.200&    0.004\\
id0003&   0.035&    2.397&  10.244&    0.004&    1.841&    0.006\\
id0004&  -1.231&    0.439&  10.281&    0.005&    1.861&    0.008\\
id0005&  -4.724&   -2.274&  10.345&    0.015&    2.099&    0.019\\
id0006&   3.425&   -6.441&  10.362&    0.005&    1.993&    0.005\\
id0007&  -2.313&   -1.276&  10.456&    0.008&    1.997&    0.011\\
id0008&  -3.966&    8.233&  10.619&    0.006&    1.776&    0.006\\
id0009&  -3.778&    2.719&  10.878&    0.008&    1.928&    0.010\\
id0010&   1.475&   -2.617&  10.965&    0.006&    1.851&    0.008\\
id0011&   2.474&    3.383&  11.011&    0.004&    1.625&    0.005\\
id0012&  -6.597&   -6.550&  11.144&    0.012&    2.190&    0.014\\
id0013&  -0.551&   -0.702&  11.151&    0.009&    1.833&    0.011\\
id0014&   3.172&    1.102&  11.412&    0.036&    1.610&    0.040\\
id0015&  -1.211&    2.081&  11.502&    0.010&    1.701&    0.011\\
id0016&   2.593&    0.360&  11.518&    0.006&    1.703&    0.007\\
id0017&  -6.799&   -3.663&  11.546&    0.014&    2.098&    0.016\\
id0018&   0.547&    0.449&  11.788&    0.009&    1.769&    0.011\\
id0019&  -5.653&   -7.197&  11.854&    0.013&    2.204&    0.014\\
id0020&  -4.541&    3.291&  11.877&    0.011&    1.876&    0.012\\
id0021&   0.997&   -1.747&  11.935&    0.009&    1.747&    0.011\\
id0022&   7.738&   -3.396&  12.010&    0.005&    1.759&    0.006\\
id0023&   2.317&    2.331&  12.054&    0.006&    1.661&    0.008\\
id0024&  -0.167&   -3.573&  12.061&    0.011&    1.889&    0.013\\
id0025&   1.743&    2.971&  12.133&    0.006&    1.651&    0.008\\
id0026&   1.314&    0.071&  12.159&    0.009&    1.735&    0.011\\
id0027&   1.744&   -0.018&  12.160&    0.008&    1.747&    0.010\\
id0028&  -1.895&    0.316&  12.187&    0.011&    1.759&    0.013\\
id0029&   0.768&    0.110&  12.209&    0.008&    1.751&    0.010\\
id0030&   3.165&   -1.081&  12.266&    0.005&    1.729&    0.006\\
id0031&  -4.579&    1.401&  12.307&    0.014&    1.853&    0.016\\
id0032&  -6.135&   -0.728&  12.392&    0.013&    1.998&    0.015\\
id0033&   0.828&    1.643&  12.412&    0.009&    1.700&    0.011\\
id0034&  -3.426&   -0.737&  12.621&    0.013&    1.822&    0.016\\
id0035&   1.027&   -5.139&  12.647&    0.009&    1.897&    0.010\\
id0036&   1.715&   -4.828&  12.647&    0.008&    1.818&    0.009\\
id0037&   1.187&   -4.556&  12.692&    0.009&    1.842&    0.010\\
id0038&  -2.836&   -1.152&  12.710&    0.014&    3.476&    0.021\\
id0039&  -1.656&   -3.784&  12.720&    0.013&    1.914&    0.015\\
id0040&   3.053&    2.314&  12.720&    0.006&    1.557&    0.008\\
id0041&   6.024&   -3.942&  12.727&    0.004&    2.230&    0.006\\
id0042&  -0.420&   -1.726&  12.765&    0.013&    1.748&    0.015\\
id0043&  -1.192&    0.045&  12.827&    0.012&    1.743&    0.015\\
id0044&   2.163&   -2.146&  12.829&    0.007&    1.744&    0.008\\
id0045&   2.313&   -1.721&  12.832&    0.007&    1.793&    0.009\\
id0046&   6.788&    6.129&  12.832&    0.003&    1.595&    0.005\\
id0047&   2.959&    0.146&  12.844&    0.006&    1.719&    0.010\\
id0048&  -4.634&    4.691&  12.846&    0.011&    1.973&    0.013\\
id0049&   1.905&    0.158&  12.847&    0.009&    1.772&    0.011\\
id0050&  -8.544&    0.891&  12.880&    0.014&    2.080&    0.015\\
id0051&  -0.657&    8.911&  12.880&    0.006&    0.682&    0.007\\
id0052&  -2.578&    3.890&  12.881&    0.011&    1.718&    0.013\\
id0053&  -1.871&    3.790&  12.949&    0.010&    1.723&    0.012\\
id0054&   2.109&    0.193&  12.950&    0.010&    1.696&    0.012\\
id0055&  -7.054&   -1.049&  12.958&    0.014&    1.949&    0.016\\
id0056&  -4.869&   -2.917&  12.984&    0.013&    1.969&    0.016\\
id0057&  -2.010&   -0.300&  12.991&    0.010&    2.008&    0.019\\
id0058&   2.836&   -5.226&  13.024&    0.006&    1.726&    0.007\\
id0059&  -1.038&    1.330&  13.030&    0.009&    1.717&    0.011\\
id0060&   0.275&    0.566&  13.041&    0.011&    1.674&    0.013\\
id0061&  -1.549&    0.542&  13.050&    0.017&    1.402&    0.033\\
id0062&   1.943&    0.635&  13.053&    0.006&    1.688&    0.008\\
id0063&   0.003&   -6.503&  13.058&    0.011&    1.932&    0.013\\
id0064&   0.726&   -5.128&  13.106&    0.009&    1.945&    0.012\\
id0065&   6.410&    1.191&  13.113&    0.026&    1.513&    0.030\\
id0066&  -6.414&   -5.701&  13.124&    0.015&    2.231&    0.018\\
id0067&  -1.375&   -0.985&  13.149&    0.009&    1.774&    0.011\\
id0068&   4.860&   -7.045&  13.196&    0.005&    1.851&    0.006\\
id0069&   2.625&    9.212&  13.197&    0.005&    2.064&    0.008\\
id0070&  -0.438&    1.008&  13.199&    0.010&    1.688&    0.012\\
id0071&   1.501&    1.336&  13.205&    0.009&    1.640&    0.010\\
id0072&  -1.958&    5.401&  13.218&    0.010&    1.771&    0.012\\
id0073&  -7.213&    2.378&  13.220&    0.014&    2.259&    0.016\\
id0074&  -2.550&   -6.959&  13.235&    0.014&    2.174&    0.017\\
id0075&   1.336&    0.731&  13.258&    0.006&    1.675&    0.009\\
id0076&  -7.333&   -0.185&  13.280&    0.013&    1.977&    0.015\\
id0077&  -7.401&    4.067&  13.341&    0.012&    2.000&    0.014\\
id0078&  -1.286&    4.213&  13.354&    0.010&    1.709&    0.012\\
id0079&   1.187&   -3.782&  13.389&    0.008&    1.706&    0.010\\
id0080&  -3.697&    3.447&  13.412&    0.012&    1.859&    0.014\\
id0081&   4.376&    3.320&  13.414&    0.004&    1.535&    0.006\\
id0082&  -1.395&    3.913&  13.452&    0.011&    1.710&    0.013\\
id0083&  -0.591&   -1.534&  13.491&    0.012&    1.762&    0.015\\
id0084&  -9.629&   -2.752&  13.569&    0.012&    1.989&    0.104\\
id0085&  -2.148&   -0.385&  13.572&    0.009&    1.845&    0.013\\
id0086&  -1.346&   -0.225&  13.574&    0.010&    1.794&    0.013\\
id0087&   5.773&    2.852&  13.575&    0.005&    1.546&    0.007\\
id0088&   3.833&    2.099&  13.579&    0.006&    1.549&    0.008\\
id0089&  -8.411&    2.968&  13.591&    0.012&    2.001&    0.014\\
id0090&   3.317&   -8.356&  13.616&    0.006&    1.949&    0.008\\
id0091&  -2.470&    0.458&  13.640&    0.012&    1.776&    0.014\\
id0092&  -0.359&    8.501&  13.649&    0.006&    0.747&    0.007\\
id0093&  -1.834&    1.881&  13.677&    0.012&    1.694&    0.015\\
id0094&  -1.536&    4.136&  13.687&    0.012&    1.725&    0.014\\
id0095&  -3.245&   -2.301&  13.697&    0.015&    1.783&    0.017\\
id0096&   2.446&    9.567&  13.735&    0.005&    1.696&    0.009\\
id0097&  -7.959&    3.029&  13.762&    0.013&    2.035&    0.015\\
id0098&  -0.655&    4.158&  13.790&    0.009&    1.714&    0.012\\
id0099&  -2.223&   -5.508&  13.795&    0.015&    2.056&    0.018\\
id0100&   2.381&    1.403&  13.798&    0.006&    1.635&    0.011\\
id0101&  -3.484&    3.213&  13.814&    0.011&    1.814&    0.014\\
id0102&  -2.207&    2.177&  13.826&    0.011&    1.775&    0.014\\
id0103&  -0.793&    3.409&  13.842&    0.009&    1.834&    0.012\\
id0104&   1.750&    1.024&  13.852&    0.007&    1.628&    0.013\\
id0105&   1.752&   -0.719&  13.864&    0.008&    1.694&    0.012\\
id0106&   0.449&   -7.577&  13.869&    0.010&    2.066&    0.013\\
id0107&   0.906&   -9.522&  13.881&    0.009&    2.117&    0.011\\
id0108&   8.076&   -5.405&  13.897&    0.005&    2.730&    0.017\\
id0109&   2.878&   -2.287&  13.905&    0.006&    1.742&    0.010\\
id0110&  -6.050&   -6.996&  13.910&    0.015&    2.132&    0.021\\
id0111&  -2.280&    2.292&  13.911&    0.016&    1.717&    0.019\\
id0112&  -0.242&   -7.042&  13.936&    0.012&    2.020&    0.016\\
id0113&  -5.510&    0.861&  13.952&    0.011&    1.865&    0.014\\
id0114&  -3.419&   -5.262&  13.954&    0.016&    2.018&    0.021\\
id0115&   0.805&   -1.893&  14.030&    0.015&    1.729&    0.020\\
id0116&   1.928&    5.524&  14.037&    0.007&    4.002&    0.060\\
id0117&  -2.224&    0.864&  14.121&    0.011&    1.804&    0.017\\
id0118&  -3.876&    3.425&  14.130&    0.013&    1.828&    0.019\\
id0119&  -3.546&    4.130&  14.139&    0.011&    1.141&    0.012\\
id0120&   0.866&   -4.469&  14.158&    0.010&    1.857&    0.016\\
id0121&   1.223&    3.237&  14.177&    0.010&    1.718&    0.013\\
id0122&   7.688&   -2.964&  14.178&    0.006&    1.760&    0.012\\
id0123&   7.147&    4.649&  14.201&    0.005&    0.647&    0.006\\
id0124&   0.225&    1.823&  14.202&    0.009&    1.743&    0.017\\
id0125&  -0.860&    1.440&  14.206&    0.013&    1.681&    0.018\\
id0126&  -0.807&    2.015&  14.217&    0.012&    1.835&    0.022\\
id0127&  -4.194&    2.911&  14.235&    0.013&    1.770&    0.018\\
id0128&  -5.292&   -3.703&  14.248&    0.016&    2.062&    0.021\\
id0129&   1.451&    2.746&  14.257&    0.009&    1.524&    0.014\\
id0130&  -1.781&    0.536&  14.270&    0.022&    1.722&    0.029\\
id0131&   2.708&   -7.186&  14.276&    0.006&    4.415&    0.076\\
id0132&   3.642&    3.657&  14.313&    0.008&    1.479&    0.012\\
id0133&   0.695&    0.901&  14.368&    0.011&    1.721&    0.017\\
id0134&  -2.663&    1.372&  14.397&    0.017&    1.782&    0.021\\
id0135&  -1.616&    4.687&  14.399&    0.009&    1.692&    0.014\\
id0136&   7.074&    5.101&  14.449&    0.007&    1.545&    0.015\\
id0137&   1.068&   -0.496&  14.475&    0.008&    1.731&    0.015\\
id0138&   3.412&    4.794&  14.504&    0.010&    2.058&    0.033\\
id0139&  -4.459&   -3.387&  14.507&    0.011&    1.870&    0.021\\
id0140&  -5.526&    0.256&  14.533&    0.014&    1.905&    0.020\\
id0141&  -2.497&    0.813&  14.565&    0.014&    1.680&    0.018\\
id0142&   6.207&   -1.525&  14.569&    0.007&    1.829&    0.016\\
id0143&  -0.481&   -2.451&  14.573&    0.013&    1.777&    0.020\\
id0144&  -4.192&    2.580&  14.580&    0.016&    1.788&    0.031\\
id0145&  -0.457&    2.511&  14.582&    0.009&    1.801&    0.022\\
id0146&  -5.147&    2.979&  14.584&    0.012&    1.769&    0.018\\
id0147&  -0.622&   -2.216&  14.611&    0.016&    1.754&    0.021\\
id0148&  -3.480&    0.505&  14.618&    0.015&    2.040&    0.023\\
id0149&  -4.471&   -7.097&  14.633&    0.036&    2.069&    0.039\\
id0150&   1.600&   -7.755&  14.643&    0.011&    1.955&    0.018\\
id0151&  -0.269&    3.630&  14.651&    0.009&    1.620&    0.014\\
id0152&  -3.252&    0.229&  14.658&    0.013&    1.847&    0.018\\
id0153&  -0.374&   -3.065&  14.695&    0.015&    1.825&    0.023\\
id0154&  -1.054&   -2.844&  14.729&    0.009&    1.782&    0.028\\
id0155&  -1.905&    1.986&  14.738&    0.015&    1.706&    0.023\\
id0156&   2.145&    3.637&  14.742&    0.012&    1.614&    0.024\\
id0157&  -3.805&    1.962&  14.761&    0.011&    2.952&    0.048\\
id0158&   4.127&   -1.340&  14.765&    0.008&    1.726&    0.017\\
id0159&  -1.011&    9.773&  14.797&    0.009&    1.983&    0.026\\
id0160&  -5.157&   -1.740&  14.804&    0.012&    3.032&    0.078\\
id0161&   6.799&    2.287&  14.842&    0.010&    1.593&    0.021\\
id0162&  -3.578&    1.984&  14.845&    0.011&    1.823&    0.021\\
id0163&   6.380&   -0.399&  14.873&    0.006&    1.944&    0.022\\
id0164&  -4.092&   -2.576&  14.899&    0.011&    2.005&    0.041\\
id0165&  -8.057&    0.698&  14.908&    0.016&    2.141&    0.036\\
id0166&  -1.119&    9.788&  14.912&    0.009&    1.574&    0.018\\
id0167&   3.940&   -0.075&  14.948&    0.009&    2.264&    0.026\\
id0168&  -1.240&    5.197&  14.962&    0.008&    1.661&    0.020\\
id0169&   1.583&   -0.683&  14.972&    0.017&    1.700&    0.029\\
id0170&   6.151&    2.920&  14.979&    0.009&    1.528&    0.018\\
id0171&  -1.167&    1.027&  14.997&    0.015&    2.177&    0.041\\
id0172&  -7.837&   -5.839&  15.007&    0.018&    3.112&    0.071\\
id0173&   2.903&   -3.966&  15.008&    0.010&    1.779&    0.021\\
id0174&   1.422&   -0.205&  15.011&    0.014&    2.273&    0.055\\
id0175&   1.296&   -7.982&  15.013&    0.011&    2.080&    0.028\\
id0176&  -2.106&   -7.463&  15.023&    0.017&    2.238&    0.029\\
id0177&  -2.406&   -2.980&  15.024&    0.011&    2.125&    0.040\\
id0178&   5.792&    2.385&  15.044&    0.010&    1.580&    0.021\\
id0179&   3.561&    5.697&  15.058&    0.012&    2.967&    0.067\\
id0180&  -1.274&    7.131&  15.060&    0.010&    1.872&    0.026\\
id0181&  -0.167&   -0.708&  15.071&    0.026&    1.502&    0.033\\
id0182&  -0.651&    3.358&  15.084&    0.014&    1.551&    0.025\\
id0183&   2.599&   -5.086&  15.097&    0.013&    2.330&    0.033\\
id0184&  -4.830&    1.231&  15.109&    0.022&    2.519&    0.053\\
id0185&   0.853&    4.741&  15.122&    0.015&    1.619&    0.026\\
id0186&  -5.013&    5.589&  15.134&    0.015&    2.925&    0.057\\
id0187&   1.985&   -1.052&  15.136&    0.010&    1.724&    0.022\\
id0188&   6.879&   -3.483&  15.143&    0.010&    2.006&    0.035\\
id0189&   0.756&   -6.453&  15.146&    0.012&    2.074&    0.026\\
id0190&   0.473&   -2.772&  15.159&    0.008&    1.892&    0.025\\
id0191&   2.299&    4.750&  15.173&    0.010&    2.932&    0.048\\
id0192&   1.081&   -3.430&  15.188&    0.010&    1.902&    0.030\\
id0193&  -6.927&   -2.706&  15.199&    0.012&    2.377&    0.038\\
id0194&  -3.092&   -6.313&  15.202&    0.015&    2.158&    0.030\\
id0195&  -8.804&   -0.157&  15.216&    0.015&    2.095&    0.035\\
id0196&  -0.448&    1.722&  15.228&    0.016&    1.881&    0.037\\
id0197&  -1.286&   -2.717&  15.229&    0.010&    1.979&    0.046\\
id0198&   7.207&   -4.669&  15.247&    0.009&    1.795&    0.021\\
id0199&  -0.657&    6.140&  15.268&    0.010&    1.679&    0.021\\
id0200&   6.956&    4.920&  15.288&    0.012&    2.235&    0.040\\
id0201&  -4.692&   -1.085&  15.296&    0.012&    1.808&    0.028\\
id0202&  -3.284&   -0.151&  15.312&    0.015&    1.797&    0.029\\
id0203&   4.486&   -7.436&  15.331&    0.011&    2.202&    0.039\\
id0204&   0.234&   10.004&  15.333&    0.010&    1.311&    0.020\\
id0205&  -6.975&    2.327&  15.338&    0.026&    3.059&    0.084\\
id0206&  -3.042&    5.744&  15.366&    0.014&    2.053&    0.037\\
id0207&  -2.688&    0.521&  15.375&    0.022&    2.086&    0.038\\
id0208&  -6.241&    7.122&  15.379&    0.015&    1.844&    0.029\\
id0209&  -0.004&    0.001&  15.396&    0.009&    2.146&    0.039\\
id0210&  -4.892&   -3.533&  15.400&    0.016&    2.133&    0.038\\
id0211&  -4.286&    3.843&  15.405&    0.012&    2.180&    0.049\\
id0212&   1.356&   -1.177&  15.408&    0.039&    1.879&    0.074\\
id0213&  -2.851&   -1.611&  15.414&    0.023&    2.568&    0.078\\
id0214&  -9.614&    1.339&  15.414&    0.096&    2.123&    0.102\\
id0215&   0.482&    0.011&  15.437&    0.016&    1.581&    0.034\\
id0216&   2.252&    4.021&  15.440&    0.013&    1.693&    0.040\\
id0217&   7.071&   -0.910&  15.444&    0.008&    1.950&    0.029\\
id0218&  -9.785&   -1.112&  15.486&    0.014&    1.838&    0.039\\
id0219&  -3.328&    6.849&  15.496&    0.014&    1.851&    0.034\\
id0220&   5.155&   -3.139&  15.502&    0.012&    2.209&    0.050\\
id0221&   2.423&   -0.196&  15.507&    0.009&    2.153&    0.047\\
id0222&  -0.066&   -6.032&  15.525&    0.015&    1.969&    0.038\\
id0223&   2.726&    7.833&  15.525&    0.012&    1.986&    0.106\\
id0224&   6.919&   -5.023&  15.526&    0.013&    1.972&    0.038\\
id0225&  -4.561&   -3.109&  15.529&    0.018&    2.004&    0.052\\
id0226&   5.492&    6.194&  15.534&    0.015&    1.928&    0.034\\
id0227&   2.740&   -9.153&  15.565&    0.009&    2.184&    0.053\\
id0228&  -3.241&   -4.224&  15.567&    0.018&    2.003&    0.038\\
id0229&  -4.513&    0.464&  15.573&    0.013&    2.812&    0.082\\
id0230&   4.064&    1.734&  15.585&    0.043&    1.479&    0.058\\
id0231&   3.382&    4.549&  15.600&    0.015&    1.910&    0.046\\
id0232&  -2.873&    0.396&  15.606&    0.014&    2.035&    0.035\\
id0233&  -1.395&    2.757&  15.628&    0.015&    1.828&    0.046\\
id0234&  -6.542&    1.096&  15.637&    0.021&    2.094&    0.046\\
id0235&   5.400&   -6.818&  15.649&    0.013&    2.398&    0.047\\
id0236&   2.428&   -6.272&  15.657&    0.016&    3.307&    0.072\\
id0237&   1.967&    3.773&  15.660&    0.017&    1.601&    0.045\\
id0238&  -0.932&   -0.032&  15.670&    0.018&    1.839&    0.059\\
id0239&   5.185&    4.245&  15.680&    0.013&    1.946&    0.044\\
id0240&  -1.880&    5.847&  15.707&    0.019&    2.042&    0.046\\
id0241&  -0.035&   -5.131&  15.717&    0.016&    1.911&    0.048\\
id0242&  -2.854&   -1.800&  15.726&    0.026&    2.270&    0.108\\
id0243&   5.552&    2.330&  15.735&    0.014&    1.681&    0.041\\
id0244&   2.039&   -1.210&  15.736&    0.017&    2.123&    0.053\\
id0245&   0.105&    8.169&  15.737&    0.011&    2.837&    0.120\\
id0246&   0.972&    2.596&  15.740&    0.032&    1.748&    0.053\\
id0247&   2.834&   -0.143&  15.744&    0.015&    2.151&    0.074\\
id0248&  -4.716&   -0.699&  15.745&    0.020&    2.033&    0.056\\
id0249&  -6.006&    7.918&  15.774&    0.012&    1.906&    0.035\\
id0250&  -7.731&    1.105&  15.786&    0.023&    2.222&    0.051\\
id0251&   8.128&   -3.621&  15.800&    0.024&    1.795&    0.055\\
id0252&   7.030&    3.315&  15.812&    0.016&    1.446&    0.038\\
id0253&   3.328&    2.797&  15.815&    0.015&    1.959&    0.069\\
id0254&  -3.436&    0.126&  15.831&    0.022&    1.998&    0.056\\
id0255&   2.466&    4.711&  15.842&    0.020&    3.188&    0.144\\
id0256&  -5.805&   -5.538&  15.848&    0.019&    2.234&    0.048\\
id0257&   8.255&   -1.454&  15.868&    0.015&    1.844&    0.042\\
id0258&  -2.892&    6.817&  15.869&    0.019&    1.838&    0.046\\
id0259&   0.217&   -2.406&  15.877&    0.016&    2.062&    0.062\\
id0260&   1.133&   -7.871&  15.888&    0.020&    2.208&    0.071\\
id0261&   6.665&    2.110&  15.896&    0.014&    0.661&    0.021\\
id0262&  -1.813&   -6.310&  15.908&    0.018&    2.291&    0.069\\
id0263&   0.253&   -8.896&  15.920&    0.022&    2.287&    0.066\\
id0264&  -3.745&   -6.847&  15.929&    0.017&    2.538&    0.083\\
id0265&  -2.605&   -1.798&  15.937&    0.071&    2.140&    0.115\\
id0266&   6.580&    4.830&  15.970&    0.015&    1.726&    0.053\\
id0267&   5.794&    1.132&  15.981&    0.017&    1.728&    0.054\\
id0268&  -6.037&   -2.792&  15.986&    0.020&    2.272&    0.073\\
id0269&  -3.382&    0.369&  15.995&    0.027&    2.108&    0.063\\
id0270&  -1.036&    2.373&  15.998&    0.026&    1.775&    0.068\\
id0271&   0.669&    8.249&  16.009&    0.016&    2.337&    0.089\\
id0272&  -0.312&    1.159&  16.021&    0.025&    1.185&    0.056\\
id0273&   6.336&   -1.678&  16.032&    0.017&    0.816&    0.024\\
id0274&  -4.904&    2.488&  16.037&    0.028&    1.776&    0.052\\
id0275&  -4.324&    6.381&  16.039&    0.028&    2.467&    0.095\\
id0276&   1.955&    9.096&  16.039&    0.015&    3.310&    0.238\\
id0277&   6.055&   -6.810&  16.052&    0.025&    2.001&    0.061\\
id0278&   6.583&   -5.200&  16.057&    0.015&    1.982&    0.057\\
id0279&  -7.009&   -3.155&  16.060&    0.026&    2.485&    0.055\\
id0280&  -4.231&    6.413&  16.062&    0.029&    2.315&    0.122\\
id0281&  -2.000&   -5.202&  16.084&    0.029&    2.799&    0.125\\
id0282&  -4.281&   -3.535&  16.090&    0.016&    1.977&    0.048\\
id0283&  -5.971&   -2.457&  16.115&    0.020&    1.830&    0.062\\
id0284&  -4.984&   -3.249&  16.116&    0.020&    1.036&    0.039\\
id0285&  -0.204&    0.459&  16.118&    0.025&    1.373&    0.058\\
id0286&  -1.081&   -2.496&  16.138&    0.023&    2.344&    0.125\\
id0287&   6.611&    0.551&  16.145&    0.017&    1.557&    0.045\\
id0288&  -1.383&    6.530&  16.149&    0.017&    1.765&    0.057\\
id0289&   5.943&   -0.540&  16.154&    0.014&    1.959&    0.066\\
id0290&   7.995&    0.267&  16.163&    0.024&    0.293&    0.030\\
id0291&  -0.730&   -2.501&  16.175&    0.025&    1.440&    0.073\\
id0292&  -0.266&   -2.945&  16.176&    0.032&    1.878&    0.062\\
id0293&  -0.686&   -7.271&  16.196&    0.024&    2.392&    0.090\\
id0294&  -2.066&    3.987&  16.204&    0.026&    2.114&    0.090\\
id0295&  -0.980&    9.187&  16.211&    0.021&    1.570&    0.064\\
id0296&  -0.530&    0.290&  16.236&    0.046&    1.896&    0.104\\
id0297&  -2.390&   -8.054&  16.248&    0.028&    2.259&    0.088\\
id0298&  -8.167&    2.639&  16.261&    0.029&    1.919&    0.075\\
id0299&  -1.022&   -3.939&  16.267&    0.020&    2.328&    0.084\\
id0300&   4.758&    1.882&  16.280&    0.020&    1.680&    0.063\\
id0301&   5.558&   -0.150&  16.281&    0.021&    2.418&    0.094\\
id0302&  -1.479&   -1.330&  16.299&    0.042&    1.943&    0.132\\
id0303&  -1.988&    2.369&  16.299&    0.030&    1.639&    0.059\\
id0304&   0.512&    5.347&  16.305&    0.028&    1.705&    0.075\\
id0305&   8.477&   -1.554&  16.309&    0.022&    2.018&    0.090\\
id0306&  -2.004&    9.312&  16.310&    0.025&    0.378&    0.030\\
id0307&  -5.065&    1.679&  16.312&    0.028&    2.104&    0.098\\
id0308&   4.944&   -0.525&  16.314&    0.018&    1.908&    0.065\\
id0309&  -0.977&    6.017&  16.325&    0.023&    1.881&    0.059\\
id0310&  -3.419&   -6.735&  16.329&    0.024&    2.785&    0.128\\
id0311&  -4.426&   -0.180&  16.340&    0.031&    2.165&    0.091\\
id0312&  -8.930&    1.109&  16.346&    0.028&    2.257&    0.088\\
id0313&   1.218&    5.624&  16.349&    0.020&    1.570&    0.051\\
id0314&  -2.466&   -7.897&  16.350&    0.019&    2.398&    0.099\\
id0315&  -0.688&    5.000&  16.364&    0.026&    1.945&    0.080\\
id0316&  -6.607&    2.850&  16.370&    0.025&    2.005&    0.067\\
id0317&   5.600&   -1.934&  16.382&    0.028&    2.061&    0.073\\
id0318&   5.955&   -0.797&  16.388&    0.017&    1.736&    0.074\\
id0319&   5.285&   -3.970&  16.423&    0.024&    2.052&    0.094\\
id0320&  -5.093&   -4.685&  16.484&    0.037&    2.857&    0.179\\
id0321&   3.461&   -2.101&  16.484&    0.026&    1.856&    0.071\\
id0322&   4.615&   -3.610&  16.493&    0.024&    2.125&    0.095\\
id0323&  -5.759&    3.164&  16.501&    0.026&    2.092&    0.077\\
id0324&   3.259&   -0.318&  16.522&    0.026&    1.787&    0.100\\
id0325&  -6.541&    3.187&  16.529&    0.027&    2.662&    0.127\\
id0326&   2.435&   -3.666&  16.545&    0.025&    1.670&    0.074\\
id0327&   2.978&   -9.335&  16.576&    0.022&    1.783&    0.067\\
id0328&   0.466&   -1.495&  16.583&    0.067&    1.581&    0.101\\
id0329&  -7.548&    6.320&  16.593&    0.029&    2.490&    0.148\\
id0330&   4.066&   -0.814&  16.594&    0.032&    1.597&    0.115\\
id0331&   5.298&    0.736&  16.602&    0.026&    1.661&    0.084\\
id0332&   2.232&   -2.787&  16.639&    0.028&    2.052&    0.140\\
id0333&   7.511&    0.632&  16.654&    0.026&    1.474&    0.074\\
id0334&   8.360&    2.662&  16.660&    0.027&    1.719&    0.091\\
id0335&  -0.271&    0.773&  16.712&    0.046&    1.263&    0.091\\
id0336&  -3.036&   -6.028&  16.714&    0.025&    2.066&    0.112\\
id0337&  -3.334&    8.828&  16.714&    0.044&    2.916&    0.157\\
id0338&   0.998&   -8.957&  16.718&    0.024&    2.283&    0.128\\
id0339&  -8.387&    1.636&  16.726&    0.026&    2.495&    0.118\\
id0340&  -8.690&    2.879&  16.732&    0.048&    1.846&    0.105\\
id0341&  -2.891&    4.370&  16.734&    0.040&    1.714&    0.092\\
id0342&   6.584&   -6.457&  16.750&    0.034&    1.731&    0.083\\
id0343&  -1.726&    2.300&  16.754&    0.035&    1.721&    0.108\\
id0344&  -5.555&   -5.459&  16.766&    0.050&    0.648&    0.063\\
id0345&   3.098&   -2.578&  16.775&    0.033&    1.619&    0.088\\
id0346&  -4.323&    6.206&  16.782&    0.028&    2.097&    0.136\\
id0347&  -3.641&   -1.107&  16.832&    0.043&    1.839&    0.102\\
id0348&   3.321&   -2.584&  16.837&    0.039&    2.083&    0.122\\
id0349&   8.412&    1.040&  16.876&    0.029&    1.752&    0.115\\
id0350&   7.851&    5.401&  16.879&    0.034&    1.017&    0.076\\
id0351&   2.607&   -3.357&  16.880&    0.038&    1.771&    0.113\\
id0352&  -6.507&    1.605&  16.886&    0.031&    1.957&    0.096\\
id0353&   3.983&    1.572&  16.887&    0.037&    1.612&    0.104\\
id0354&  -6.891&    3.352&  16.896&    0.032&    2.994&    0.329\\
id0355&   0.170&   -1.331&  16.936&    0.045&    1.631&    0.104\\
id0356&  -7.227&    3.446&  16.950&    0.032&    1.935&    0.101\\
id0357&   3.868&   -3.492&  16.978&    0.033&    2.082&    0.150\\
id0358&   7.446&    2.555&  16.982&    0.029&    1.397&    0.090\\
id0359&   4.309&   -4.269&  17.009&    0.045&    2.226&    0.184\\
id0360&   1.302&    4.909&  17.018&    0.038&    1.306&    0.092\\
id0361&  -2.174&    4.629&  17.021&    0.042&    2.022&    0.158\\
id0362&   3.333&    4.248&  17.076&    0.046&    1.355&    0.130\\
id0363&   2.100&    6.486&  17.117&    0.095&    1.351&    0.125\\
id0364&   2.809&    8.602&  17.118&    0.053&    2.008&    0.178\\
id0365&   0.087&   -6.062&  17.121&    0.050&    2.344&    0.207\\
id0366&   1.055&    4.741&  17.132&    0.075&    1.416&    0.105\\
id0367&   5.169&   -6.118&  17.135&    0.032&    1.871&    0.162\\
id0368&   7.140&   -0.419&  17.137&    0.040&    2.154&    0.187\\
id0369&  -1.937&   -6.048&  17.141&    0.053&    2.396&    0.187\\
id0370&  -6.360&   -3.015&  17.141&    0.044&    3.571&    0.444\\
id0371&  -1.988&   -3.227&  17.155&    0.087&    1.721&    0.173\\
id0372&  -7.392&   -5.114&  17.166&    0.044&    2.185&    0.210\\
id0373&  -8.268&   -0.982&  17.166&    0.047&    2.104&    0.177\\
id0374&   3.922&    6.630&  17.181&    0.036&    1.795&    0.148\\
id0375&   4.596&    6.833&  17.181&    0.032&    2.390&    0.281\\
id0376&  -8.634&    0.167&  17.196&    0.055&    1.910&    0.156\\
id0377&   6.033&    6.457&  17.207&    0.082&    1.658&    0.167\\
id0378&  -3.615&   -7.043&  17.252&    0.034&    2.534&    0.244\\
id0379&  -2.737&    7.874&  17.271&    0.043&    1.409&    0.109\\
id0380&  -2.119&    5.214&  17.289&    0.054&    1.947&    0.162\\
id0381&   7.909&    0.725&  17.310&    0.044&    1.848&    0.186\\
id0382&  -6.028&    2.026&  17.310&    0.045&    2.701&    0.294\\
id0383&   3.969&    9.196&  17.343&    0.060&    1.754&    0.136\\
id0384&  -3.872&    4.950&  17.355&    0.058&    2.002&    0.240\\
id0385&   8.482&    5.213&  17.359&    0.049&    1.988&    0.220\\
id0386&  -5.518&   -0.683&  17.360&    0.050&    2.118&    0.195\\
id0387&  -1.939&   -8.657&  17.398&    0.046&    2.532&    0.251\\
id0388&   5.169&   -3.505&  17.409&    0.051&    1.478&    0.135\\
id0389&  -0.338&   -4.483&  17.410&    0.061&    1.567&    0.157\\
id0390&   4.865&   -2.094&  17.481&    0.061&    2.453&    0.337\\
id0391&   5.389&    7.068&  17.484&    0.054&    2.297&    0.305\\
id0392&  -5.070&    6.652&  17.499&    0.053&    1.727&    0.158\\
id0393&  -1.980&   -3.614&  17.510&    0.078&    1.297&    0.132\\
id0394&  -8.024&   -2.261&  17.515&    0.078&    2.366&    0.281\\
id0395&  -9.574&    0.329&  17.518&    0.047&    3.088&    0.544\\
id0396&  -4.076&   -0.389&  17.531&    0.089&    2.275&    0.282\\
id0397&   2.563&   -7.347&  17.556&    0.062&    0.175&    0.094\\
id0398&   6.430&    6.857&  17.564&    0.054&    0.534&    0.082\\
id0399&   3.557&   -8.590&  17.573&    0.061&    1.775&    0.170\\
id0400&   6.044&    6.811&  17.580&    0.052&    1.167&    0.167\\
id0401&   8.823&    0.041&  17.590&    0.057&    1.489&    0.181\\
id0402&   3.607&   -1.340&  17.614&    0.072&    1.663&    0.221\\
id0403&  -1.224&    8.143&  17.628&    0.081&    2.411&    0.417\\
id0404&   3.084&    7.874&  17.647&    0.060&    1.926&    0.221\\
id0405&  -2.233&    9.026&  17.657&    0.070&    1.229&    0.153\\
id0406&  -5.529&    7.619&  17.778&    0.079&    1.588&    0.225\\
id0407&   1.220&    1.051&  17.842&    0.098&    0.951&    0.150\\
id0408&   0.470&   -7.871&  17.845&    0.080&    0.586&    0.110\\
id0409&   1.175&    6.604&  17.893&    0.058&    1.367&    0.210\\
id0410&   7.685&   -4.464&  17.950&    0.090&    2.225&    0.469\\
id0411&  -4.246&   -6.333&  17.955&    0.091&    2.421&    0.425\\
id0412&   0.494&   -3.373&  17.969&    0.082&    1.150&    0.151\\
id0413&   3.595&   -3.688&  18.034&    0.094&    2.312&    0.481\\
id0414&   0.495&    9.876&  18.069&    0.090&    1.761&    0.286\\
id0415&   6.741&    0.065&  18.107&    0.111&    1.266&    0.244\\
id0416&  -5.782&   -7.968&  18.341&    0.096&    1.160&    0.231\\
id0417&  -6.974&   -2.266&  18.354&    0.147&    1.567&    0.311\\
id0418&  -3.265&   -3.706&  18.360&    0.190&    2.417&    0.572\\
id0419&  -0.259&   -6.722&  18.493&    0.198&    1.422&    0.401\\
id0420&   0.820&    5.119&  18.559&    0.187&    1.442&    0.499\\
id0421&  -6.373&   -2.179&  18.643&    0.154&    1.038&    0.288\\
id0422&  -9.857&    1.767&  18.846&    0.157&    1.472&    0.369\\
id0423&  -7.953&    0.074&  18.920&    0.172&    1.258&    0.437\\
id0424&  -7.118&   -2.223&  19.131&    0.235&    1.462&    0.566\\
id0425&  -5.280&    0.409&  19.520&    0.387&    0.097&    0.444\\
id0426&  -7.437&    1.096&  19.767&    0.513&    0.455&    0.641\\
id0427&   8.219&   -2.979&  19.781&    0.483&   -0.637&    0.511\\
\hline    
\end{longtable}
\begin{list}{}{}
\item[$^{\mathrm{a}}$] In this catalog the star ID is designated by sorting the data in order of increasing $K_{S}$ magnitude.
\item[$^{\mathrm{b}}$] Positions with respect to $\alpha = 17^{h}\;45^{m}\;50.798^{s},\; \delta = -28^{\circ}\;49^{'}\;25.606^{''}\;$ (J2000).
\item[$^{\mathrm{c}}$] NACO $K_{S}$ band magnitude.
\item[$^{\mathrm{d}}$] The 1{$\sigma$} uncertainty in $K_{S}$.
\item[$^{\mathrm{e}}$] NACO $(H-K_{S})$ color index.
\item[$^{\mathrm{f}}$] The 1{$\sigma$} uncertainty in $(H-K_{S})$.
\end{list}
}

\addtocounter{table}{1} 
\onllongtab{3}{
\begin{longtable}{c c c c c c c c c}
\caption{Photometry of $JHK_{S}$ stars. The catalog contains point sources within $10''$ from the cluster center.}
\label{ol143}\\
Star ID $^{a}$  & $\Delta RA$ $^{b}$ $['']$ & $\Delta DEC$ $^{b}$ $['']$ & $K_{S}$ $^{c}$ & $\sigma_{K_{S}}$ $^{d}$ & $(H-K_{S})$ $^{e}$ & $\sigma_{(H-K_{S})}$ $^{f}$ & $(J-H)$ $^{g}$ & $\sigma_{(J-H)}$ $^{h}$\\ 
\hline 
\endfirsthead
\caption{Continued.} \\ 
\hline 
Star ID $^{a}$  & $\Delta RA$ $^{b}$ $['']$ & $\Delta DEC$ $^{b}$ $['']$ & $K_{S}$ $^{c}$ & $\sigma_{K_{S}}$ $^{d}$ & $(H-K_{S})$ $^{e}$ & $\sigma_{(H-K_{S})}$ $^{f}$ & $(J-H)$ $^{g}$ & $\sigma_{(J-H)}$ $^{h}$\\ 
\hline
\endhead
\hline
\endfoot    
\hline
\endlastfoot
id0001 &     -1.897&   -2.321&   10.068&    0.006&    1.957&    0.009&    3.203&    0.009 \\
id0002 &      2.601&    5.739&    10.216&    0.003&    2.200&    0.004&    4.048&    0.007 \\
id0003 &      0.035&    2.397&    10.244&    0.004&    1.841&    0.006&    3.021&    0.006 \\
id0004 &     -1.231&    0.439&    10.281&    0.005&    1.861&    0.008&    3.064&    0.007 \\
id0005 &     -4.724&   -2.274&    10.345&    0.015&    2.099&    0.019&    3.418&    0.013 \\
id0006 &      3.425&   -6.441&    10.362&    0.005&    1.993&    0.005&    3.220&    0.003 \\
id0007 &     -2.313&   -1.276&    10.456&    0.008&    1.997&    0.011&    3.214&    0.009 \\
id0008 &     -3.966&    8.233&    10.619&    0.006&    1.776&    0.006&    3.034&    0.003 \\
id0009 &     -3.778&    2.719&    10.878&    0.008&    1.928&    0.010&    3.161&    0.008 \\
id0010 &      1.475&   -2.617&    10.965&    0.006&    1.851&    0.008&    3.015&    0.007 \\
id0011 &      2.474&    3.383&    11.011&    0.004&    1.625&    0.005&    2.830&    0.006 \\
id0012 &     -6.597&   -6.550&    11.144&    0.012&    2.190&    0.014&    3.533&    0.011 \\
id0013 &     -0.551&   -0.702&    11.151&    0.009&    1.833&    0.011&    3.018&    0.009 \\
id0014 &      3.172&    1.102&    11.412&    0.036&    1.610&    0.040&    2.813&    0.019 \\
id0015 &     -1.211&    2.081&    11.502&    0.010&    1.701&    0.011&    2.899&    0.007 \\
id0016 &      2.593&    0.360&    11.518&    0.006&    1.703&    0.007&    2.896&    0.008 \\
id0017 &     -6.799&   -3.663&    11.546&    0.014&    2.098&    0.016&    3.425&    0.012 \\
id0018 &      0.547&    0.449&    11.788&    0.009&    1.769&    0.011&    2.993&    0.013 \\
id0019 &     -5.653&   -7.197&    11.854&    0.013&    2.204&    0.014&    3.590&    0.016 \\
id0020 &     -4.541&    3.291&    11.877&    0.011&    1.876&    0.012&    3.148&    0.010 \\
id0021 &      0.997&   -1.747&    11.935&    0.009&    1.747&    0.011&    2.898&    0.009 \\
id0022 &      7.738&   -3.396&    12.010&    0.005&    1.759&    0.006&    3.017&    0.010 \\
id0023 &      2.317&    2.331&    12.054&    0.006&    1.661&    0.008&    2.813&    0.009 \\
id0024 &     -0.167&   -3.573&    12.061&    0.011&    1.889&    0.013&    3.142&    0.012 \\
id0025 &      1.743&    2.971&    12.133&    0.006&    1.651&    0.008&    2.765&    0.009 \\
id0026 &      1.314&    0.071&    12.159&    0.009&    1.735&    0.011&    2.911&    0.011 \\
id0027 &      1.744&   -0.018&    12.160&    0.008&    1.747&    0.010&    2.933&    0.011 \\
id0028 &     -1.895&    0.316&    12.187&    0.011&    1.759&    0.013&    2.976&    0.013 \\
id0029 &      0.768&    0.110&    12.209&    0.008&    1.751&    0.010&    2.944&    0.014 \\
id0030 &      3.165&   -1.081&    12.266&    0.005&    1.729&    0.006&    3.004&    0.012 \\
id0032 &     -6.135&   -0.728&    12.392&    0.013&    1.998&    0.015&    3.361&    0.021 \\
id0033 &      0.828&    1.643&    12.412&    0.009&    1.700&    0.011&    2.996&    0.013 \\
id0034 &     -3.426&   -0.737&    12.621&    0.013&    1.822&    0.016&    3.107&    0.019 \\
id0035 &      1.027&   -5.139&    12.647&    0.009&    1.897&    0.010&    3.099&    0.015 \\
id0036 &      1.715&   -4.828&    12.647&    0.008&    1.818&    0.009&    3.001&    0.015 \\
id0037 &      1.187&   -4.556&    12.692&    0.009&    1.842&    0.010&    3.040&    0.021 \\
id0039 &     -1.656&   -3.784&    12.720&    0.013&    1.914&    0.015&    3.230&    0.020 \\
id0040 &      3.053&    2.314&    12.720&    0.006&    1.557&    0.008&    2.704&    0.012 \\
id0041 &      6.024&   -3.942&    12.727&    0.004&    2.230&    0.006&    4.367&    0.075 \\
id0042 &     -0.420&   -1.726&    12.765&    0.013&    1.748&    0.015&    3.037&    0.017 \\
id0044 &      2.163&   -2.146&    12.829&    0.007&    1.744&    0.008&    2.921&    0.016 \\
id0045 &      2.313&   -1.721&    12.832&    0.007&    1.793&    0.009&    3.113&    0.021 \\
id0046 &      6.788&    6.129&    12.832&    0.003&    1.595&    0.005&    2.917&    0.020 \\
id0047 &      2.959&    0.146&    12.844&    0.006&    1.719&    0.010&    2.917&    0.021 \\
id0048 &     -4.634&    4.691&    12.846&    0.011&    1.973&    0.013&    3.488&    0.032 \\
id0049 &      1.905&    0.158&    12.847&    0.009&    1.772&    0.011&    2.947&    0.019 \\
id0051 &     -0.657&    8.911&    12.880&    0.006&    0.682&    0.007&    1.167&    0.005 \\
id0052 &     -2.578&    3.890&    12.881&    0.011&    1.718&    0.013&    2.975&    0.017 \\
id0053 &     -1.871&    3.790&    12.949&    0.010&    1.723&    0.012&    2.941&    0.016 \\
id0054 &      2.109&    0.193&    12.950&    0.010&    1.696&    0.012&    2.896&    0.018 \\
id0056 &     -4.869&   -2.917&    12.984&    0.013&    1.969&    0.016&    3.334&    0.024 \\
id0057 &     -2.010&   -0.300&    12.991&    0.010&    2.008&    0.019&    3.060&    0.029 \\
id0058 &      2.836&   -5.226&    13.024&    0.006&    1.726&    0.007&    2.921&    0.017 \\
id0059 &     -1.038&    1.330&    13.030&    0.009&    1.717&    0.011&    2.780&    0.016 \\
id0060 &      0.275&    0.566&    13.041&    0.011&    1.674&    0.013&    2.978&    0.017 \\
id0061 &     -1.549&    0.542&    13.050&    0.017&    1.402&    0.033&    2.842&    0.035 \\
id0062 &      1.943&    0.635&    13.053&    0.006&    1.688&    0.008&    3.057&    0.025 \\
id0063 &      0.003&   -6.503&    13.058&    0.011&    1.932&    0.013&    3.113&    0.029 \\
id0064 &      0.726&   -5.128&    13.106&    0.009&    1.945&    0.012&    3.161&    0.029 \\
id0065 &      6.410&    1.191&    13.113&    0.026&    1.513&    0.030&    2.688&    0.021 \\
id0066 &     -6.414&   -5.701&    13.124&    0.015&    2.231&    0.018&    3.482&    0.051 \\
id0067 &     -1.375&   -0.985&    13.149&    0.009&    1.774&    0.011&    3.172&    0.027 \\
id0068 &      4.860&   -7.045&    13.196&    0.005&    1.851&    0.006&    3.072&    0.027 \\
id0070 &     -0.438&    1.008&    13.199&    0.010&    1.688&    0.012&    3.012&    0.028 \\
id0072 &     -1.958&    5.401&    13.218&    0.010&    1.771&    0.012&    3.135&    0.028 \\
id0073 &     -7.213&    2.378&    13.220&    0.014&    2.259&    0.016&    4.139&    0.093 \\
id0074 &     -2.550&   -6.959&    13.235&    0.014&    2.174&    0.017&    3.953&    0.159 \\
id0075 &      1.336&    0.731&    13.258&    0.006&    1.675&    0.009&    2.821&    0.023 \\
id0076 &     -7.333&   -0.185&    13.280&    0.013&    1.977&    0.015&    3.320&    0.038 \\
id0077 &     -7.401&    4.067&    13.341&    0.012&    2.000&    0.014&    3.230&    0.033 \\
id0078 &     -1.286&    4.213&    13.354&    0.010&    1.709&    0.012&    2.753&    0.025 \\
id0079 &      1.187&   -3.782&    13.389&    0.008&    1.706&    0.010&    2.959&    0.025 \\
id0081 &      4.376&    3.320&    13.414&    0.004&    1.535&    0.006&    2.758&    0.023 \\
id0082 &     -1.395&    3.913&    13.452&    0.011&    1.710&    0.013&    2.804&    0.022 \\
id0083 &     -0.591&   -1.534&    13.491&    0.012&    1.762&    0.015&    2.932&    0.030 \\
id0085 &     -2.148&   -0.385&    13.572&    0.009&    1.845&    0.013&    3.160&    0.040 \\
id0087 &      5.773&    2.852&    13.575&    0.005&    1.546&    0.007&    2.730&    0.022 \\
id0088 &      3.833&    2.099&    13.579&    0.006&    1.549&    0.008&    2.785&    0.025 \\
id0090 &      3.317&   -8.356&    13.616&    0.006&    1.949&    0.008&    3.196&    0.044 \\
id0091 &     -2.470&    0.458&    13.640&    0.012&    1.776&    0.014&    3.156&    0.041 \\
id0092 &     -0.359&    8.501&    13.649&    0.006&    0.747&    0.007&    1.270&    0.008 \\
id0093 &     -1.834&    1.881&    13.677&    0.012&    1.694&    0.015&    2.750&    0.041 \\
id0095 &     -3.245&   -2.301&    13.697&    0.015&    1.783&    0.017&    3.175&    0.041 \\
id0096 &      2.446&    9.567&    13.735&    0.005&    1.696&    0.009&    3.134&    0.046 \\
id0098 &     -0.655&    4.158&    13.790&    0.009&    1.714&    0.012&    3.058&    0.033 \\
id0099 &     -2.223&   -5.508&    13.795&    0.015&    2.056&    0.018&    3.182&    0.051 \\
id0100 &      2.381&    1.403&    13.798&    0.006&    1.635&    0.011&    2.867&    0.039 \\
id0101 &     -3.484&    3.213&    13.814&    0.011&    1.814&    0.014&    3.043&    0.041 \\
id0104 &      1.750&    1.024&    13.852&    0.007&    1.628&    0.013&    2.807&    0.033 \\
id0105 &      1.752&   -0.719&    13.864&    0.008&    1.694&    0.012&    2.678&    0.033 \\
id0107 &      0.906&   -9.522&    13.881&    0.009&    2.117&    0.011&    3.357&    0.094 \\
id0109 &      2.878&   -2.287&    13.905&    0.006&    1.742&    0.010&    2.978&    0.041 \\
id0112 &     -0.242&   -7.042&    13.936&    0.012&    2.020&    0.016&    3.537&    0.088 \\
id0114 &     -3.419&   -5.262&    13.954&    0.016&    2.018&    0.021&    3.499&    0.067 \\
id0117 &     -2.224&    0.864&    14.121&    0.011&    1.804&    0.017&    2.976&    0.054 \\
id0119 &     -3.546&    4.130&    14.139&    0.011&    1.141&    0.012&    1.856&    0.012 \\
id0120 &      0.866&   -4.469&    14.158&    0.010&    1.857&    0.016&    2.920&    0.072 \\
id0121 &      1.223&    3.237&    14.177&    0.010&    1.718&    0.013&    3.151&    0.060 \\
id0122 &      7.688&   -2.964&    14.178&    0.006&    1.760&    0.012&    3.386&    0.080 \\
id0123 &      7.147&    4.649&    14.201&    0.005&    0.647&    0.006&    1.212&    0.007 \\
id0128 &     -5.292&   -3.703&    14.248&    0.016&    2.062&    0.021&    3.700&    0.155 \\
id0129 &      1.451&    2.746&    14.257&    0.009&    1.524&    0.014&    2.951&    0.055 \\
id0132 &      3.642&    3.657&    14.313&    0.008&    1.479&    0.012&    2.768&    0.041 \\
id0133 &      0.695&    0.901&    14.368&    0.011&    1.721&    0.017&    3.228&    0.083 \\
id0134 &     -2.663&    1.372&    14.397&    0.017&    1.782&    0.021&    3.156&    0.068 \\
id0137 &      1.068&   -0.496&    14.475&    0.008&    1.731&    0.015&    2.754&    0.049 \\
id0138 &      3.412&    4.794&    14.504&    0.010&    2.058&    0.033&    4.301&    0.337 \\
id0139 &     -4.459&   -3.387&    14.507&    0.011&    1.870&    0.021&    3.526&    0.157 \\
id0140 &     -5.526&    0.256&    14.533&    0.014&    1.905&    0.020&    3.209&    0.100 \\
id0141 &     -2.497&    0.813&    14.565&    0.014&    1.680&    0.018&    3.187&    0.073 \\
id0144 &     -4.192&    2.580&    14.580&    0.016&    1.788&    0.031&    2.735&    0.077 \\
id0147 &     -0.622&   -2.216&    14.611&    0.016&    1.754&    0.021&    3.155&    0.081 \\
id0151 &     -0.269&    3.630&    14.651&    0.009&    1.620&    0.014&    2.975&    0.061 \\
id0159 &     -1.011&    9.773&    14.797&    0.009&    1.983&    0.026&    2.025&    0.067 \\
id0168 &     -1.240&    5.197&    14.962&    0.008&    1.661&    0.020&    3.091&    0.090 \\
id0178 &      5.792&    2.385&    15.044&    0.010&    1.580&    0.021&    2.453&    0.061 \\
id0190 &      0.473&   -2.772&    15.159&    0.008&    1.892&    0.025&    3.124&    0.175 \\
id0204 &      0.234&   10.004&    15.333&    0.010&    1.311&    0.020&    2.317&    0.059 \\
id0215 &      0.482&    0.011&    15.437&    0.016&    1.581&    0.034&    3.451&    0.248 \\
id0261 &      6.665&    2.110&    15.896&    0.014&    0.661&    0.021&    1.159&    0.025 \\
id0273 &      6.336&   -1.678&    16.032&    0.017&    0.816&    0.024&    1.373&    0.033 \\
id0290 &      7.995&    0.267&    16.163&    0.024&    0.293&    0.030&    0.607&    0.021 \\
id0306 &     -2.004&    9.312&    16.310&    0.025&    0.378&    0.030&    0.748&    0.022 \\
id0323 &     -5.759&    3.164&    16.501&    0.026&    2.092&    0.077&    2.702&    0.399 \\
id0344 &     -5.555&   -5.459&    16.766&    0.050&    0.648&    0.063&    0.838&    0.049 \\
id0398 &      6.430&    6.857&    17.564&    0.054&    0.534&    0.082&    2.185&    0.358 \\
\hline             
\end{longtable}    
\begin{list}{}{}   
\item[$^{\mathrm{a}}$] In this catalog the star ID is designated by sorting the data in order of increasing $K_{S}$ magnitude.
\item[$^{\mathrm{b}}$] Positions with respect to $\alpha = 17^{h}\;45^{m}\;50.798^{s},\; \delta = -28^{\circ}\;49^{'}\;25.606^{''}\;$ (J2000).
\item[$^{\mathrm{c}}$] NACO $K_{S}$ band magnitude.
\item[$^{\mathrm{d}}$] The 1{$\sigma$} uncertainty in $K_{S}$.
\item[$^{\mathrm{e}}$] NACO $(H-K_{S})$ color index.
\item[$^{\mathrm{f}}$] The 1{$\sigma$} uncertainty in $(H-K_{S})$.
\item[$^{\mathrm{g}}$] NACO $(J-H)$ color index.
\item[$^{\mathrm{h}}$] The 1{$\sigma$} uncertainty in $(J-H)$.
\end{list}
}

\subsubsection{The NACO Natural Photometric System}

As noted by Selman (\cite{selman04}), to avoid systematic effects in transforming broad-band photometry of reddened stars into a standard photometric system (e.g. Johnson, 2MASS, HST/NICMOS), one must work in the natural system defined by the instrument. This occurs because extinction distorts the spectral energy distributions (SED) in ways that are not matched by standard stars unless, of course, the standard stars span the same range in spectral types and extinction as the program stars. Since the SEDs of early-type stars and the extinction law are flatter in the IR, a priory one would expect this problem to be less severe for broad-band $JHK_{S}$ photometry.  Unfortunately this is not the case of the highly reddened stars of the Arches cluster.

If one observes with passbands that differ from that of the standard system, big color terms can be derived from unreddened standards. The danger of this procedure is to assume, via an extrapolation, that these color terms can be applied to standardize reddened program stars. On the basis of synthetic photometry we have reached the conclusion that, at the extinction of the Arches cluster, the color terms differ significantly from the ones obtained with unreddened standards. Therefore, the extrapolation leads to a systematic shift of the zero points in $J$, $H$, and $K_{S}$ (Fig. 2.7 of Selman \cite{selman04} illustrates the effect in the optical regime).

\begin{figure}
\centering
\includegraphics[angle=0,width=6cm]{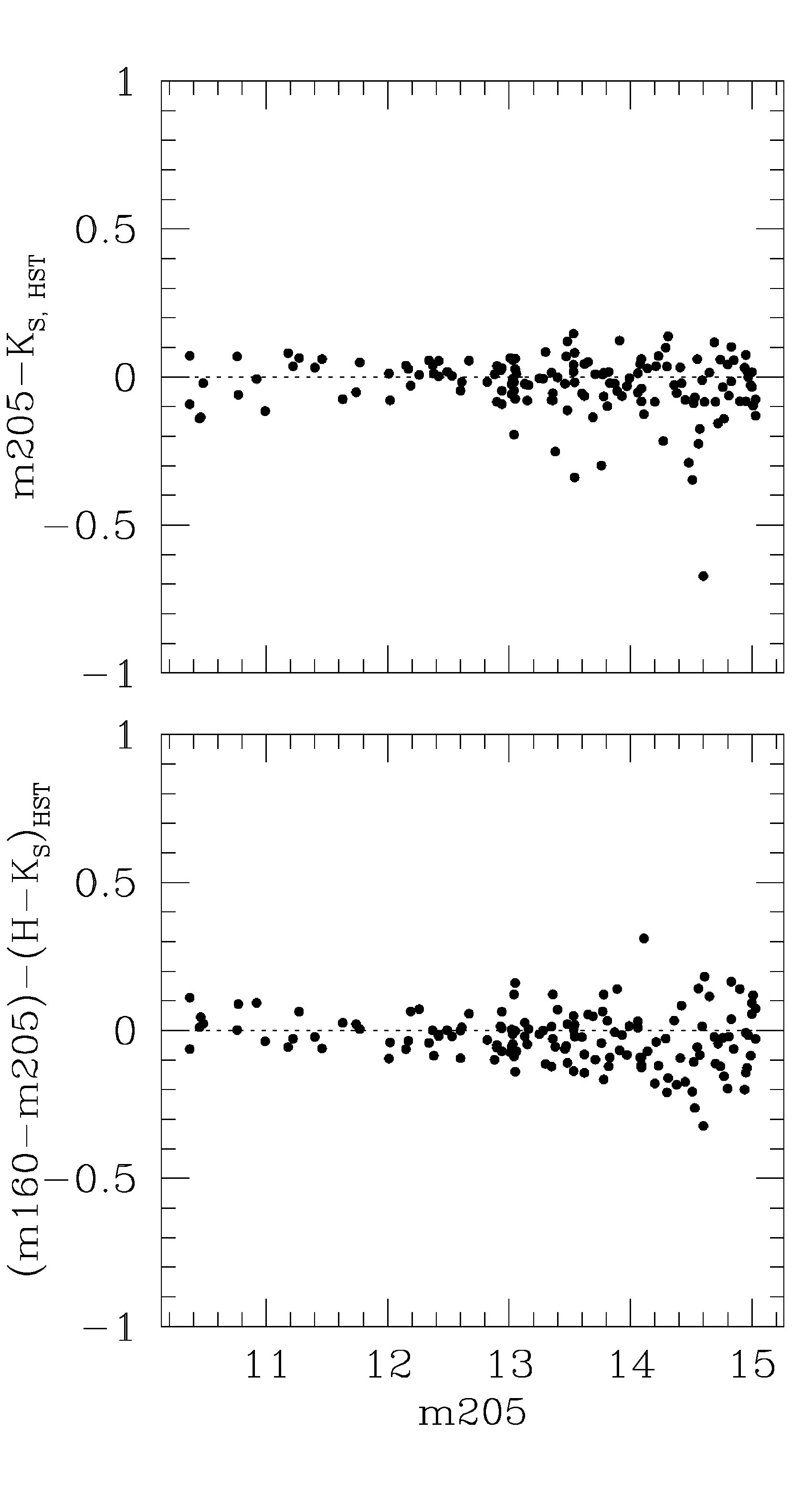}
\caption{Differences between the magnitudes of Figer et al. (\cite{figer02}) and our NACO photometry converted into the HST/NICMOS system. The comparison is derived from 150 common stars (confirmed by visual inspection) in both catalogs.}
\label{comparison}
\end{figure}

But to circumvent this problem by working in the {\it natural} photometric system have a serious drawback: it becomes difficult to compare our observations with previous investigations.  Fortunately, Figer et al. (\cite{figer02})  published their photometry catalog for bright stars in the Arches, so using these stars as local standards we can transform our observations to their photometric system and thus perform a detailed comparison for 150 stars in common (as shown in Fig.\ref{comparison}). The rms scatter, 0.08 mag. in magnitude and 0.09 mag. in color, is consistent with the photometric errors, but there is a clear trend (specially in color) of the differences becoming systematically negative for fainter stars (see Figure~\ref{comparison}). This underlies the problem of transforming between photometric systems discussed above, and reaffirms our conclusion about the importance of working in the natural photometric system of NACO.

\subsubsection{Isochrone Conversion}

To proceed as described above, it is necessary to transform the stellar models (that will be compared with our observations) into the NACO system as well. In the work of Lejeune \& Schaerer (\cite{lejeune}), Geneva isochrones are placed in the observational plane of the Johnson system (Bessell \& Brett, \cite{bessell}). Thus we need to compute the transforming equations from the Bessell \& Brett system into the NACO natural system.

In our case, only one photometric HST/NICMOS standard star (S361-D) was observed during our Service Mode run using the same instrument configuration (i.e. camera, wavefront sensor, and dichroic) of our science images. The airmass of the multiple pointings on S361-D was $1.03$, with a $\Delta X \leq 0.08$ compared to our $HK_{S}$ science frames. Achieved Strehl ratios were $3\%$ in $J$, $12\%$ in $H$, and $25\%$ in $K_{S}$; also very similar to the AO performance in the Arches cluster field.

We have verified that at zero reddening the color terms of the transformation are negligible by using model atmospheres and the filter band-passes of Bessell \& Brett (\cite{bessell}) and NACO. For early type stars these terms contribute less than 0.01~mag to the transformed magnitudes and therefore have been ignored. Thus the standard star measurements are only used to estimate the photometric zero points, for which one star is sufficient (although it is always safer to have more than one!).

Thus, to transform the $JHK$ magnitudes of the isochrones of Lejeune \& Schaerer (\cite{lejeune}) into the NACO natural photometric system, the following zero points were obtained,

\[ \begin{array}{l}
 J_{NACO} = J-(21.49\pm0.04)\\
 H_{NACO} = H-(21.32\pm0.02)\\
 K_{S,NACO} = K-(20.37\pm0.04)\\
 \end{array}\]

Uncertainties in the zero points of $(H-K_{S})_{NACO}$ and $(J-H)_{NACO}$  will have an effect in the determination of our IMFs and their fitted slopes. We study this issue on the basis of numerical simulations in Appendix~A.

\subsection{Incompleteness Corrections}
\label{Artificial Stars Experiments}

\begin{figure*}
\centering
\includegraphics[width=13.cm]{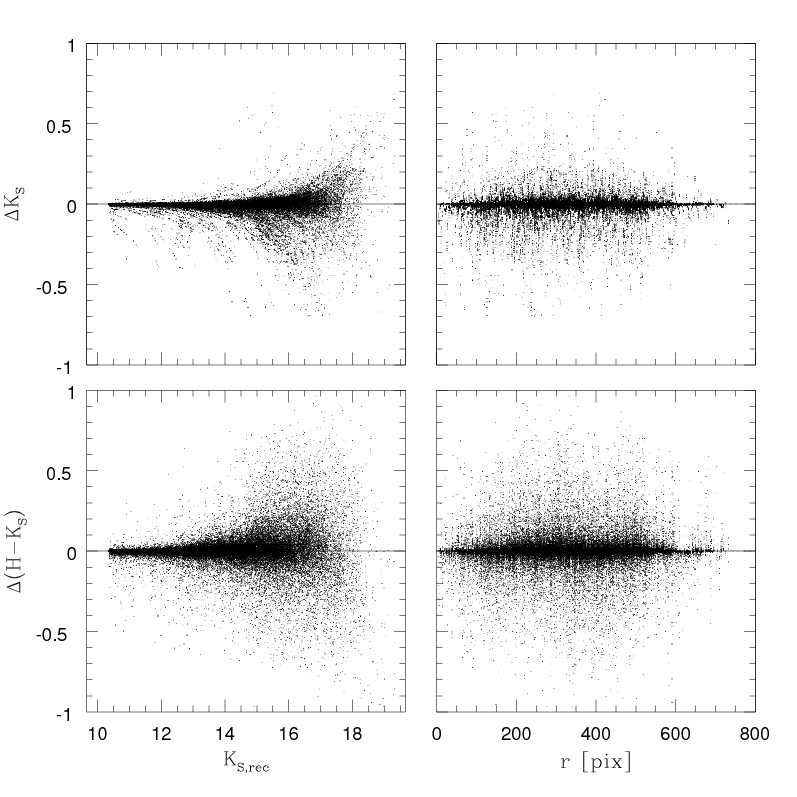}
\caption{Results of Monte Carlo experiments. Left panels shows the differences between colors and magnitudes of added and recovered stars as a function of recovered $K_{S}$ magnitude. The right panels plot these differences as a function of distance to the cluster center (in pixels).}
\label{completeness2}
\end{figure*}

We used Monte Carlo experiments to study the completeness limits of our photometry (Stetson \cite{stetson92}). They consist of adding artificial stars spanning the real instrumental colors and magnitudes of our data at random positions in $K_{S}$ frames. This automatically sets their $H$ positions as we already know the frame offset from DAOMATCH. Each experiment consists in the creation of 126 pairs of artificial $HK_{S}$ images, and in each pair a maximum of 30 synthetic stars are added by DAOPHOT's ADDSTAR in order to preserve the original crowding properties. We then run the same procedure described above to perform photometry on the artificial images and count the stars recovered in \textit{both} filters ($J$ band experiments are not required as for the IMF derivation multi-color data is used only when available). The recovery fraction determines the incompleteness correction as a function of magnitude. Fig.~\ref{completeness2} shows the result of 50 experiments, including more than $93.500$ recovered stars. For the whole range of recovered $K_{S}$ magnitudes ($K_{S,rec}$) the mean values of $\Delta K_{S}$ and $\Delta (H-K_{S})$ remain bound to $\pm 0.015$, while a $98\%$ ($99.8\%$) of stars with $K_{S,rec}\geq14(10.5)$ have $\vert\Delta K_{S}\vert\leq 0.3$. As expected, the scatter increases for fainter stars, but there is no apparent systematic trend of recovered magnitude or color as a function of distance to the cluster center.

\begin{figure*}
\centering
\includegraphics[width=13cm]{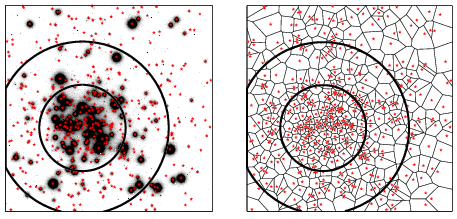}  
\caption{Voronoi diagram built from stars brighter than $K_{S}=17$ in the NACO field of approximately $24''\times24''$. Adopting a distance to the Galactic center of $8\;kpc$ , circles represents projected distances of $0.2$ and $0.4\;pc$ from the cluster center. Note the smaller area of Voronoi cells at the cluster core, indicating a higher local stellar density (crowding).}
\label{voron}
\end{figure*}

\begin{figure*}
\centering
\includegraphics[width=15cm]{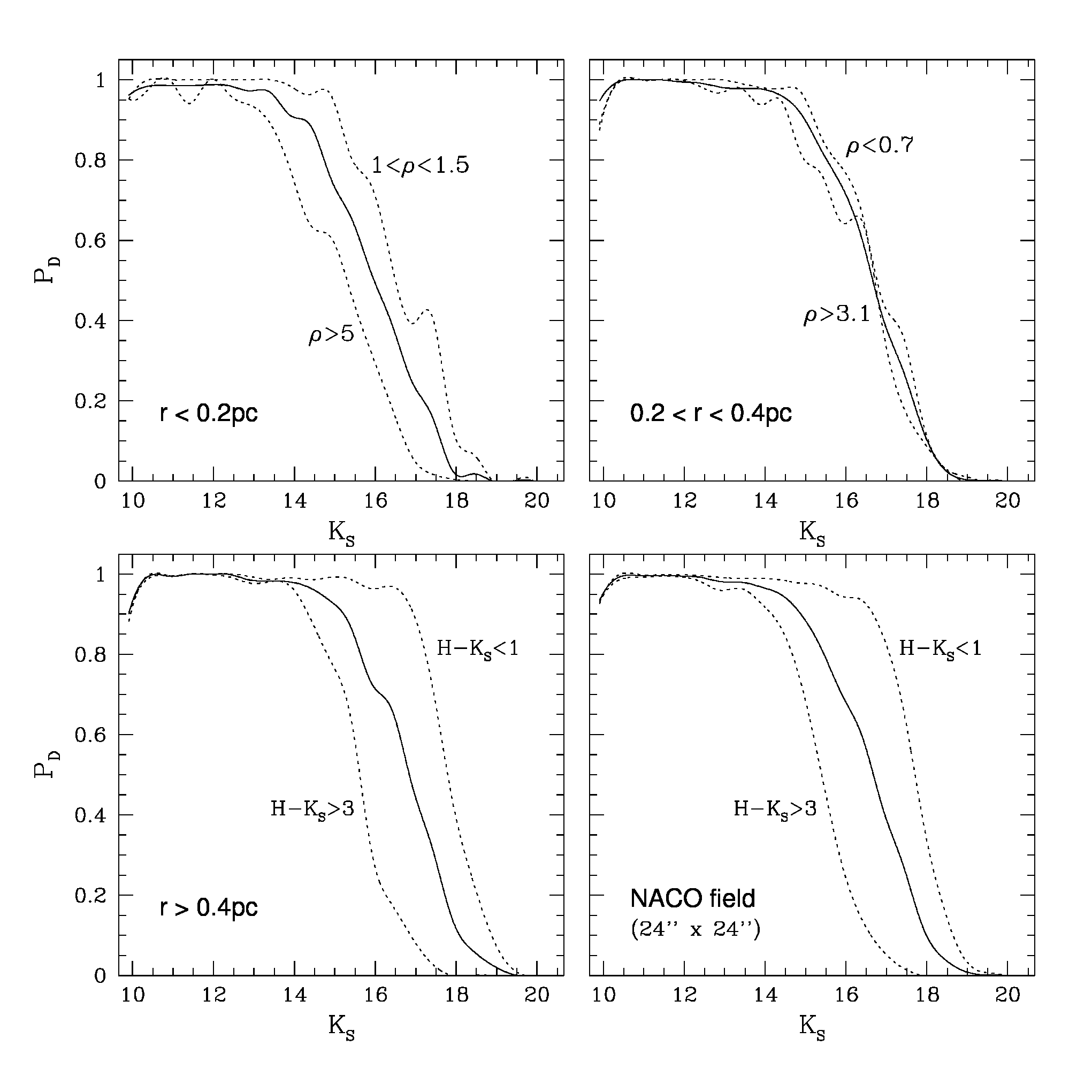}
\caption{Detection probabilities computed from Monte Carlo experiments in different radial bins. The solid lines represent the recovery fraction as a function of magnitude only. In the upper panels the dotted lines represent probabilities marginalized by local stellar density $\rho$ [$star/arcsec^{2}$]. In the bottom panels dotted lines show probabilities marginalized by $(H-K_{S})$ color.}
\label{completeness}
\end{figure*}

While the lack of radial gradients in the recovered magnitudes and colors is reassuring,  it is also unexpected given the sharp increase in stellar density towards the cluster center (Selman et al. \cite{selman99}). To further investigate this issue we included in the Monte Carlo experiments a reliable estimate of the local stellar density, obtained from Voronoi diagrams (De Berg et al. \cite{de_berg}; Aurenhammer \cite{aurenhamer}). In our case the task is to subdivide the (2-D) frames containing $N$ stars into $N$ cells in a one to one relation between cells and star numbers. The construction of cells can be understood if we notice that for a particular star there is a boundary including the space lying closer to it than to any other star in the frame. This boundary defines one {\it Voronoi cell}, i.e. a convex and possibly unbounded polygon that can be called a \textit{nearest neighbor} region. It follows from the definition of Voronoi diagram that the sides of the cells are always  straight lines. The example in Fig~\ref{voron} shows the Voronoi {\it tessellation} built from stars detected in our $K_{S}$ image.

The inverse of the area of each Voronoi cell is our estimate for the local stellar density ($\rho$). We avoid the problem of unbounded (infinite area) cells, which arise for stars near the edges of a frame, by just specifying the frame edge as an additional boundary. For $\rho$ thus defined we can then compute the recovery fraction of artificial stars as a function of three variables: magnitude, color, and local stellar density. In general $\rho$ will be very well correlated with crowding, although, as shown in Fig.\ref{voron},  close pairs of stars in low density regions will generate large Voronoi cells. In these cases while the local density measured by the size of the cells remains  low, the stars are locally crowded. Still the probability that one of our artificial stars will fall on top of such a pair is very small, so we do not expect these effects to significantly influence our statistical crowding corrections.

The completeness limits are brighter in the cluster core than further out in both bands.  Fig.~\ref{completeness}  shows detection probabilities (i.e. the normalized recovery fractions) for the $\sim24''\times24''$ NACO field of Arches used in this work, and for two different radial cuts chosen to match the subdivisions used in previous investigations.  The upper panels show that the local density must be considered when applying completeness corrections to our data (especially in the core), and underlies the need of a large number of Monte Carlo simulations mentioned above. We performed 50 experiments, which summed an approximate total of 189000 added synthetic stars. Detection probabilities for each radial subdivision are ordered in a 3-D array considering magnitudes, colors, and local densities (21 x 6 x 7 bins), each element being the ratio of recovered to added synthetic stars of $(H-K_{S})$ colors consistent with the trend observed in the Arches. The correction factor for real stars characterized by ($K_{S}$, $H-K_{S}$, $\rho$) is determined by linear interpolation of the corresponding 3-D array. Table~\ref{completeness_mag} shows the 50\% completeness limits for H and $K_{S}$ in the defined radial bins.

\begin{table*}
\caption{Completeness limits of the photometry at different annuli of the Arches cluster.} 
\label{completeness_mag}      
\centering                          
\begin{tabular}{c c c c c}        
\hline\hline                 
50\% completeness limit & $r \leq 0.2~pc$ & $0.2 < r \leq 0.4~pc$ & $r > 0.4~pc$ & NACO field \\ 
\hline                        
   H       &  18.2   & 18.7   & 18.9   & 18.7   \\      
   $K_{S}$ &  16.0   & 16.7   & 16.8   & 16.7   \\ 
\hline                                   
\end{tabular}
\end{table*}

\begin{figure*}
\centering
\includegraphics[width=15cm]{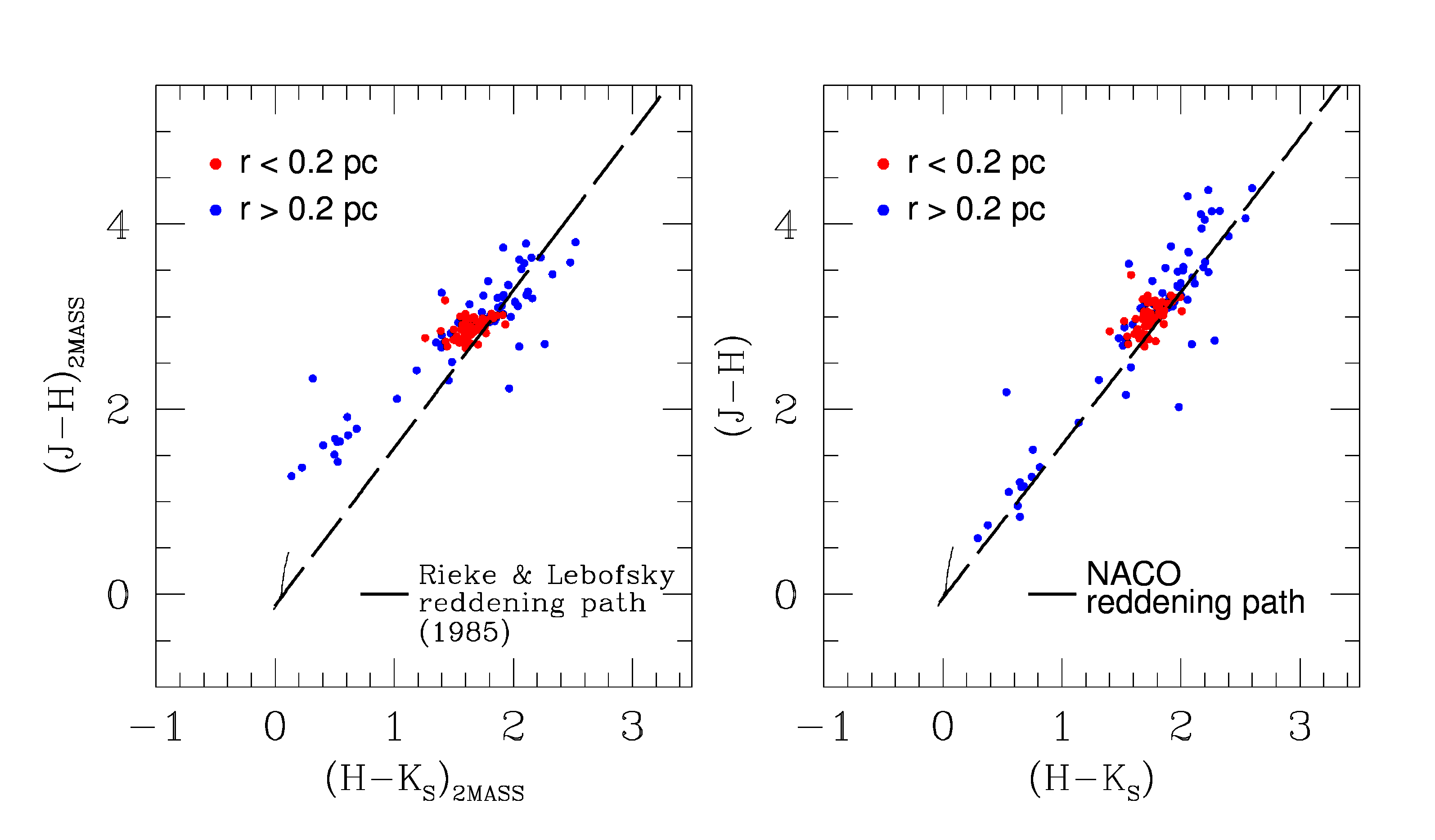}
\caption{ \textbf{Left:} NACO colors transformed into the 2MASS system. The reddening path for the normal Rieke \& Lebofsky extinction law (\cite{rieke}) is shown together with our data transformed to the standard photometric system, as described in the text. \textbf{Right:} NACO natural colors. The slope of the reddening path here has been calculated using the model atmospheres of Kurucz (\cite{kurucz}) and the average galactic extinction curve of Fitzpatrick (\cite{fitzpatrick2}).}
\label{twocolor}
\end{figure*}

\section{Data Analysis}
\subsection{The Reddening Law}
\label{Reddening Law}

Selman (\cite{selman04}) and Selman and Melnick (\cite{selmanymelnick}) showed that it is not possible to transform broad-band UBV photometry of highly reddened stars to the standard Johnson (or indeed to any) photometric system using unreddened photometric standards. We have already seen in Section 2.3.1 that the same applies to our NIR observations of the Arches cluster. In this Section we illustrate with an example the risks of proceeding outside the NACO natural photometric system.

The Two-color diagram of SGB02 (their Figure~10) shows a significant tilt between the data points and the Rieke \& Lebofsky  (\cite{rieke}) standard reddening law that they cannot explain. In fact, it is mentioned that even slight changes in their color transformation equations result in even larger discrepancies between the data points and the reddening path. We find the same effect when we transform our data to the standard photometric system following the prescription of SGB02: first we transform to the HST/NICMOS system using Figer's photometry (\cite{figer02}) as local standards and then we use transformation equations from Brandner et al. (\cite{brandner}; derived from mainly unreddened main-sequence stars) to go from HST/NICMOS to the $2MASS$ system, which provides a close match to the standard (Johnson) photometric system. The result is shown in the left panel of Fig.~\ref{twocolor} where a significant tilt between the data and the reddening path is clearly seen.

Using an extended grid of simulations, Selman (\cite{selman04}) concluded that the solution is to work in the natural photometric system of the instrument, and to use models to calculate the reddening path in this system from the appropriate (in our case the Galactic) reddening law. Fitzpatrick (\cite{fitzpatrick2}) presents a monochromatic extinction curve based on $2MASS\;JHK_{S}$ photometry. The curve is especially well determined in the $1.25-2.2\;\mu m$ region, and represents an improvement over previous work (Fitzpatrick, \cite{fitzpatrick}). We used this average Galactic extinction law (with $R_{V}=3.1$) to redden the models up to $A_{V}=40$ mag. and to obtain in this way the reddening parameters for the NACO passband. We found $S^{NACO}=\frac{E(J-H)}{E(H-K_{S})}=1.66$ for the color-color slope, and $R^{NACO}_{K_{S}}=\frac{A_{K_{S}}}{E(H-K_{S})}=1.61$ for the ratio of total to selective absorption. Both Rieke \& Lebofsky (\cite{rieke}) and NACO reddening vectors are very similar in the color-color plane, as illustrated by the slopes in Fig.~\ref{twocolor} ($S=1.7$ for Johnson filters). With the observations and the stellar models already in the NACO natural system, the right panel of the Figure shows that the tilt is no longer apparent, which constitutes a good sanity check of our procedures.

\begin{figure*}
\begin{center}
\includegraphics[width=16cm]{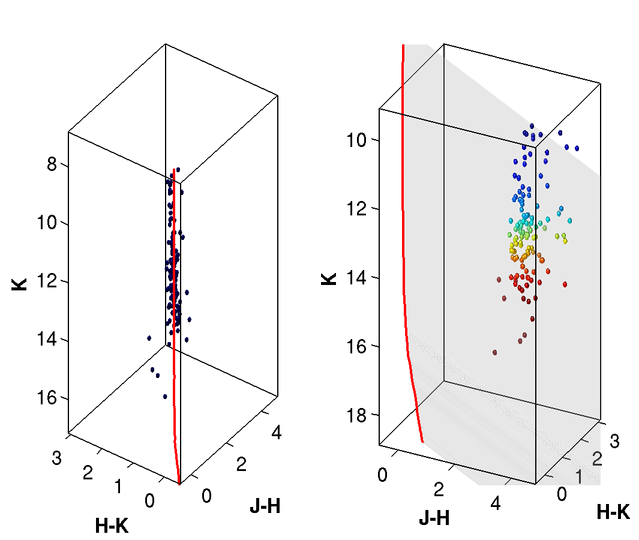}
\end{center}
\caption{\textbf{Left:} Color-magnitude Stereogram. This is the projection viewed from the reddening vector direction, i.e. the Reddening free view. \textbf{Right:} Theoretical Surface for a $2.5\;Myr$ Geneva isochrone with solar metallicity (in red). The surface does indeed go through the cloud of stars: some stars are below, some above the {\it Theoretical Surface.}}
 \label{cms_free}
\end{figure*}

The values of our reddening parameters, specially $R^{NACO}_{K_{S}}$, differ from the standard Rieke \& Lebofsky ($R_{K}=1.78$ for Johnson filters). They also differ from the recent results of Nishiyama et al. (\cite{nishiyama}): $S=1.72$ and $R_{K_{S}}=1.44$, which were derived using red clump stars in the region $\vert l\vert \lesssim 2^{\circ}$ and $0.5^{\circ} \lesssim \vert b\vert \lesssim 1^{\circ}$. In Appendix~B we present a detailed discussion of the reddening paths for different photometric systems using Kurucz (\cite{kurucz}) model atmospheres and Fitzpatrick's (\cite{fitzpatrick2}) parameterization of the Galactic extinction law. To end this Section we define the three dimensional NACO reddening vector, $\textbf{R}_{NACO}=(1,S^{NACO},R^{NACO}_{K_{S}})^{T}=(1,1.66,1.61)^{T}$, that will be used throughout the paper.

\subsection{Cluster Age and Metallicity}

In order to determine the physical parameters of the cluster stars we need to compare the observations with stellar models of a given age and metallicity. In principle it should be possible to leave these as free parameters to be determined from the observations. In practice, however, even if age and metallicity are accurately known, it is still extremely challenging to derive physical parameters of young massive stars from broad-band photometry. Crowding and variable extinction make this challenge even more difficult.

Published results show a good agreement in the spectral types derived for emission-line stars in the Arches cluster. Both Blum et al. (\cite{blum}) and Figer et al. (\cite{figer02}) agree in assigning $WN7-WN9$ spectral types to the brightest stars in the cluster. The presence of WC stars is discarded because of the absence of the emission line $He\;I\;2.06\;\mu m$ or the triplet $C\;IV\;2.08\;\mu m$ (see Figure 6 of Figer et al. \cite{figer02}). This absence of carbon dominant Wolf Rayets allows, from evolutionary models and in agreement with the finding of late type nitrogen-rich Wolf Rayets, to constrain the cluster age to less than $3.5\;Myr$.

In spite of the agreement in spectral classification, Figer et al. determine an age range of $2-3\;Myr$ for the cluster, while  Blum et al. obtain $2-4.5\;Myr$ arguing that the absence of carbon rich Wolf Rayets (the presence of which would indicate the older age) may be only an artifact of the resolution and completeness of the photometry as WCs are intrinsically fainter than WNLs. However, the more recent results of Najarro et al. (\cite{najarro}) and Martins et al.  (\cite{martins}) have established a younger upper limit of the age interval. Najarro et al. constrained the age to $2-2.5\;Myr$ from the high resolution ($R\sim23.000$) spectroscopy of five massive stars ($3\;$WNLs) and Martins et al. obtained a range $2-4\;Myr$ for $O$ supergiants and $2.5\pm0.5\;Myr$ for the Wolf-Rayets. Although at a lower resolution ($R\sim4.000$), Martins et al. had a wider spectral coverage for a larger sample of $28$ stars.

Solar Fe abundances have been measured for the Galactic center and the Quintuplet clusters using late type supergiants (Carr et al. \cite{carr}; Ramirez et al. \cite{ramirez}). Moreover, the result of Najarro et al. (\cite{najarro}) constrained the metallicity of the Arches cluster to be solar while Martins et al. (\cite{martins}) find a slightly super solar metallicity ($Z=1.3-1.4 Z_{\odot}$) for the lightest metals. Therefore, and in agreement with the forenamed spectroscopic studies, we will adopt an age of $2.5\;Myr$ and solar abundances in the comparison of our observations to theoretical models. This age choice is also in agreement with our photometry, as the tip of the reddened $2.5\;Myr$ Geneva isochrone matches the brightest stars of the cluster (see Fig.~\ref{cms_free}, left).

\subsection{The Bayesian Method}

We will use the Bayesian method introduced by Selman et. al (\cite{selman99}) to derive the physical parameters (initial masses, $M_{ini}$; effective temperatures, $T_{eff}$; surface gravities, $log\;g$), as well as the reddening of our stars from the $JHK_{S}$ photometry. Briefly, we use the work of Lejeune \& Schaerer (\cite{lejeune}) who computed  model stellar atmospheres for each point of Geneva isochrones and  convolved these synthetic spectra with the $JHK$ response functions of Bessell \& Brett (\cite{bessell}) to build theoretical isochrones in the Johnson system. In our case the photometric quantities are measured in the NACO natural system, so in order to locate the unreddened theoretical isochrones in the instrumental $3-D$ space it is necessary to shift the original colors of Lejeune \& Schaerer using the transformation equations as explained in Section 2.3.2.

Assuming that the Arches cluster was formed in a single (instantaneous) burst of star formation, the position of a star will be shifted from the unreddened theoretical isochrone along the reddening vector. Because in very young clusters the reddening can vary significantly from star to star, some stars will be near the isochrone while others  will be far. In the absence of observational errors the locus of the observations define a surface in the 3-D color-color-magnitude space. A corresponding set of {\it Theoretical Surfaces} is generated from the synthetic isochrones by applying variable amounts of extinction along the NACO reddening vector $\textbf R_{NACO}$ defined above. We then use the Bayes  theorem to assign relative probabilities to the physical models (points of the theoretical isochrone translated to absolute magnitudes) assuming uniform but positive priors for the reddening, and a Gaussian model for the photometric errors. From the {\it Theoretical Surface} coming out of a $2.5\;Myr$ Geneva isochrone with standard mass loss rate and solar metallicity, we read for each star the most probable color excess $E(H-K_{S})$ and physical parameters: $M_{ini}$, $T_{eff}$, and $log\;g$. The main advantage of this approach is that by using colors and luminosities together it is possible to circumvent the problem of the near-degenerate NIR colors of young massive stars.

The color-magnitude stereogram (CMS; Selman et al. \cite{selman99}; Selman and Melnick, \cite{selmanymelnick}), provides a useful tool to visualize the data and the unreddened theoretical isochrones in the color-color-magnitude space. The Y-Z and X-Z projections are the $(J-H)-K_{S}$ and $(H-K_{S})-K_{S}$ CMDs respectively, while the X-Y projection is just the $(H-K_{S})-(J-H)$ color-color diagram. We do not take into account here the effects of binarity, intrinsic reddening, or stellar rotation, so it is assumed that all the scatter in the data is caused by photometric uncertainties and differential extinction. The quality of our observations can be seen in Fig.~\ref{cms_free}: the 3D surface is indeed formed by the program stars as it is seen in the left panel that shows the \emph{reddening free} view of the CMS, i.e. an orthographic projection perpendicular to the direction of the reddening vector. The right panel shows the {\it Theoretical Surface} for a $2.5\;Myr$ Geneva isochrone and the severe differential extinction affecting Arches.

For the $2.5\;Myr$ isochrone one star falls above the Theoretical Surface in the CMS. At this age, Lejeune \& Schaerer (\cite{lejeune}) give colors and magnitudes only for stars below $M_{ini}\sim105\;M_{\odot}$ corresponding to the end of the hydrogen-burning phase of evolution. In order to estimate the mass of this star we extrapolated the mass-flux relation, obtaining a value of $M_{ini}=119.3\;M_{\odot}$ from the Bayesian method. This upper limit for the most massive stars we observe in Arches agrees nicely with the uppermost initial mass of the Geneva models (Schaller et al. \cite{schaller}), i.e. stars of $120\;M_{\odot}$ that live up to $3\;Myr$.

Table~\ref{bayes_info603}, published electronically, presents the catalog with all the observed data and physical parameters derived using the Bayesian method.

\begin{figure}[!ht]
\begin{center}
\includegraphics[width=8cm]{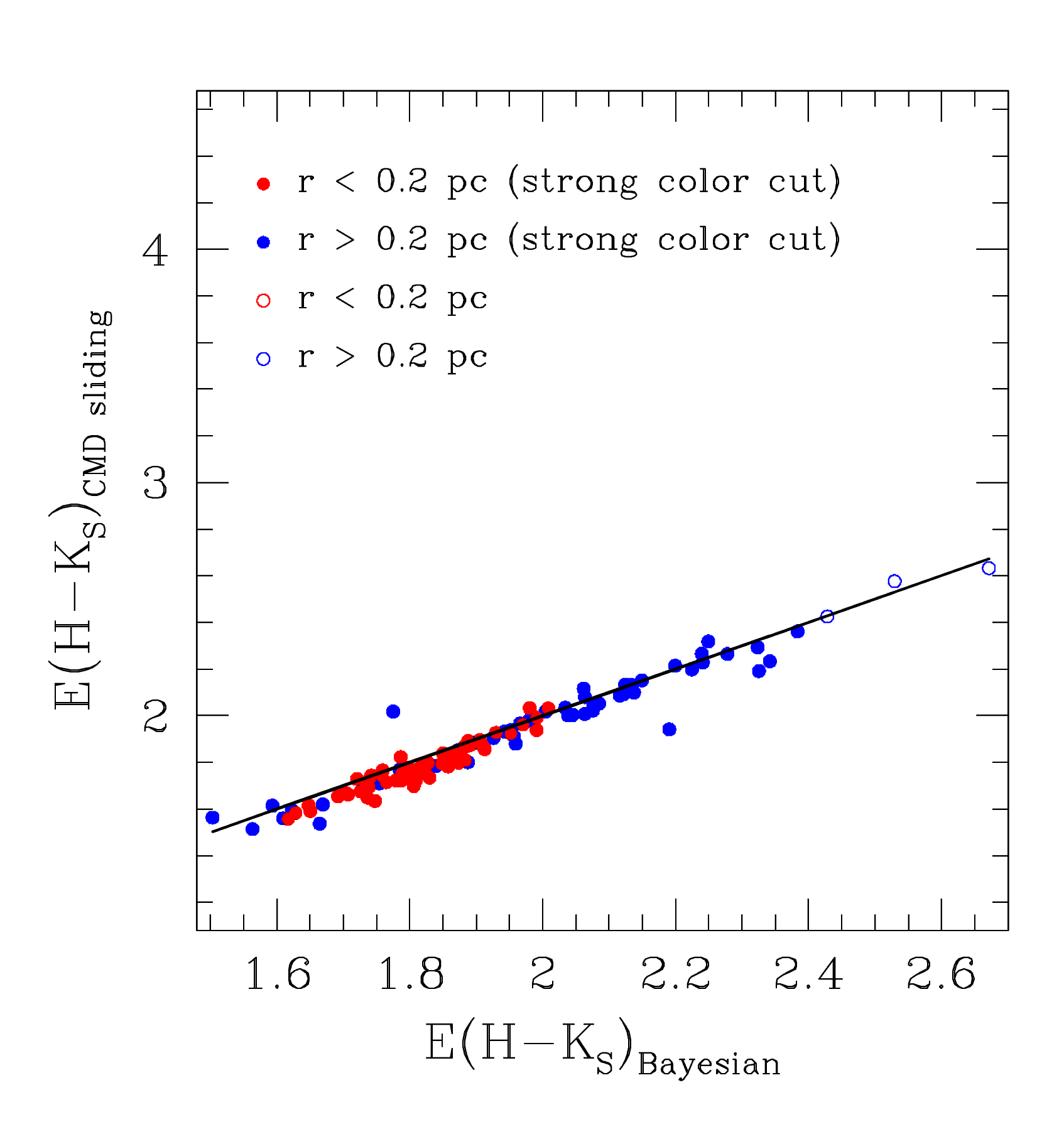}
\end{center}
\caption{Comparison between the color excesses derived from the Bayesian and CMD Sliding methods for stars with $JHK_{S}$ photometry. The color distinguishes stars inside (red) and outside (blue)  the core of the cluster. Solid circles represent stars within the strong color cut shown in Fig.\ref{cmd}; open circles show the (red) objects rejected by the strong color cut criterion. The straight line corresponds to X=Y.}
\label{excesses}
\end{figure}

\addtocounter{table}{1} 
\onllongtab{5}{
\begin{longtable}{c c c c c c c c c c c}
\caption{Physical parameters for $JHK_{S}$ stars derived from the Bayesian method. The catalog contains point sources within $10''$ from the cluster center and within the strong color cut as shown in Fig.\ref{cmd}.}
\label{bayes_info603}\\ 
Star ID $^{a}$  & $K_{S}$ $^{b}$ & $(H-K_{S})$ $^{c}$ & $(J-H)$ $^{d}$ & $M_{V}$ $^{e}$& $log\;L$ $^{f}$& $log\;T_{eff}$ $^{g}$& $log\;g$ $^{h}$& $E(H-K_{S})$ $^{i}$& $M_{ini}$ $^{j}$ & $M_{act}$ $^{k}$\\ 
\hline 
\endfirsthead
\caption{Continued.} \\ 
\hline 
Star ID $^{a}$  & $K_{S}$ $^{b}$ & $(H-K_{S})$ $^{c}$ & $(J-H)$ $^{d}$ & $M_{V}$ $^{e}$& $log\;L$ $^{f}$& $log\;T_{eff}$ $^{g}$& $log\;g$ $^{h}$& $E(H-K_{S})$ $^{i}$& $M_{ini}$ $^{j}$ & $M_{act}$ $^{k}$\\ 
\hline
\endhead
\hline
\endfoot    
\hline
\endlastfoot
id0001 &    10.068 &   1.957 &   3.203 &    -8.408 &   6.258 &   4.411 &   2.628 &   1.991 & 101.23 &  70.51\\
id0002 &    10.216 &   2.200 &   4.048 &    -8.840 &   6.371 &   4.386 &   2.436 &   2.342 & 119.73 &  73.81\\
id0003 &    10.244 &   1.841 &   3.021 &    -8.102 &   6.184 &   4.431 &   2.767 &   1.882 &  89.23 &  68.18\\
id0004 &    10.281 &   1.861 &   3.064 &    -8.102 &   6.184 &   4.431 &   2.767 &   1.905 &  89.23 &  68.18\\
id0005 &    10.345 &   2.099 &   3.418 &    -8.344 &   6.238 &   4.414 &   2.658 &   2.124 &  97.73 &  69.96\\
id0006 &    10.362 &   1.993 &   3.220 &    -8.149 &   6.193 &   4.427 &   2.742 &   2.005 &  90.73 &  68.55\\
id0007 &    10.456 &   1.997 &   3.214 &    -8.082 &   6.180 &   4.433 &   2.777 &   2.008 &  88.73 &  68.05\\
id0008 &    10.619 &   1.776 &   3.034 &    -7.805 &   6.119 &   4.460 &   2.936 &   1.888 &  81.23 &  66.15\\
id0009 &    10.878 &   1.928 &   3.161 &    -7.601 &   6.080 &   4.481 &   3.050 &   1.970 &  77.23 &  65.05\\
id0010 &    10.965 &   1.851 &   3.015 &    -7.413 &   6.050 &   4.501 &   3.153 &   1.892 &  74.23 &  63.98\\
id0011 &    11.011 &   1.625 &   2.830 &    -7.157 &   6.009 &   4.528 &   3.292 &   1.736 &  70.23 &  62.32\\
id0012 &    11.144 &   2.190 &   3.533 &    -7.686 &   6.095 &   4.472 &   3.004 &   2.199 &  78.73 &  65.49\\
id0013 &    11.151 &   1.833 &   3.018 &    -7.221 &   6.019 &   4.522 &   3.256 &   1.884 &  71.23 &  62.76\\
id0014 &    11.412 &   1.610 &   2.813 &    -6.810 &   5.952 &   4.556 &   3.443 &   1.748 &  65.23 &  59.72\\
id0015 &    11.502 &   1.701 &   2.899 &    -6.810 &   5.952 &   4.556 &   3.443 &   1.810 &  65.23 &  59.72\\
id0016 &    11.518 &   1.703 &   2.896 &    -6.747 &   5.940 &   4.560 &   3.465 &   1.794 &  64.23 &  59.11\\
id0017 &    11.546 &   2.098 &   3.425 &    -7.221 &   6.019 &   4.522 &   3.256 &   2.134 &  71.23 &  62.76\\
id0018 &    11.788 &   1.769 &   2.993 &    -6.573 &   5.903 &   4.571 &   3.531 &   1.852 &  61.23 &  57.19\\
id0019 &    11.854 &   2.204 &   3.590 &    -7.086 &   5.998 &   4.534 &   3.327 &   2.241 &  69.23 &  61.86\\
id0020 &    11.877 &   1.876 &   3.148 &    -6.655 &   5.922 &   4.566 &   3.498 &   1.957 &  62.73 &  58.17\\
id0021 &    11.935 &   1.747 &   2.898 &    -6.363 &   5.853 &   4.580 &   3.597 &   1.814 &  57.73 &  54.59\\
id0022 &    12.010 &   1.759 &   3.017 &    -6.333 &   5.845 &   4.581 &   3.605 &   1.839 &  57.23 &  54.19\\
id0023 &    12.054 &   1.661 &   2.813 &    -6.121 &   5.787 &   4.588 &   3.667 &   1.739 &  53.73 &  51.35\\
id0024 &    12.061 &   1.889 &   3.142 &    -6.452 &   5.876 &   4.577 &   3.570 &   1.952 &  59.23 &  55.77\\
id0025 &    12.133 &   1.651 &   2.765 &    -6.027 &   5.753 &   4.589 &   3.692 &   1.726 &  51.73 &  49.63\\
id0026 &    12.159 &   1.735 &   2.911 &    -6.151 &   5.796 &   4.587 &   3.659 &   1.815 &  54.23 &  51.76\\
id0027 &    12.160 &   1.747 &   2.933 &    -6.151 &   5.796 &   4.587 &   3.659 &   1.822 &  54.23 &  51.76\\
id0028 &    12.187 &   1.759 &   2.976 &    -6.183 &   5.804 &   4.586 &   3.650 &   1.848 &  54.73 &  52.17\\
id0029 &    12.209 &   1.751 &   2.944 &    -6.094 &   5.778 &   4.589 &   3.674 &   1.819 &  53.23 &  50.92\\
id0030 &    12.266 &   1.729 &   3.004 &    -6.048 &   5.762 &   4.589 &   3.686 &   1.819 &  52.23 &  50.06\\
id0032 &    12.392 &   1.998 &   3.361 &    -6.333 &   5.845 &   4.581 &   3.605 &   2.076 &  57.23 &  54.19\\
id0033 &    12.412 &   1.700 &   2.996 &    -5.949 &   5.726 &   4.590 &   3.712 &   1.830 &  50.23 &  48.32\\
id0034 &    12.621 &   1.822 &   3.107 &    -5.849 &   5.690 &   4.591 &   3.737 &   1.913 &  48.23 &  46.56\\
id0035 &    12.647 &   1.897 &   3.099 &    -5.849 &   5.690 &   4.591 &   3.737 &   1.943 &  48.23 &  46.56\\
id0036 &    12.647 &   1.818 &   3.001 &    -5.742 &   5.651 &   4.592 &   3.762 &   1.873 &  46.23 &  44.78\\
id0037 &    12.692 &   1.842 &   3.040 &    -5.718 &   5.641 &   4.592 &   3.768 &   1.889 &  45.73 &  44.34\\
id0039 &    12.720 &   1.914 &   3.230 &    -5.873 &   5.699 &   4.591 &   3.730 &   1.991 &  48.73 &  47.00\\
id0040 &    12.720 &   1.557 &   2.704 &    -5.327 &   5.481 &   4.591 &   3.856 &   1.651 &  38.73 &  37.93\\
id0041 &    12.727 &   2.230 &   4.367 &    -6.303 &   5.837 &   4.582 &   3.614 &   2.278 &  56.73 &  53.79\\
id0042 &    12.765 &   1.748 &   3.037 &    -5.636 &   5.610 &   4.592 &   3.787 &   1.858 &  44.23 &  42.99\\
id0044 &    12.829 &   1.744 &   2.921 &    -5.469 &   5.545 &   4.593 &   3.824 &   1.810 &  41.23 &  40.25\\
id0045 &    12.832 &   1.793 &   3.113 &    -5.578 &   5.589 &   4.593 &   3.799 &   1.876 &  43.23 &  42.08\\
id0046 &    12.832 &   1.595 &   2.917 &    -5.231 &   5.437 &   4.590 &   3.876 &   1.669 &  37.23 &  36.51\\
id0047 &    12.844 &   1.719 &   2.917 &    -5.411 &   5.522 &   4.593 &   3.836 &   1.790 &  40.23 &  39.33\\
id0048 &    12.846 &   1.973 &   3.488 &    -5.873 &   5.699 &   4.591 &   3.730 &   2.063 &  48.73 &  47.00\\
id0049 &    12.847 &   1.772 &   2.947 &    -5.469 &   5.545 &   4.593 &   3.824 &   1.828 &  41.23 &  40.25\\
id0051 &    12.880 &   0.682 &   1.167 &    -3.719 &   4.631 &   4.518 &   4.099 &   0.761 &  18.58 &  18.49\\
id0052 &    12.881 &   1.718 &   2.975 &    -5.469 &   5.545 &   4.593 &   3.824 &   1.825 &  41.23 &  40.25\\
id0053 &    12.949 &   1.723 &   2.941 &    -5.354 &   5.496 &   4.592 &   3.850 &   1.808 &  39.23 &  38.40\\
id0054 &    12.950 &   1.696 &   2.896 &    -5.327 &   5.481 &   4.591 &   3.856 &   1.786 &  38.73 &  37.93\\
id0056 &    12.984 &   1.969 &   3.334 &    -5.693 &   5.631 &   4.592 &   3.774 &   2.046 &  45.23 &  43.89\\
id0057 &    12.991 &   2.008 &   3.060 &    -5.551 &   5.578 &   4.593 &   3.805 &   1.981 &  42.73 &  41.62\\
id0058 &    13.024 &   1.726 &   2.921 &    -5.231 &   5.437 &   4.590 &   3.876 &   1.786 &  37.23 &  36.51\\
id0059 &    13.030 &   1.717 &   2.780 &    -5.169 &   5.407 &   4.588 &   3.889 &   1.758 &  36.23 &  35.56\\
id0060 &    13.041 &   1.674 &   2.978 &    -5.296 &   5.467 &   4.590 &   3.863 &   1.806 &  38.23 &  37.45\\
id0061 &    13.050 &   1.402 &   2.842 &    -5.045 &   5.344 &   4.585 &   3.915 &   1.664 &  34.23 &  33.66\\
id0062 &    13.053 &   1.688 &   3.057 &    -5.198 &   5.422 &   4.589 &   3.882 &   1.780 &  36.73 &  36.03\\
id0063 &    13.058 &   1.932 &   3.113 &    -5.469 &   5.545 &   4.593 &   3.824 &   1.966 &  41.23 &  40.25\\
id0064 &    13.106 &   1.945 &   3.161 &    -5.441 &   5.534 &   4.593 &   3.830 &   1.979 &  40.73 &  39.79\\
id0065 &    13.113 &   1.513 &   2.688 &    -5.009 &   5.328 &   4.584 &   3.921 &   1.665 &  33.73 &  33.18\\
id0066 &    13.124 &   2.231 &   3.482 &    -5.825 &   5.680 &   4.591 &   3.743 &   2.239 &  47.73 &  46.12\\
id0067 &    13.149 &   1.774 &   3.172 &    -5.264 &   5.452 &   4.590 &   3.869 &   1.874 &  37.73 &  36.98\\
id0068 &    13.196 &   1.851 &   3.072 &    -5.231 &   5.437 &   4.590 &   3.876 &   1.898 &  37.23 &  36.51\\
id0070 &    13.199 &   1.688 &   3.012 &    -5.080 &   5.360 &   4.585 &   3.908 &   1.788 &  34.73 &  34.13\\
id0072 &    13.218 &   1.771 &   3.135 &    -5.198 &   5.422 &   4.589 &   3.882 &   1.872 &  36.73 &  36.03\\
id0073 &    13.220 &   2.259 &   4.139 &    -5.900 &   5.708 &   4.591 &   3.724 &   2.324 &  49.23 &  47.44\\
id0074 &    13.235 &   2.174 &   3.953 &    -5.718 &   5.641 &   4.592 &   3.768 &   2.225 &  45.73 &  44.34\\
id0075 &    13.258 &   1.675 &   2.821 &    -4.908 &   5.277 &   4.581 &   3.941 &   1.735 &  32.23 &  31.75\\
id0076 &    13.280 &   1.977 &   3.320 &    -5.381 &   5.510 &   4.593 &   3.843 &   2.038 &  39.73 &  38.87\\
id0077 &    13.341 &   2.000 &   3.230 &    -5.296 &   5.467 &   4.590 &   3.863 &   2.034 &  38.23 &  37.45\\
id0078 &    13.354 &   1.709 &   2.753 &    -4.813 &   5.224 &   4.577 &   3.959 &   1.742 &  30.73 &  30.31\\
id0079 &    13.389 &   1.706 &   2.959 &    -4.875 &   5.260 &   4.580 &   3.947 &   1.786 &  31.73 &  31.27\\
id0081 &    13.414 &   1.535 &   2.758 &    -4.567 &   5.091 &   4.566 &   3.995 &   1.616 &  27.23 &  26.91\\
id0082 &    13.452 &   1.710 &   2.804 &    -4.750 &   5.188 &   4.574 &   3.970 &   1.758 &  29.73 &  29.34\\
id0083 &    13.491 &   1.762 &   2.932 &    -4.813 &   5.224 &   4.577 &   3.959 &   1.819 &  30.73 &  30.31\\
id0085 &    13.572 &   1.845 &   3.160 &    -4.875 &   5.260 &   4.580 &   3.947 &   1.908 &  31.73 &  31.27\\
id0087 &    13.575 &   1.546 &   2.730 &    -4.418 &   5.007 &   4.558 &   4.017 &   1.624 &  25.23 &  24.97\\
id0088 &    13.579 &   1.549 &   2.785 &    -4.418 &   5.007 &   4.558 &   4.017 &   1.628 &  25.23 &  24.97\\
id0090 &    13.616 &   1.949 &   3.196 &    -4.942 &   5.294 &   4.582 &   3.934 &   1.984 &  32.73 &  32.23\\
id0091 &    13.640 &   1.776 &   3.156 &    -4.750 &   5.188 &   4.574 &   3.970 &   1.861 &  29.73 &  29.34\\
id0092 &    13.649 &   0.747 &   1.270 &    -3.013 &   4.268 &   4.470 &   4.156 &   0.815 &  14.08 &  14.06\\
id0093 &    13.677 &   1.694 &   2.750 &    -4.494 &   5.050 &   4.562 &   4.005 &   1.740 &  26.23 &  25.94\\
id0095 &    13.697 &   1.783 &   3.175 &    -4.750 &   5.188 &   4.574 &   3.970 &   1.884 &  29.73 &  29.34\\
id0096 &    13.735 &   1.696 &   3.134 &    -4.456 &   5.029 &   4.560 &   4.011 &   1.754 &  25.73 &  25.45\\
id0098 &    13.790 &   1.714 &   3.058 &    -4.494 &   5.050 &   4.562 &   4.005 &   1.801 &  26.23 &  25.94\\
id0099 &    13.795 &   2.056 &   3.182 &    -4.875 &   5.260 &   4.580 &   3.947 &   2.064 &  31.73 &  31.27\\
id0100 &    13.798 &   1.635 &   2.867 &    -4.314 &   4.951 &   4.553 &   4.031 &   1.703 &  24.08 &  23.85\\
id0101 &    13.814 &   1.814 &   3.043 &    -4.567 &   5.091 &   4.566 &   3.995 &   1.868 &  27.23 &  26.91\\
id0104 &    13.852 &   1.628 &   2.807 &    -4.270 &   4.928 &   4.551 &   4.036 &   1.707 &  23.63 &  23.41\\
id0105 &    13.864 &   1.694 &   2.678 &    -4.265 &   4.923 &   4.550 &   4.037 &   1.721 &  23.53 &  23.32\\
id0107 &    13.881 &   2.117 &   3.357 &    -4.942 &   5.294 &   4.582 &   3.934 &   2.149 &  32.73 &  32.23\\
id0109 &    13.905 &   1.742 &   2.978 &    -4.357 &   4.977 &   4.555 &   4.023 &   1.798 &  24.58 &  24.33\\
id0112 &    13.936 &   2.020 &   3.537 &    -4.782 &   5.206 &   4.576 &   3.965 &   2.076 &  30.23 &  29.82\\
id0114 &    13.954 &   2.018 &   3.499 &    -4.782 &   5.206 &   4.576 &   3.965 &   2.085 &  30.23 &  29.82\\
id0117 &    14.121 &   1.804 &   2.976 &    -4.222 &   4.897 &   4.547 &   4.044 &   1.850 &  23.03 &  22.83\\
id0119 &    14.139 &   1.141 &   1.856 &    -3.105 &   4.317 &   4.478 &   4.148 &   1.184 &  14.58 &  14.55\\
id0120 &    14.158 &   1.857 &   2.920 &    -4.242 &   4.908 &   4.548 &   4.041 &   1.888 &  23.23 &  23.03\\
id0121 &    14.177 &   1.718 &   3.151 &    -4.074 &   4.818 &   4.539 &   4.062 &   1.788 &  21.58 &  21.42\\
id0122 &    14.178 &   1.760 &   3.386 &    -4.121 &   4.843 &   4.542 &   4.057 &   1.823 &  22.03 &  21.86\\
id0123 &    14.201 &   0.647 &   1.212 &    -2.342 &   3.902 &   4.419 &   4.201 &   0.751 &  10.83 &  10.83\\
id0128 &    14.248 &   2.062 &   3.700 &    -4.531 &   5.071 &   4.564 &   4.000 &   2.116 &  26.73 &  26.43\\
id0129 &    14.257 &   1.524 &   2.951 &    -3.712 &   4.627 &   4.517 &   4.099 &   1.617 &  18.53 &  18.44\\
id0132 &    14.313 &   1.479 &   2.768 &    -3.564 &   4.552 &   4.507 &   4.111 &   1.564 &  17.48 &  17.41\\
id0133 &    14.368 &   1.721 &   3.228 &    -3.885 &   4.718 &   4.528 &   4.082 &   1.794 &  19.88 &  19.76\\
id0134 &    14.397 &   1.782 &   3.156 &    -3.972 &   4.763 &   4.533 &   4.074 &   1.858 &  20.63 &  20.49\\
id0137 &    14.475 &   1.731 &   2.754 &    -3.705 &   4.624 &   4.517 &   4.100 &   1.759 &  18.48 &  18.39\\
id0138 &    14.504 &   2.058 &   4.301 &    -4.278 &   4.931 &   4.551 &   4.036 &   2.122 &  23.68 &  23.46\\
id0139 &    14.507 &   1.870 &   3.526 &    -3.954 &   4.754 &   4.533 &   4.076 &   1.926 &  20.48 &  20.34\\
id0140 &    14.533 &   1.905 &   3.209 &    -3.972 &   4.763 &   4.533 &   4.074 &   1.953 &  20.63 &  20.49\\
id0141 &    14.565 &   1.680 &   3.187 &    -3.645 &   4.592 &   4.512 &   4.105 &   1.765 &  18.03 &  17.95\\
id0144 &    14.580 &   1.788 &   2.735 &    -3.637 &   4.589 &   4.512 &   4.106 &   1.787 &  17.98 &  17.90\\
id0147 &    14.611 &   1.754 &   3.155 &    -3.685 &   4.613 &   4.515 &   4.101 &   1.823 &  18.33 &  18.24\\
id0151 &    14.651 &   1.620 &   2.975 &    -3.425 &   4.480 &   4.498 &   4.123 &   1.692 &  16.53 &  16.47\\
id0159 &    14.797 &   1.983 &   2.025 &    -3.381 &   4.457 &   4.495 &   4.127 &   1.775 &  16.23 &  16.18\\
id0168 &    14.962 &   1.661 &   3.091 &    -3.160 &   4.345 &   4.481 &   4.144 &   1.730 &  14.88 &  14.85\\
id0178 &    15.044 &   1.580 &   2.453 &    -2.842 &   4.177 &   4.458 &   4.167 &   1.594 &  13.18 &  13.16\\
id0190 &    15.159 &   1.892 &   3.124 &    -3.285 &   4.408 &   4.489 &   4.134 &   1.930 &  15.63 &  15.59\\
id0204 &    15.333 &   1.311 &   2.317 &    -2.166 &   3.799 &   4.404 &   4.211 &   1.372 &  10.08 &  10.08\\
id0215 &    15.437 &   1.581 &   3.451 &    -2.532 &   4.010 &   4.435 &   4.189 &   1.648 &  11.68 &  11.67\\
id0261 &    15.896 &   0.661 &   1.159 &    -0.338 &   2.655 &   4.219 &   4.289 &   0.721 &   4.73 &   4.73\\
id0273 &    16.032 &   0.816 &   1.373 &    -0.437 &   2.721 &   4.230 &   4.286 &   0.859 &   4.93 &   4.93\\
id0290 &    16.163 &   0.293 &   0.607 &     0.778 &   1.884 &   4.084 &   4.313 &   0.362 &   2.96 &   2.96\\
id0306 &    16.310 &   0.378 &   0.748 &     0.781 &   1.881 &   4.083 &   4.313 &   0.448 &   2.95 &   2.95\\
id0323 &    16.501 &   2.092 &   2.702 &    -2.092 &   3.756 &   4.397 &   4.216 &   2.062 &   9.78 &   9.78\\
id0344 &    16.766 &   0.648 &   0.838 &     1.363 &   1.492 &   4.011 &   4.315 &   0.544 &   2.35 &   2.35\\
id0398 &    17.564 &   0.534 &   2.185 &     2.406 &   0.937 &   3.905 &   4.312 &   0.606 &   1.72 &   1.72\\
\hline 
\end{longtable}
\begin{list}{}{}
\item[$^{\mathrm{a}}$] In this catalog the star ID is designated by sorting the data in order of increasing $K_{S}$ magnitude.
\item[$^{\mathrm{b}}$] NACO $K_{S}$ band magnitude.
\item[$^{\mathrm{c}}$] NACO $(H-K_{S})$ color index.
\item[$^{\mathrm{d}}$] NACO $(J-H)$ color index.
\item[$^{\mathrm{e}}$] Absolute $V$ band magnitude.
\item[$^{\mathrm{f}}$] Log of the luminosity in $[L_{\odot}]$.
\item[$^{\mathrm{g}}$] Log of the effective temperature in $[K]$.
\item[$^{\mathrm{h}}$] Log of the surface gravity in $[N/kg]$.
\item[$^{\mathrm{i}}$] Color excess.
\item[$^{\mathrm{j}}$] Initial mass in $[M_{\odot}]$.
\item[$^{\mathrm{k}}$] Present-Day mass in $[M_{\odot}]$.
\end{list} 
}

\begin{figure*}
\centering
\includegraphics[width=15cm]{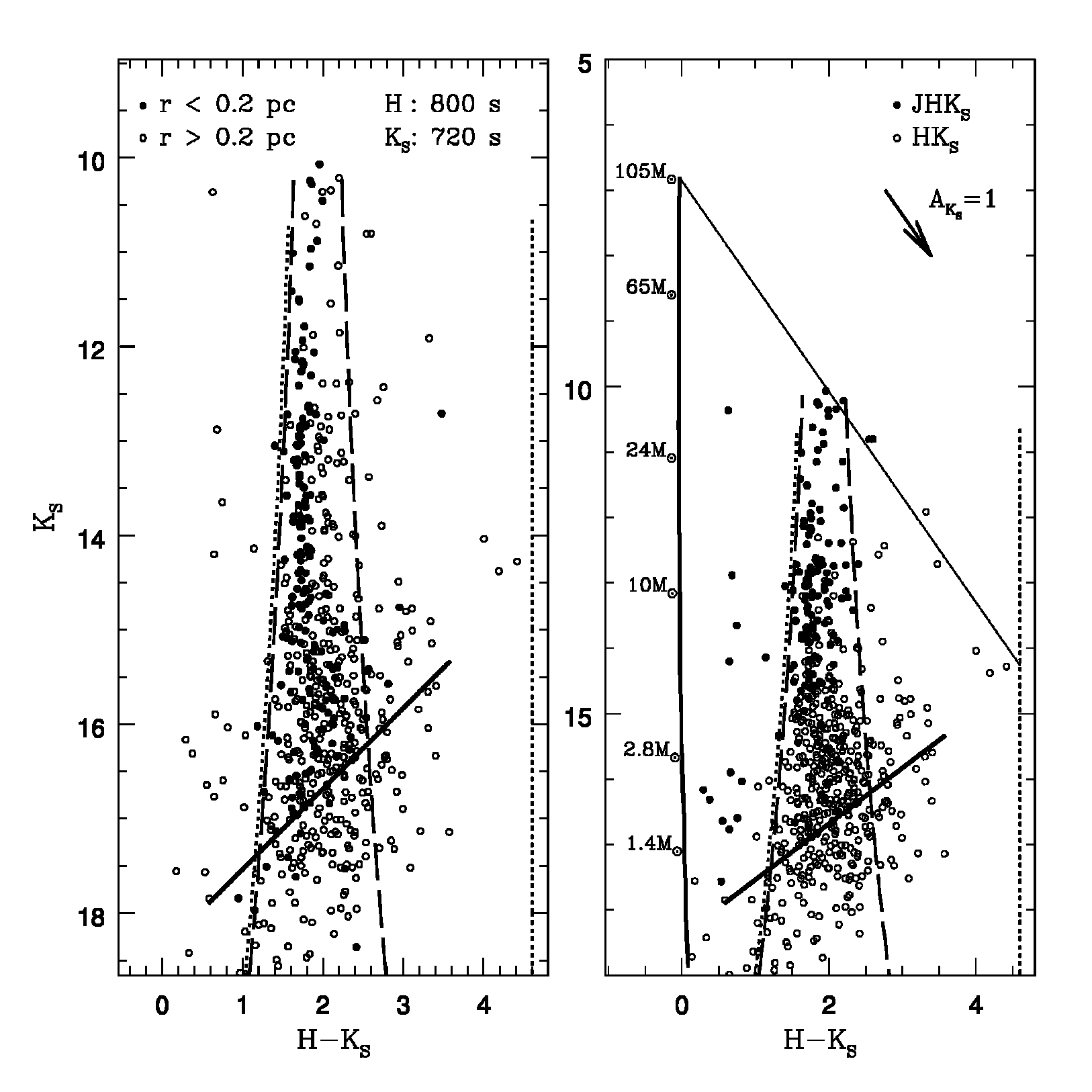}
\caption{\textbf{Left:} Color-magnitude diagram for stars in the NACO natural photometric system. \textbf{Right:} Same as before, but also showing a $2.5\;Myr$ unreddened isochrone and the projection of the reddening vector, both in the NACO system. The $50\%$ completeness limit is shown as a solid line in both panels. Dash and dotted lines represent two proposed color cuts to reject field contamination in our sample. The dotted line allows for the redder objects to be included while the dash line one is considered through the text as the strong color selection criterion.}
\label{cmd}
\end{figure*}

\subsection{Stellar Physical Parameters and Reddening from $HK_{S}$ Photometry only}

Our Bayesian method can only be applied to the brightest stars in the cluster for which we have complete $JHK_{S}$ information.  Therefore, to deal with the bulk of the stars (for which we only have $HK_{S}$ data) we have investigated an additional procedure. It consists in ``sliding'' each star along the reddening vector in the color-magnitude diagram until the theoretical isochrone is reached. The intersection point determines the intrinsic color and luminosity for each star. This method to correct for individual reddening makes use of the ratio of total to selective absorption ($R_{K_{S}}^{NACO}=1.61$) obtained from the average Galactic extinction curve in Section 3.1. As for the comparison between the two approaches, it is clear that none can discriminate extinction from the effects of e.g., rotation, binarity, or intrinsic NIR excess. However, it is important to note that one drawback of the CMD Sliding is not dealing with the photometric uncertainties: the final dereddened location for the program stars is the isochrone itself. The Bayesian approach, on the other hand, finds the most likely point in the {\it Theoretical Surface} for each star and then reads its reddening value. This difference led to, in some cases, underestimate the $E(H-K_{S})$ derived from the CMD Sliding with respect to their Bayesian counterparts. Still, there is an excellent correspondence between the two, as shown in Fig.~\ref{excesses}

\subsection{Cluster Membership} 
\label{Color Magnitude Diagram}

\begin{figure*}
\begin{center}
\includegraphics[width=15cm]{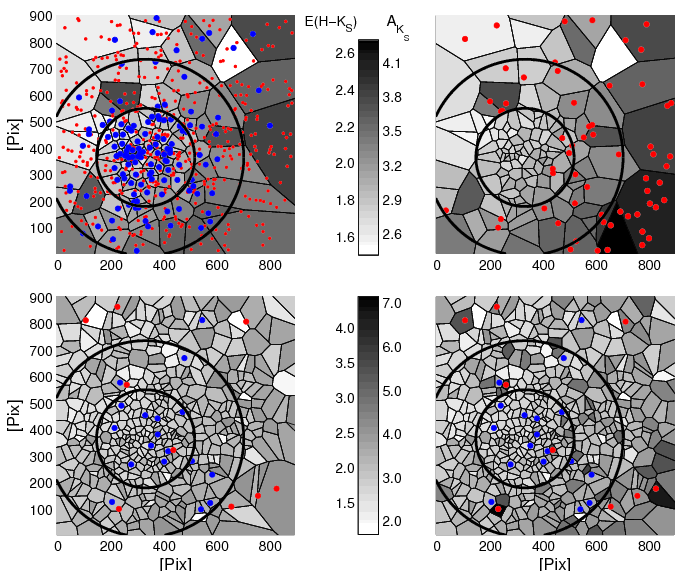}
\end{center} 
\caption{ {\bf(a) Top left}. Voronoi tessellation of the reddening values obtained with the Bayesian method  ($JHK_{S}$ stars are represented by blue dots). Several stars with only  $HK_{S}$ data (red dots) are seen to fall on each Voronoi cell.  Only stars within the strong color cut shown in Fig.\ref{cmd} are included in this diagram. {\bf (b) Top right}. Voronoi tessellation of stars with multi-color data without applying the strong color selection criterion. The red dots represent the (red) stars that are excluded by the strong color cut. {\bf (c) Bottom-left}. Here the tessellation is drawn from the CMD Sliding method, and applying the strong color selection criterion. The blue (red) dots represent massive stars ($> 65 M_{\odot}$) that are included (excluded) by the color cut. Note that these excluded red stars may also be foreground or background contamination. {\bf (d) Bottom-right}. Same as previous panel, but without applying the strong color cut to determine the Voronoi tessellation.  The reddening distributions corresponding to the two bottom figures are presented in Fig.~12 for the three indicated annular regions between projected distances of 0.2 and 0.4 pc at a Galactic center distance modulus of $14.52$ mag.}
\label{voron_redd_maps}
\end{figure*}

The color-magnitude (CMD) diagram for our data in the {\it natural} NACO photometric system is displayed in Fig.~\ref{cmd}. Shown in the right hand plot are our photometric completeness limit, the $2.5\;Myr$ isochrone, and the reddening path for the most luminous stars in the isochrone. Following previous investigations (e.g. SGB02; SBG05), we introduce a strong color cut to improve the rejection of putative non cluster members. This is indicated in Fig.~\ref{cmd} by the dashed lines that widen towards fainter magnitudes, where the photometric uncertainties increase. Approximately 105 stars would be discarded from our catalogs by this criterion.

Fig.~\ref{cmd} exposes a potentially serious problem of completeness. Due to the large and variable extinction to the cluster, some of the stars that we consider as field contamination may be very massive cluster members. While the bluer ones are likely to be foreground stars, the reddest can be foreground or background objects, but also heavily obscured main sequence cluster members. In fact all the {\it red} stars above the completeness limit of our photometry, that are discarded by the strong color cut criterion, would be more massive than $16\;M_{\odot}$ if membership could be proved.

In addition to the photometric information, we can examine the spatial positions of these groups of stars when they are plotted over Voronoi reddening maps as in Fig.~\ref{voron_redd_maps}. To understand this Figure we note that the tessellations of the top and bottom panels are drawn from $JHK_{S}$ and $HK_{S}$ data respectively, i.e. the reddening values have been derived using the Bayesian and CMD Sliding methods. On the other hand, the left panels pick the stars by taking into account the strong color selection criterion while the right ones allows the previously banned {\it red} stars. This is why the right panels are, naturally, more patchy.

Fig.~\ref{voron_redd_maps} (b) shows somewhat surprisingly that the redder stars (with masses hypothetically $> 16\;M_{\odot}$) tend to concentrate in highly extincted regions in the outskirts of the cluster, although a significant fraction of these are located at radii $<\;0.4 pc$. This is again illustrated in the two bottom panels of the Figure, which shows that this group could represent as much as a third of the total number of stars more massive than $65 M_{\odot}$ in the cluster. Clearly, if these stars are cluster members, our photometry (and indeed all published investigations) is incomplete even for the most massive stars thus rendering futile any attempt to derive an IMF from photometry alone.

Since the inclusion of these stars skews the extinction distribution towards higher values (upper panels of Fig.~\ref{incl example}), if the true distribution of extinction toward the cluster is Gaussian, as expected from the central limit theorem if the extinction is caused by a large number of dust clouds along the line of sight to the cluster, one could argue that most of these stars are foreground to the cluster.  This argument, however, is clearly too weak to be used to settle the delicate issue of cluster membership. Unfortunately, without further information such as spectroscopy or proper motions, we have no way of telling whether these are cluster members or foreground contamination. Thus, to build an IMF from our photometry, we have no other choice than to overpass the potential completeness problem and follow previous work by applying the strong color cut.

\subsection{Magnitude Limit Effect} 
\label{Magnitude Limit Effect}

Even when we impose the strong color selection the extinction towards the Arches is found to vary between $2.13 < A_{K_{S}} < 4.14$ magnitudes, which agrees with the range derived by SGB02 ($1.9 < A_{K} < 4.1$ magnitudes). Thus before we determine the IMF of the cluster, we need to consider the important systematic effect arising from the large extinction variation of 2 mag. in $K_{S}$ and 3.2 mag. in $H$: the Magnitude Limit Effect (MLE; Selman et al., \cite{selman99}). 

\begin{figure*}[!ht]
\begin{center}
\includegraphics[width=15cm]{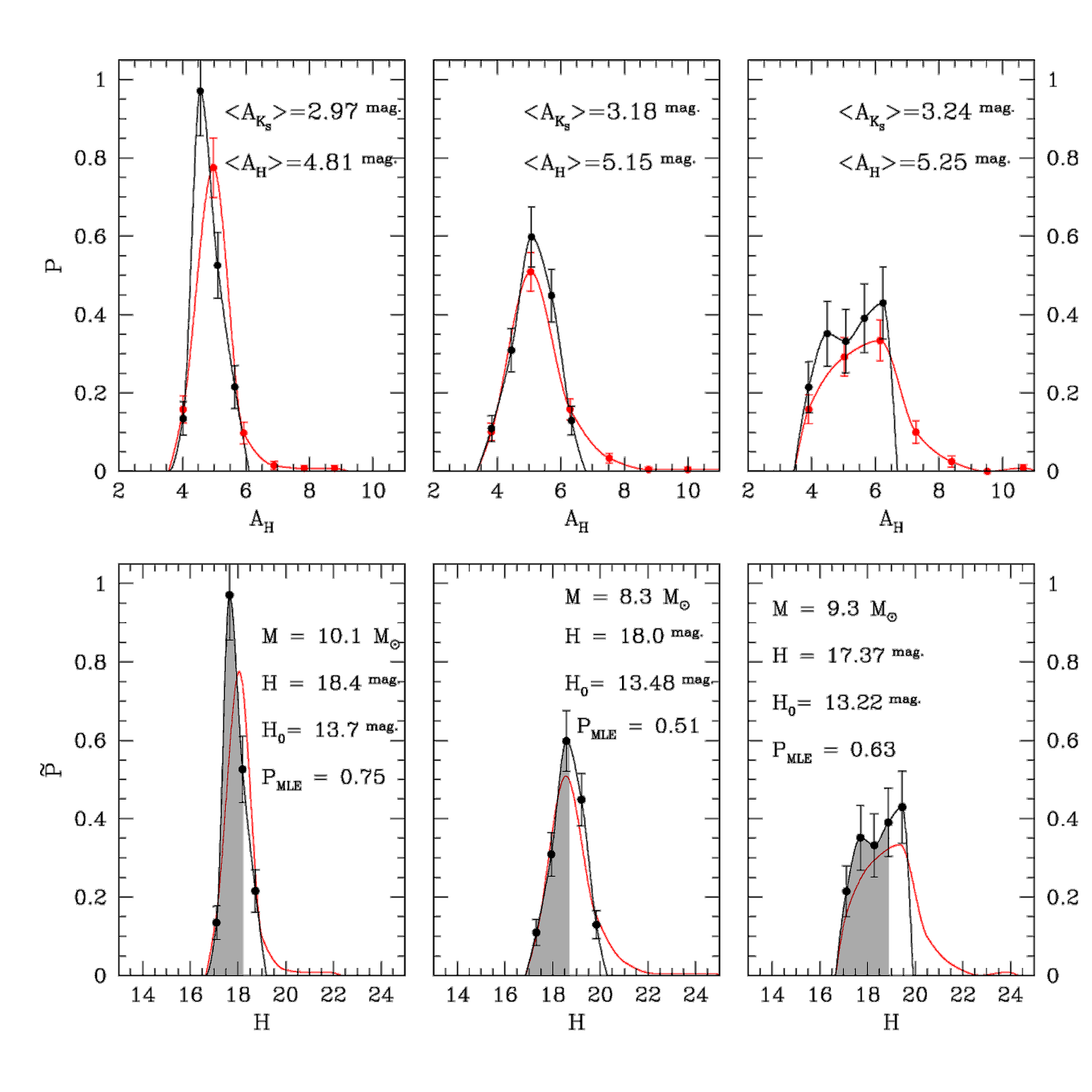}
\end{center}
\caption{In the upper panels the normalized extinction distributions $P(A_{H})$ are shown for the three radial bins considered in the text: $r<0.2\;pc$ (left), $0.2<r<0.4\;pc$ (middle), and $r>0.4\;pc$ (right). The extinction distributions of stars within the strong color cut of Figure~\ref{cmd} are represented by black lines; the distributions including the red objects excluded by the strong selection criterion are drawn by red lines. The bottom panels illustrate the Magnitude Limit Effect. These show how the correction is applied for stars at the three forenamed radial divisions. Apparent ($H$) and zero reddening ($H_{0}$) magnitudes are indicated in each panel together with the initial mass and inclusion probability. As described in the text, the correction factor for each star corresponds to the shaded area under the curve.}
\label{incl example}
\end{figure*}

Depending on their individual extinction, stars of a given mass (and luminosity) may or may not be included in our photometry catalog . Although previous investigations have noted these large star-to-star extinction variations towards the Arches, they have chosen to apply an average extinction correction (Kim et al. \cite{kim_arches}), or to subtract a radial term for radii $> 0.2 pc$, assuming the reddening in the cluster core to be uniform (SGB02, SBG05). Our data show that the radial trend results from the area of high extinction located SW of the cluster center. This is seen in our Voronoi reddening maps as a clear asymmetry in the NE-SW direction and in their Figure 6 as a widening scatter in color for stars as they increase their distance to the cluster's center. 

As described by Selman et al. (\cite{selman99})  this insidious effect can be corrected statistically if one knows the distribution of extinctions of the cluster stars. Briefly, from these distributions one computes the probability that each star is included in the photometric catalog. The distributions of extinction for the Arches cluster in each annulus, shown in the upper panels of Fig.~\ref{incl example}, were determined only from stars within the strong color cut selection criterion and above the completeness limits of Table~\ref{completeness_mag}. Then we use the NACO reddening vector to transform color excesses to absorption,

\begin{displaymath}
 A_{K_{S}}=R_{K_{S}}^{NACO}\times E(H-K_{S})  
\end{displaymath}
\begin{displaymath}
 A_{H}=\frac{(R_{K_{S}}^{NACO}+1)\times A_{K_{S}}}{R_{K_{S}}^{NACO}}
\end{displaymath}

\begin{figure*}
\centering
\includegraphics[width=15cm]{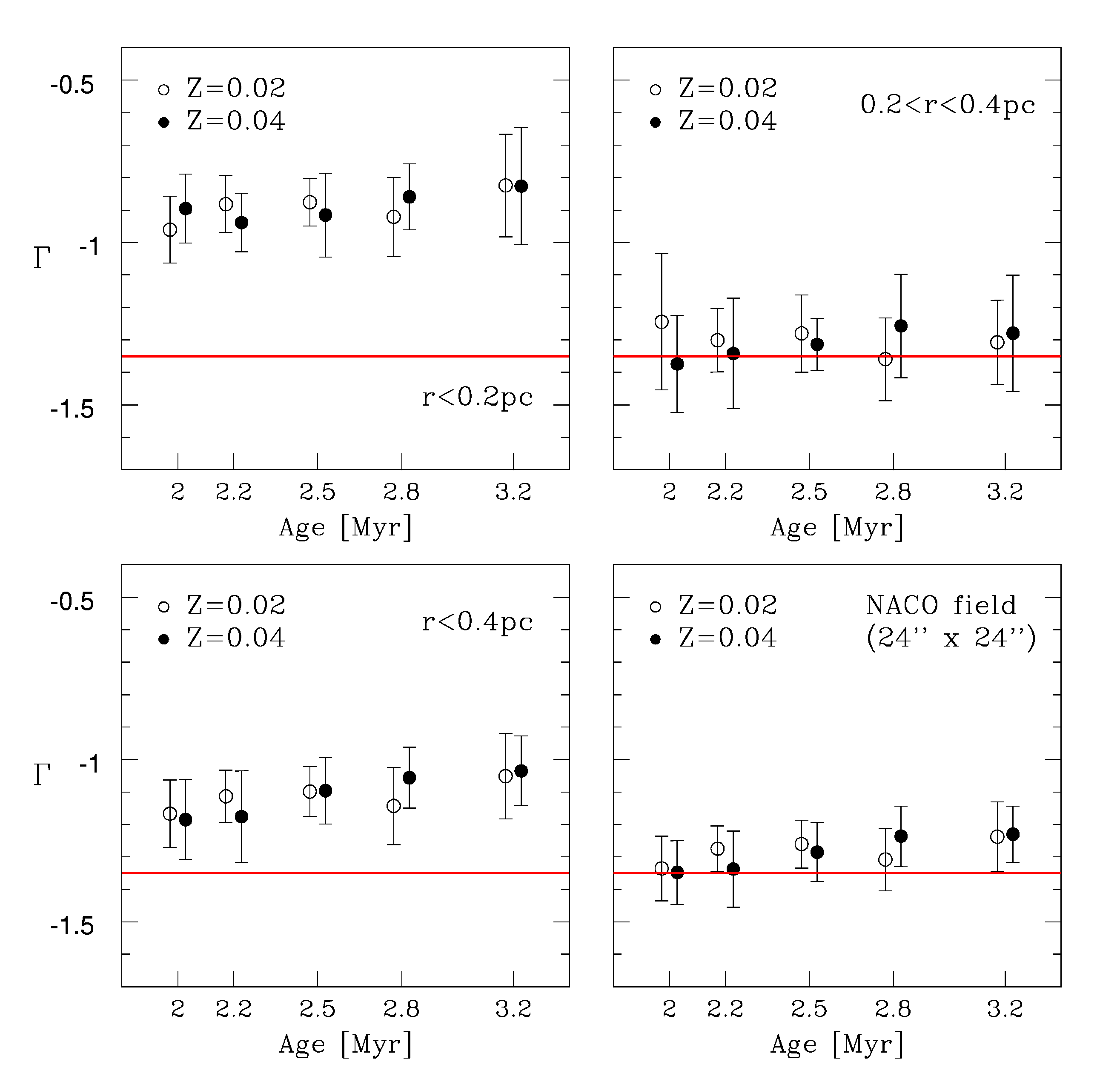}
\caption{Each panel show our results at a given distance from the cluster center. Open (solid) circles correspond to the IMF slope determined by comparison with stellar models of solar (double-solar) metallicity. More specifically, circles stand for the average slope computed from several different binnings. Accordingly, the error bars show the dispersion of these values, i.e. our estimation for the uncertainty in $\Gamma$ due to binning (the procedure is explained in Appendix~C). The x-axis represents the age of these models, ranging from $2.0$ to $3.2\;Myr$. Red lines represent the Salpeter slope as a reference.} 
\label{resumen3}
\end{figure*}

The normalized extinction distributions $P(A_{H})$ and $P(A_{K_{S}})$ can be used as probability densities and, since extinction decreases with increasing wavelength, $P(A_{H})$ is the broader distribution and dominates the magnitude limit effect.  To compute the magnitude limit correction it is simpler to express the probability distribution of extinction in terms of $H$ magnitude instead of absorption.  For each star the Bayesian or CMD Sliding methods give the absolute magnitude, which can be converted to a zero reddening apparent magnitude $H_{0}$ by adding the assumed distance modulus of the cluster, $(m-M)_{0}=14.52$. Thus, for each star we transform $P(A_{H})$ into  $\tilde{P}(H)= P(A_{H}+H_{0})$ and, assuming that there are no systematic trends in the distribution of extinction with stellar luminosity, the probability that the star $j$ is included in the catalog is given by, 

\begin{displaymath}
  P_{MLE_{j}}=1-\int_{H_{lim}}^{\infty}\tilde{P}_{j}(H)~dH
\end{displaymath}
where $MLE$ denotes that the correction is due to the Magnitude Limit Effect and $H_{lim}$ is the completeness limit of the corresponding radial bin, given in Table \ref{completeness_mag}.

Bottom panels of Fig.~\ref{incl example} presents an illustration of the MLE. For each star we use the extinction distribution of the annular ring where it is located and the Figure shows the extinction distribution as a function of magnitude. The distributions obtained using the CMD Sliding method are shown including all stars (red line), and including only stars satisfying the strong color selection discussed in the previous section (black line), but only the latter was used to calculate the inclusion corrections. Clearly, even at magnitudes brighter than the completeness limit the corrections can be significant. Therefore, if not corrected, the magnitude limit effect may severely deplete the lower mass bins of the IMF thus artificially flattening the slope.

\section{Results}

\subsection{The Initial Mass Function of the Arches cluster}

As discussed in the preceding sections, in order to interpret the data we made the simplifying assumptions that all the stars in the Arches cluster formed $2.5\;Myr$ ago, have solar metallicity, and are located at a distance modulus of $14.52$ magnitudes. The physical parameters and individual reddening for cluster members were derived using the CMD Sliding or the Bayesian method, depending on the availability of $J$-band information. It is noted that only stars within the strong color cut were considered to build the IMFs at different radii. This is explained in Section 3.5, where we emphasize that to proceed otherwise implies that even the most massive stellar bins of the IMF would be affected by incompleteness.
 
Linear fits to the data have been obtained with the routine \textbf{fitexy} (Press et al. \cite{press}), that considers uncertainties in both coordinates. In regards to y-axis errors, we note that the $j-th$ bin of the IMF can be represented as,
\begin{displaymath}
 log \xi_{j}=log\left(\frac{N_{j}}{\overline{P_{D}}\times \overline{{P_{MLE}}}\times C}\right)
\end{displaymath}
where $\overline{P_{D}}$ and $\overline{P_{MLE}}$ are the average corrections for stars in that particular mass bin, $N$ is the number of stars actually observed, and $C$ is a constant grouping the size of the logarithmic mass interval and the area normalization of the IMF. Thus error bars are computed considering the poissonian uncertainty in the counts together with a scaling due to the incompleteness and MLE corrections.

The typical formal fitting uncertainties for our IMF slopes, $\Gamma$, are $\sim 0.15$. But to give a better estimate of the real errors, we must also consider how the size and starting points of the logarithmic mass bins affect our results. To clear this issue, in Appendix~C we applied several different binnings to our data to obtain the slopes. The dispersion of these values, our estimate of uncertainty in the IMF slope due to a particular binning, is $0.09$\footnote{Ma\'iz Apell\'aniz \& \'Ubeda (\cite{jesus}) use bins of variable width, such that each contains an equal number of stars. In this way, they minimize the biases in the IMF slope determination by assigning the same statistical weight to every bin. Such numerical biases play, however, only a minimum role in our particular situation for the Arches, i.e. a few uniform-size bins and hundreds of stars. As can be seen in their Table 1, numerical experiments indicate that for \textit{our case} the bias is not significant and the slope $\Gamma$ flattens only by $0.014-0.015$ with respect to its input value. This is negligible compared with our systematic errors.}. Furthermore, to be safe in view of the spread in age and metallicity determinations of previous literature, we built IMFs using different isochrones. Results are presented in Fig.~\ref{resumen3} and Table~\ref{resumen3table}, where we again estimate an uncertainty of less than $0.1$ owing to the choice of stellar models that are compared with our observations. Therefore we judge the total error in $\Gamma$ due to fitting, binning, and the choice of isochrone, to be $\pm0.2$.

The IMF in each annulus is derived for stars above $10\;M_{\odot}$, using our preferred model, the $2.5\;Myr$ isochrone of solar metallicity (Lejeune \& Schaerer \cite{lejeune}). Taking into account our estimation for the real errors, the slopes range from $\Gamma=-0.88\pm0.20$ for $r<0.2\;pc$, to $\Gamma=-1.28\pm0.20$ in the $0.2-0.4\;pc$ annulus. The IMF slope for all the stars we observe in the $\sim24''\times24''$ NACO field (displayed in Fig.\ref{voron}) of the Arches cluster is $\Gamma=-1.26\pm0.20$. But because of possible problems of field contamination discussed below,  we consider that our best estimate for the global IMF is obtained for the region $r<0.4\;pc$ for which we derive a slope of $\Gamma=-1.1\pm0.2$.

\begin{figure}
\includegraphics[width=9.5cm]{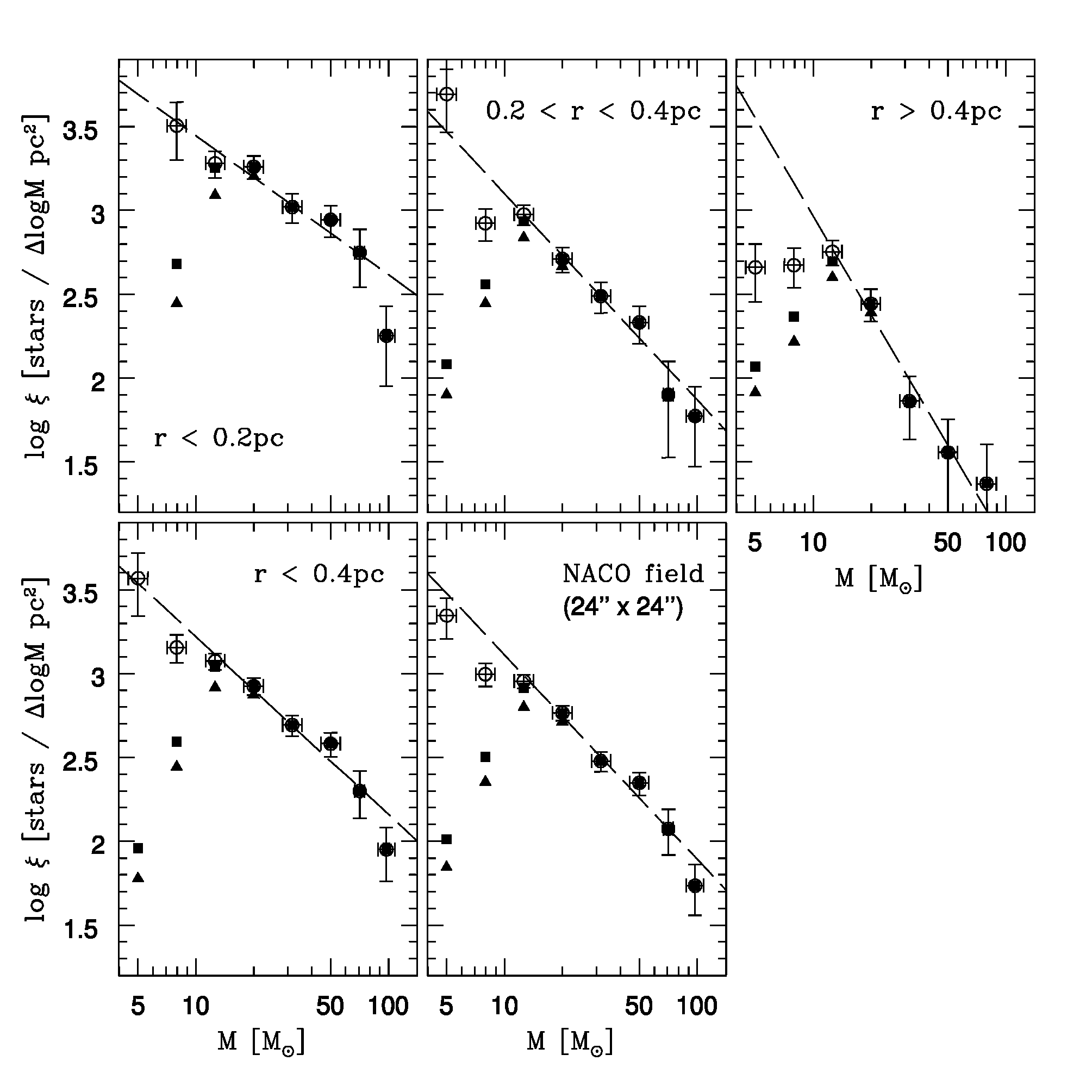}
\caption{An illustration on how the MLE can mimic turnovers in the IMF. A $2.5\;Myr$ isochrone of solar metallicity and no enhanced mass loss (Lejeune \& Schaerer \cite{lejeune}) was used to derive the cluster IMF at different radii. A particular binning with $\Delta logM=0.2\;dex$ has been chosen. Triangles represents raw data; squares and circles are the detection and MLE corrected counts respectively. Note how the raw and detection corrected counts fall suddenly for stars just below $10~M_{\odot}$.}
\label{temporal}
\end{figure}

Finally, it is worth noting that at the low mass end of our IMF (particularly in the mass bin $4-6.3\;M_{\odot}$) the values of $log\;\xi$ are quite uncertain because of the differential extinction corrections that reach average correction factors as high as $30$. This is exemplified in Fig.~\ref{temporal} where the IMF of the Arches cluster is shown for a particular binning choice. Filled triangles show the raw results; filled squares show the data corrected for incompleteness; and open circles with error bars represent the corrections for incompleteness and MLE. These high uncertainties led us to determine fits to the data only for stars more massive than $10\;M_{\odot}$, for which the MLE corrections are not important. This underlines the fact that the MLE is a steep function of magnitude; but also illustrate that, when ignored, it can mimic turnovers in the IMF even for stars within the completeness limit derived from artificial stars experiments.

\subsection{Field Contamination}

Because of the high density towards the Galactic center, the steep IMF for the outer region of the cluster ($\Gamma=-1.95$ for $r > 0.4\;pc$) may be an indication of non-negligible field contamination. Although the low mass limit for the linear fit is $10\;M_{\odot}$, this slope is rather uncertain. To settle this issue in a more quantitatively way we compare our observations with the results of Kim et al. (\cite{kim_arches}). They observed three control fields separated by $\sim 60''$ from the center of the Arches cluster, i.e. twice the value of the tidal radius of $25''-30''$ derived from the comparison of HST observations and N-body simulations (Kim et al. \cite{kim2}).

\begin{figure*}
\centering
\includegraphics[width=15cm]{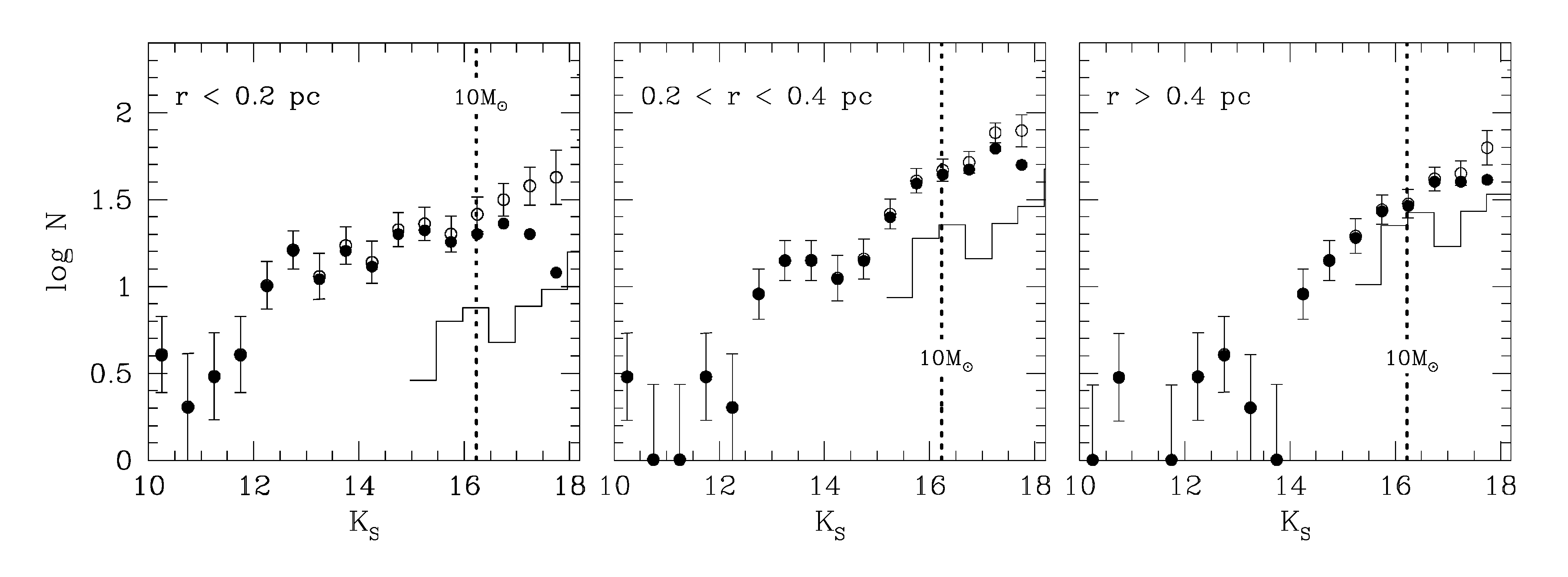}
\caption{A comparison between our $K_{S}$-band LFs of the Arches cluster and the incompleteness corrected average LF of three control fields from Kim et al. (\cite{kim_arches}). Filled circles represent our raw data, while open circles are corrected by incompleteness. Vertical dotted lines mark the $10\;M_{\odot}$ position for the $2.5\;Myr$ isochrone of solar metallicity, considering an average reddening of $A_{K_{S}}=3.1$ mag. From left to right the three annular regions used in this work are shown: $r<0.2\;pc$, $0.2<r<0.4\;pc$, and $r>0.4\;pc$.}
\label{field_c}
\end{figure*}

Reddened by the mean $K_{S}$ extinction of each annular region, the average of the three control fields incompleteness corrected $K_{S}$-band Luminosity Functions (LF) is presented in Fig.~\ref{field_c}. We can see from the Figure that within $0.4 pc$ the incompleteness corrected $K_{S}$-band LFs of the Arches are not significantly affected by field contamination, especially for masses above $10 M_{\odot}$ ($K_{S} < 16.23$ mag. for a $\langle A_{K_{S}}\rangle=3.1$ mag.). The situation is different for $r > 0.4\;pc$, where the field star component can be as high as $50\%$ of the cluster counts. However we must note that this estimate is an upper limit, as Kim et al. (\cite{kim_arches}) used only $K_{S}$-band data while we used stars with both $H$ and $K_{S}$ information. This allowed us to use the strong color cut criterion to reduce the number of non cluster members thus significantly lowering contamination in our IMF determination.

\begin{table}
\caption{IMF slopes plotted in Fig.~\ref{resumen3}. For each annular region the IMF is obtained from the comparison of our observations with a set of isochrones (Lejeune \& Schaerer \cite{lejeune}) with solar metallicity ranging from $2$ to $3.2$~Myr. The values between parenthesis correspond to slopes derived using double solar metallicity models.} 
\label{resumen3table}            
\centering                          
\begin{tabular}{c c c c c}        
\hline\hline                 
 $\tau^{\mathrm{a}}$   & $\Gamma_{r \leq 0.2pc}$ & $\Gamma_{0.2 < r \leq 0.4pc}$ & $\Gamma_{r < 0.4pc}$ & $\Gamma_{NACO~field}$ \\ 
\hline                        

   2       &  -0.96 (-0.90)   & -1.24 (-1.37)   & -1.17 (-1.19)   & -1.34 (-1.35)   \\
   2.2     &  -0.88 (-0.94)   & -1.30 (-1.34)   & -1.11 (-1.18)   & -1.27 (-1.34)   \\
   2.5     &  -0.88 (-0.92)   & -1.28 (-1.31)   & -1.10 (-1.10)   & -1.26 (-1.29)   \\
   2.8     &  -0.92 (-0.86)   & -1.36 (-1.26)   & -1.14 (-1.06)   & -1.31 (-1.24)   \\
   3.2     &  -0.82 (-0.83)   & -1.31 (-1.28)   & -1.05 (-1.04)   & -1.24 (-1.23)   \\
\hline                                   
\end{tabular}
\begin{list}{}{}
\item[$^{\mathrm{a}}$] Age of the isochrone in Myr.
\end{list}
\end{table}

\section{Discussion}
\subsection{Comparison with previous investigations}

From beginning to end the analysis of our high-quality NACO observations of the Arches cluster turned out to be a tortuous road.  High extinction forced us to work in the natural photometric system of NACO for which the reddening parameters had to be calculated.  Even for known (or assumed) age and metallicity, the existing models are limited by the availability of suitable stellar atmosphere models for massive stars. The extinction within the strong color cut criterion can vary as much as $2.13 < A_{K_{S}} < 4.14$ mag., hiding even some of the most massive clusters stars beyond the limit of photometric completeness. It is therefore hardly surprising if at the end of the road we find results that are not totally consistent with previous work.

In particular, our IMFs are steeper than previous investigations, and while we also observe the flattening towards the center of the cluster reported by previous authors, this flattening is much less dramatic in our data, which, however, span only the more massive mass range ($>10M_{\odot}$). Ranging from $\Gamma=-0.88\pm0.20$\ for $r<0.2\;pc$, to $\Gamma=-1.28\pm0.20$ in the $0.2-0.4\;pc$ annulus, the global IMF slope of $\Gamma=-1.1\pm0.2$ is consistent within the uncertainties with the Salpeter value, $\Gamma=-1.35$.

In the mass range in common with previous investigations ($1.0 < log(M/M_{\odot}) < 1.8$), the discrepancy in the central part is very large: while in our data the IMF of the core is clearly flatter than in the outer parts, it is still significantly steeper than claimed by SBG05 who obtained $\Gamma=-0.26\pm0.07$ for $r<0.2\;pc$. For the $0.2-0.4\;pc$ annulus, SBG05 show a broken power law with $\Gamma=-1.21$ ($16 <  M/M_{\odot} < 60$) that flattens to $\Gamma=-0.69$ ($6 < M/M_{\odot} < 16$). At least above $10M_{\odot}$ there is no sign of such flattening in our data.

Similar discrepancies occur when we compare our results with the work of Kim et al. (\cite{kim_arches}), that is the first study to observe control fields to estimate the background contamination. From the $K_{S}$-band LF only, they find a PDMF slope in an annular region $5'' - 9''$ (our $0.2- 0.4\;pc$ annulus corresponds to $5'' - 10''$) of $\Gamma=-0.91\pm0.08$ in the mass range $1.3<M/M_{\odot}<50$ by using an average extinction value of $A_{K_{S}}=3.1\pm0.19$ mag. Since their counts actually show an \textit{excess} of stars in the lowest mass bins, we must conclude that uncertainties in the background subtraction must also affect the counts in these bins (see also Portegies Zwart et al. \cite{portegies_zwart}).  Thus, if anything, their slope gets even flatter when the lower mass bins are removed.  If their data are restricted to the mass range in common with our work, $M>10M_{\odot}$, the slope flattens to $\Gamma=-0.8$, which is inconsistent with our steeper result in the $0.2-0.4\;pc$ region.

It is important to note that the discrepancies in $\Gamma$ are not due to differences in the assumed age of the cluster. We fixed this parameter to be $2.5\;Myr$, but similar, and even steeper slopes are obtained when we use the $2\;Myr$ models of Lejeune and Shaerer (\cite{lejeune}). This is shown in Fig.~\ref{resumen3} and Table~\ref{resumen3table}, where we experimented with isochrones of different ages and metallicities. To shed light into the reasons why our work is inconsistent with SBG05 and Kim et al. (\cite{kim_arches}), we list below some possible sources of inconsistency:
\begin{enumerate}
	\item  \title{\textbf{Photometric Calibration and Individual Dereddening:}}
	SBG05 chose to work in the NICMOS photometric system. Their NACO observations are transformed by applying zero points derived from the HST/NICMOS photometry of Figer et al. (\cite{figer99}). This procedure has the advantage of having a great number of local standards, and, as it uses HST photometry, also avoids possible uncertainties in aperture corrections for varying PSFs. On the other hand, the theoretical Geneva isochrone is reddened to the mean extinction of the cluster and then also transformed to the HST/NICMOS system by using color equations derived from unreddened standards and red 2MASS sources in the cluster field. As red sources are calibrated against red sources, it appears that this procedure is not affected by the systematic effects described in Section 2.3.1. Regrettably there are no details in their paper that allow us to judge these transformation equations.

	The important difference with our modus operandi, i.e. working in the NACO natural system, is that with their calibration and procedures it is not possible to derive individual reddenings. Thus their MF is computed from the average extinction value $A_{V}=25.2$ (or $A_{K_{S}}=2.82$ mag., SBG05). To account for the high extinction variation in the field, SGB02 and SBG05 postulate a radial reddening gradient to analyze their data outside 0.2 pc (see SGB02's Figure 6) while in the innermost region the extinction is assumed as uniform. We have found that: 1) Inside the cluster core, for stars within the strong color cut, the variation is as high as $2.25 < A_{K_{S}} < 3.72$ mag., with standard deviation of 0.26 mag., and 2) This procedure artificially changes colors and magnitudes of stars that don't follow the proposed radial trend, as can be seen from the asymmetries in the Voronoi reddening maps of Fig.~\ref{voron_redd_maps}. The need for individual dereddening in the field of the Arches cluster is evident. To get an idea of the differences that arise from these procedures we also computed IMF slopes from the average extinction. A radial reddening correction for radii larger than $0.2 pc$, following SGB02 and SBG05, was also applied. Results are shown in Table ~\ref{individual_vs_mean}.

	\begin{table}
	 \caption{Slopes of the Arches cluster IMF derived from average dereddening are shown in the two first rows. A $2.5\;Myr$ theoretical isochrone with solar metallicity (Lejeune and Shaerer, \cite{lejeune}) was used to derive these results.} 
	 \label{individual_vs_mean}
	 \centering
	 \renewcommand{\footnoterule}{}  
	 \begin{tabular}{c c c c } 
	 \hline\hline
	 \quad  & $r < 0.2 pc$ &  $0.2 < r < 0.4 pc$ & $r < 0.4 pc$ \\
	 \hline
	$\Gamma^{a}$  &$-0.60\pm0.17$ & $-1.15\pm0.20$    &$-0.81\pm0.19$ \\ 
	 \hline
	$\Gamma^{b}$  &$-0.60\pm0.17$ & $-1.23\pm0.23$    &$-0.96\pm0.24$ \\ 
	 \hline
	$\Gamma^{c}$  &$-0.88\pm0.20$ & $-1.28\pm0.20$    &$-1.1\pm0.2$ \\ 
	 \hline
	\end{tabular}
	\begin{list}{}{}
	\item[$^{\mathrm{a}}$] IMF slope derived without radial correction.
	\item[$^{\mathrm{b}}$] IMF slope derived with a radial correction for $r > 0.2 pc$ to take into account the differential extinction (as proposed in SGB02, SBG05).
	\item[$^{\mathrm{c}}$] This work. Individual derredening results described in Section 4.1.
	\end{list} 
	\end{table}

	At least qualitatively, Table ~\ref{individual_vs_mean} reproduce earlier work in the sense that when the radial correction is applied, it steepens the IMF slope. But then we conclude that without individual dereddening, the IMF at all radii flatten with respect to the results obtained in Section 4.1. This is especially true for the $r < 0.2pc$ region ($\Gamma=-0.6$ versus $\Gamma=-0.88$), where all previous investigations considered the extinction as uniform. The inadequacy of assuming average extinction values in annuli for the Arches cluster field is further explored in Appendix~D.

	\item  \title{\textbf{IMF versus PDMF:}}
	Another source of deviation from SGB02, SBG05 and Kim et al. (\cite{kim_arches}) is the correction for stellar evolution. We derived initial masses while they obtain the present day mass for the program stars. Following their approach to re-do our work, we conclude that our slopes are systematically steeper than the ones obtained in previous investigations by $\sim 0.1-0.15$.

	\item  \title{\textbf{Extinction Corrections:}}
	It is important to note that the Magnitude Limit Effect reduces the photometric completeness of previous investigations. In the same region studied by Kim et al. (\cite{kim_arches}) we find, from multi-color data, a variation of $\Delta E(H-K_{S})=1.43.$ mag., or $\Delta A_{K_{S}}=2.56$ mag. using the standard Rieke and Lebofsky (\cite{rieke}) law. Thus, if not considered, the MLE would reduce the completeness of their data from $1.3M_{\odot}$ to $4.7M_{\odot}$.

	This systematic effect arises in the presence of high differential reddening. To ignore it can result in a flattening of the IMF or mimic a turnover. This was illustrated in Fig.~\ref{temporal}, where the incompleteness-corrected data experiments a flattening at $\sim 10M_{\odot}$ and then a sudden decline at slightly lower masses.
\end{enumerate}

To end this Section, we conclude that the deviations from previous work can be explained from a combination of the above factors, being the individual reddening the most important. \textit{Our strong conclusion is that down to a limit of $10 M_{\odot}$, the global IMF of the Arches cluster can be fit by a single power law of slope consistent with the Salpeter value}.

\subsection{Cluster Properties and Mass Segregation}

\begin{table*}
 \caption{Physical parameters of the Arches cluster} 
 \label{king_t}
 \centering
 \renewcommand{\footnoterule}{}  
 \begin{tabular}{l c c l}
 \hline\hline
 Parameter & & Value & Comments \\
 \hline
 Core radius		&$R_{c}$  			& $0.14\pm0.05\;pc$	& Derived by fitting the empirical King density law \\
& & &to the observed $\Sigma(R)$\\ 
 Tidal radius		&$R_{t}$  			& $\sim1\;pc$ 		& Derived by Kim et al. (\cite{kim2}) \\
 Central concentration	&$c=log\frac{R_{t}}{R_{c}}$	& $0.84$		& Comparable to the less concentrated galactic globulars\\
& & &according to Harris (\cite{harris})\\
 Central surface density&$\Sigma_{0}$ 			& $2.2(\pm0.4)\times10^{3}\;stars\;pc^{-2}$		& Derived by fitting the empirical King density law \\
& & &to the observed $\Sigma(R)$\\
 Central volume density	&$\rho_{0}$ 			& $8.0(\pm1.5)\times10^{3}\;stars\;pc^{-3}$		& \\
 Central mass density	&$\rho_{m,0}$ 			& $2.0(\pm0.4)\times10^{5}\;M_{\sun}\;pc^{-3}$		& For an average mass of $25.4\;M_{\odot}$ in the $>10\;M_{\odot}$ range \\
 Cluster mass		&$M_{cl,1}$ 			& $\sim2.0(\pm0.6)\times10^{4}\;M_{\sun}$			& For an extrapolation of the photometric mass down to $1\;M_{\odot}$\\
\quad &$M_{cl,2}$&$\sim3.1(\pm0.6)\times10^{4}\;M_{\sun}$&For an extrapolation down to $0.08\;M_{\odot}$ using a Kroupa IMF (\cite{kroupa})\\
 Predicted & & &\\
velocity dispersion$^{a}$	&$\sigma=(\frac{0.4GM_{cl}}{R_{hm}})^{1/2}$	& $9\;km\;s^{-1}$			& Using $M_{cl,1}$ and a half mass radius of $0.4\;pc$ derived in SGB02\\
 \hline
\end{tabular}
\begin{list}{}{}
\item[$^{\mathrm{a}}$] Equation $4-80b$ of Binney \& Tremaine (\cite{binney}).
\end{list} 
\end{table*}

\begin{figure}[!ht]
\begin{center}
\includegraphics[width=8.5cm]{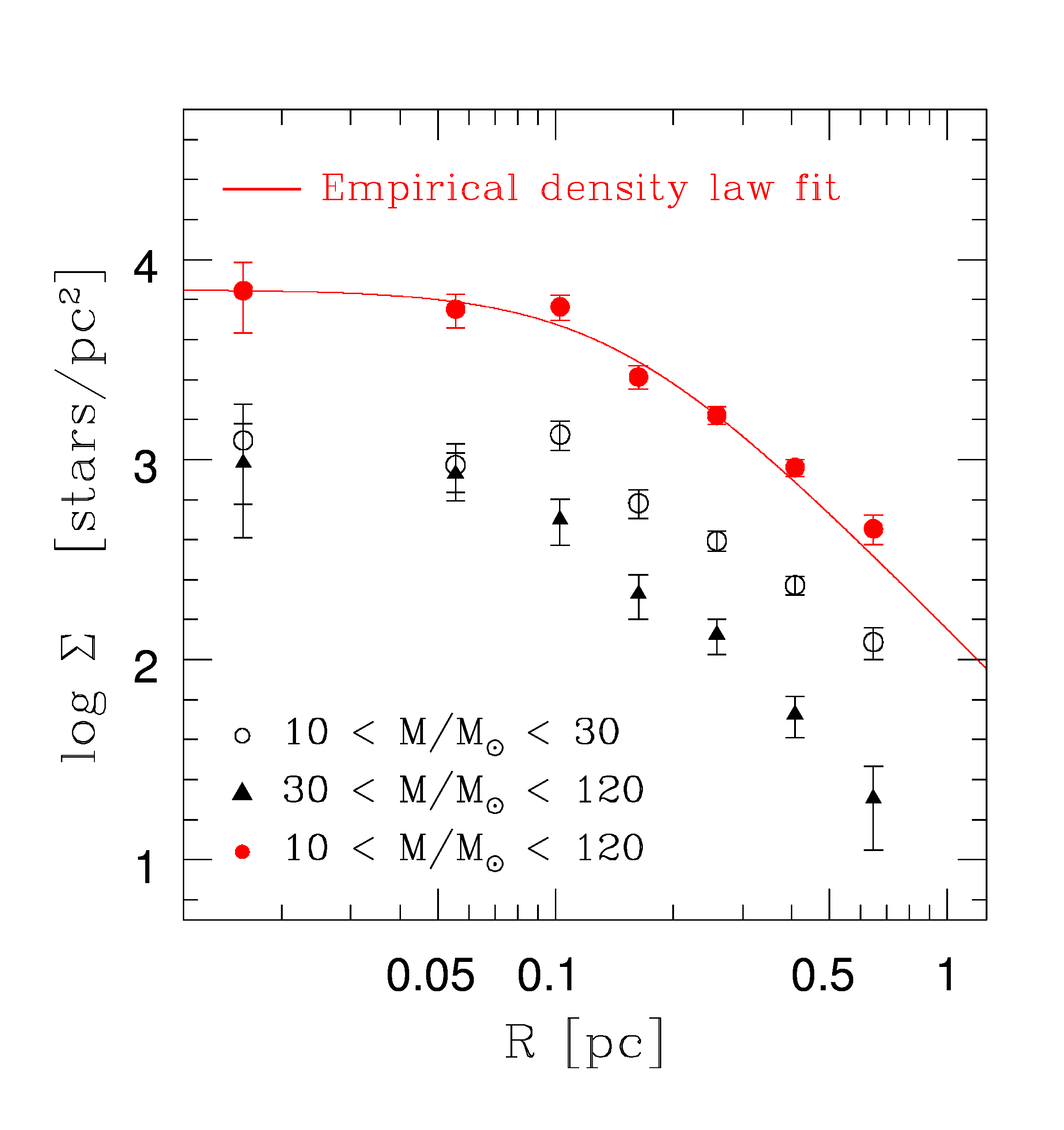}
\end{center}
\caption{Radial density profiles for stars at different mass bins, with error bars reflecting the corrections for systematic errors and Poisson statistics. The red line, with an offset of +0.5 dex for clarity, represents the fit of the empirical King density law (\cite{king}) to the surface density profile built from $10-120\;M_{\odot}$ stars .}
\label{rho_profile}
\end{figure}

To obtain physical parameters for the Arches we compared the radial surface density profile with the empirical King density formula for the inner parts of a concentrated cluster (King \cite{king}):  

\begin{displaymath}
 \Sigma(R)= \frac{\Sigma_{0}}{1+(R/R_{c})^2} \hspace{0.5cm}  [\frac{stars}{pc^{2}}]
\end{displaymath}
where $R$ is the projected distance in parsecs, $\Sigma_{0}$ is the central surface density and $R_{c}$ is the core radius, defined observationally as the distance at which the surface density profile fall at half of its central value. Despite the fact that at Arches's current evolutionary status the cluster is not in near-thermodynamic equilibrium, the comparison with the empirical King formula is useful to derive volume quantities and can be justified with the goodness of the fit to our data. Fig.~\ref{rho_profile} represents the surface density profile constructed from stars within the completeness limit of the photometry along with the fitting function as a red line. The model has an $R^{2}$ statistic of $0.95$, i.e. it explains $95\%$ of the variance in our data.

\begin{table*}
 \caption{Using the prescriptions of Oey \& Clarke (\cite{oeyclarke}) we compute $p(M_{max}|M_{up})$, i.e. the probability of observing a maximum stellar mass $M_{max}$ for a given $M_{up}$.} 
 \label{oey_probs}
 \centering
 \renewcommand{\footnoterule}{}  
 \begin{tabular}{c c c c c c c}
 \hline\hline
$\Gamma^a$ & $M_{max}$ & $N (> 10 M_{\odot})^b$ & $p(M_{max}|M_{up})$ & $p(M_{max}|M_{up})$ &  $p(M_{max}|M_{up})$ & $p(M_{max}|M_{up})$\\
\cline{4-4} \cline{5-5} \cline{6-6} \cline{7-7} 
\quad & \quad & \quad & $M_{up}=200M_{\odot}$ & $M_{up}=150M_{\odot}$ &  $M_{up}=135M_{\odot}$ & $M_{up}=120M_{\odot}$\\
 \hline
-1.1 & $120 M_{\odot}$   & 343 & $10^{-5}$ & $0.006$ & $0.06$ & 1 \\
 \hline
\end{tabular}
 \begin{list}{}{}
 \item[$^{\mathrm{a}}$] Slope of the IMF within 0.4 pc.
 \item[$^{\mathrm{b}}$] Number of stars within 0.4 pc.
 \end{list}
 \end{table*}

From the fit to $\Sigma(R)$ it is possible to go from the observed surface radial profile to true physical volume densities. Assuming an spherically symmetric Arches cluster and that the volume number density profile $\rho(r)$ remains bound as $r\rightarrow\infty$, we can solve the Abel's integral equation (see e.g. Arfken \& Weber \cite{arfken}). From this procedure we get a core radius of $R_{c}=0.14\pm0.05\;pc$ and a central mass density of $\rho_{0}=\frac{\Sigma_{0}}{2R_{c}}=2.0(\pm0.4)\times10^{5}\;M_{\sun}\;pc^{-3}$. This latter value is similar to that obtained for the denser globular clusters in the Milky Way according to the Harris (\cite{harris}) catalog.

A variation of what we have done so far would be to restrict the fit of the empirical density law to stars above $30\;M_{\odot}$. In that case the core radius decreases to $R_{c}=0.10\pm0.01$, which implies that more massive stars are more concentrated in the inner parts of the cluster. But besides the observed radial trend in the IMF slope and the mass-dependent core radius, we can also compare the surface density profiles built from stars at two different mass bins, $10<M/M_{\odot}<30$ and $30<M/M_{\odot}<120$, as shown in Fig.~\ref{rho_profile}. A $\chi^{2}$ test rejects the hypothesis that the two sets of counts come from the same parent distribution at a $99.4\%$ confidence level. This adds to the discussion of several authors (e.g. Portegies Zwart et al. \cite{portegies_zwart} and references therein) about the possibility that the flattening of the IMF towards the center of Arches may provide the best indication yet for mass segregation in a young starburst cluster

The total mass of the cluster can be obtained by means of the integration of the IMF extrapolated towards lower masses. For this purpose and taking into consideration the indications of mass segregation that we have discussed, we will use the IMF slopes derived for $r<0.2pc$ and $0.2<r<0.4pc$, i.e. $\Gamma=-0.88$ and $\Gamma=-1.28$ respectively. In the range $M > 20 M_{\sun}$ (the theoretical minimum of $O$ stars) we find a total mass of $5570 M_{\sun}$ distributed in $135\;O$ stars. Extrapolating the IMF to $1 M_{\sun}$, gives a mean mass of $5.5 M_{\sun}$ and $3.4 M_{\sun}$ for the core and first annulus respectively, and a total mass $M_{cl}\sim 2.0(\pm0.6)\times10^{4} M_{\sun}$ within a projected radius of $0.4\;pc$. This value is significantly larger than the earlier photometric estimates of $10800 M_{\sun}$ and $12000 M_{\sun}$ for extrapolations down to $1 M_{\sun}$ and $0.1 M_{\sun}$ (Figer et al. \cite{figer99}), due to the flat $\Gamma=-0.6$ they used. Although our photometric total mass estimate is larger than previous work, a discrepancy still persist with respect to the upper dynamical limit of $7\times10^{4} M_{\sun}$ for stars within $0.23 pc$ (Figer et al. \cite{figer02}). One possible explanation for this may be the effect of unaccounted binaries in the dynamic mass determination (see e.g. Bosch et al. \cite{bosch} and Bosch et al. \cite{bosch2}).

Theoretically, and in comparison with the solar neighborhood, a more massive low mass limit for the IMF could be expected as a consequence of the extreme environmental conditions prevailing in the Galactic center. High clouds densities, high core temperatures, and the presence of magnetic fields should all play a role in the process of star formation. As noted by SBG05, their claimed turnover at an unusually high mass of $6-7\;M_{\odot}$ could be understood due to the environmental dependencies of current star formation theories. However, Kim et al. (\cite{kim_arches}) did not find evidence of a turnover down to $1.3\;M_{\odot}$. If the universality of the IMF holds, even for clusters in the Galactic center, the turnover would be expected at $\sim0.5\;M_{\odot}$. In that case, extrapolating down to $0.08 M_{\sun}$ with a Kroupa IMF (\cite{kroupa}) would increase the total mass of the cluster to $M_{cl}\sim 3.1(\pm0.6)\times10^{4}$, while the number of stars would be a factor $8$ larger, for a total of $\sim43400$ members.

Table~\ref{king_t} summarizes the physical parameters of the Arches cluster derived in this Section, including an estimation of the internal velocity dispersion.

\subsection{Upper Mass Cutoff and Maximum Stellar Mass in the Arches cluster}

Figer (\cite{figer05}) used his observations of  Arches to investigate the very important question of whether there is a physical upper limit to the mass of stars.  If there is no such limit, statistically we would expect to observe at least one star as massive as $914M_{\odot}$ in Arches (Figer estimates about $1100M_{\odot}$ with a flatter IMF slope of $\Gamma=-0.9$ ). The most massive stars we observe, however, have initial masses of $\sim100-120M_{\odot}$.  Thus, on the basis of Monte Carlo experiments Figer concluded that the Arches provides strong evidence of the existence of a physical upper mass limit to the IMF of $150M_{\odot}$.

The Arches cluster is sufficiently massive to allow us to place rather stringent statistical limits on the maximal mass of stars. Integrating our average IMF within $0.4 pc$ to $914M_{\odot}$, we find that we are missing $21$ stars if the maximum initial mass is $120 M_{\odot}$. Oey \& Clarke (\cite{oeyclarke}) used simple probabilistic arguments to discuss the problem of the upper-mass cutoff of the IMF on the basis of observations of OB associations younger than $3\;Myr$ in the Milky Way and the Magellanic Clouds.  Making the assumption of a coeval population of stars that have not yet exploded as supernovae, they found evidence of a universal upper mass cutoff around $120-200 M_{\odot}$.  Using their Equation 5 we can compute the probability of observing a maximum stellar mass for a given upper mass limit of the IMF ($M_{up}$). The results are shown in Table~\ref{oey_probs}.

Table~\ref{oey_probs} shows that the probability of the IMF being derived from a population with an upper mass limit significantly larger than the observed maximum mass in the Arches cluster is negligible ($1\%$ for $M_{up}=145M_{\odot}$). The important result of a cutoff for the Arches cluster is robust, and is in good agreement with the work of Figer (\cite{figer05}). But this conclusion does depend on the maximum observed mass. We should remind ourselves that our upper mass limits depend on the age of the cluster. For example, if we use the $2\;Myr$ Geneva theoretical isochrone to derive the IMF, there would be 11 stars within our strong color selection criterion that lie above the {\it Theoretical Surface}. For that age the models predict that these 11 stars would be more massive than $120 M_{\odot}$. Still, even if we estimate a maximum observed mass of $150 M_{\odot}$ for these stars, the probability that the IMF upper mass cut-off is e.g. $180M_{\odot}$ remains bound to less than 0.05.

We conclude with one remark. With no SN remnants detected in the Arches cluster, it is unlikely that stars more massive than the limits given in Table~\ref{oey_probs} have exploded as supernovae. If they existed they must have ended their evolution as Black-Holes or White Dwarfs as strong mass loss causes their final masses for $Z=0.02$ to be rather small according to the models of massive stars with rotation of Meynet \& Maeder (\cite{meynetymaeder}).

\section{Conclusions}
\begin{enumerate}
	\item It is in general not possible to transform broad-band JH$K_{S}$ photometry of highly reddened stars to a standard photometric system. In particular this means that the Rieke-Lebofsky extinction law (\cite{rieke}) cannot be applied to highly reddened stars when the filter passbands deviate significantly from the standards of Johnson. The solution that we have adopted in this paper, to work in the natural photometric system of our instrument (NACO), works well and leads to robust results.

	\item Even if one works in the natural photometric system, correcting for extinction in the Arches field is a delicate issue. The (mostly line of sight) extinction varies by several magnitudes from star to star: $2.13 < A_{K_{S}} < 4.14$ mag. and $3.45 < A_{H} < 6.71$ mag. This severely affects the completeness of mass bins and, if not considered, the \textit{Magnitude Limit Effect} can mimic features such as changes of slope, or turn overs in the IMF.

	\item The large extinction variations toward the cluster imply that even the most massive bins may be severely affected by incompleteness. Indeed we find a group of very red stars located in the outskirts of the cluster (concentrated in the direction of the Galactic center) that could be low-mass foreground stars, but could also be reddened high-mass members of the cluster.  This fundamental incompleteness would affect all mass bins and therefore remains a serious uncertainty on the mass function of the cluster.

	\item Applying color-selection criteria consistent with previous investigations, working in the natural photometric system of our instrument, and doing a very careful correction for completeness and differential extinction, we find an IMF slope within $0.4\;pc$ of $\Gamma=-1.1\pm0.2$ that is consistent within the errors with the Salpeter law.  We confirm the strong flattening of the IMF towards the cluster center ($\Gamma=-0.88$) in comparison with the $0.2-0.4\;pc$ annulus ($\Gamma=-1.28$). Our slope for the innermost region of the cluster is significantly steeper than previously thought.

	\item The principal source of deviation from the work of SGB02, SBG05 and Kim et al.(\cite{kim_arches}) is that we obtained physical parameters for cluster members using individual dereddening, whether in the form of CMD Sliding or the Bayesian approach. This is precisely one of the advantages of working in the natural photometric system and deriving the extinction law for our set of filters. The IMF slope in the innermost region of the cluster ($r<0.2pc$) is significantly flattened when the extinction for that area is considered uniform.

	\item We can now go back to the question posed at the beginning of this paper: Can an IMF consistent with the Salpeter law be ruled out for the Arches cluster? The answer is \textit{no}. Although for a long time this cluster was considered one of the best examples against the universality of the IMF, we follow the trend of previous work and give a step further: Down to a limit of $10 M_{\odot}$, the global ($r<0.4pc$) IMF of the Arches is reasonably fit by a power law of slope consistent (within the large uncertainties) with Salpeter.

	\item Our estimation for the total mass of the Arches cluster within projected $0.4\;pc$ is $\sim 2.0(\pm0.6)\times10^{4} M_{\sun}$, significantly larger than the previous photometric determination of $10800 M_{\sun}$ (Figer et al. \cite{figer99}). Thus we are able to place stringent limits on the physical upper mass cutoff of the IMF.  Since there are no detected supernova remnants in the cluster, if stars more massive than $130-150 M_{\odot}$ exist, they must live less than $2.5\;Myr$ and end their lives without exploding as supernovae.
\end{enumerate}

\begin{acknowledgements}
We would like to thank the anonymous referee for a very careful reading that led to important clarifications. P.E. thanks Andrea Stolte for discussions and many useful suggestions, and also acknowledges support from a Fulbright-CONICYT (Chile) fellowship for graduate studies. Additionally, part of this work was supported by ESO's Director's General Discretionary Fund.
\end{acknowledgements}

\begin{appendix}

\section{Isochrone Conversion Uncertainties and Effects on $\Gamma$}

In Section 2.3.2 we obtained equations to transform Geneva Isochrones from Johnson into the NACO natural photometric system. Our goal in this Appendix is to determine how uncertainties in the zero points of these transformations affect our final results, i.e. the IMF slope.

In order to isolate the zero point effect, we have constructed our simulations as follows. First, a theoretical isochrone is placed in the observational space of the Color-magnitude stereogram (CMS; Selman et al. \cite{selman99}). Then, from random sampling of the isochrone, a set of $N=10^{4}$ stars following a Salpeter law are obtained, with masses $\geq 10M_{\odot}$. We simulate real observations by assigning randomly to each star a reddening value drawn from the normalized observed extinction distributions of the Arches cluster. Photometric uncertainties in $J$, $H$, and $K_{S}$ for each simulated star are also assigned by sampling a $3D$ Gaussian distribution built to fit real DAOPHOT errors in each filter (see Section 2.3).

\begin{figure}
\begin{center}
\includegraphics[width=8cm]{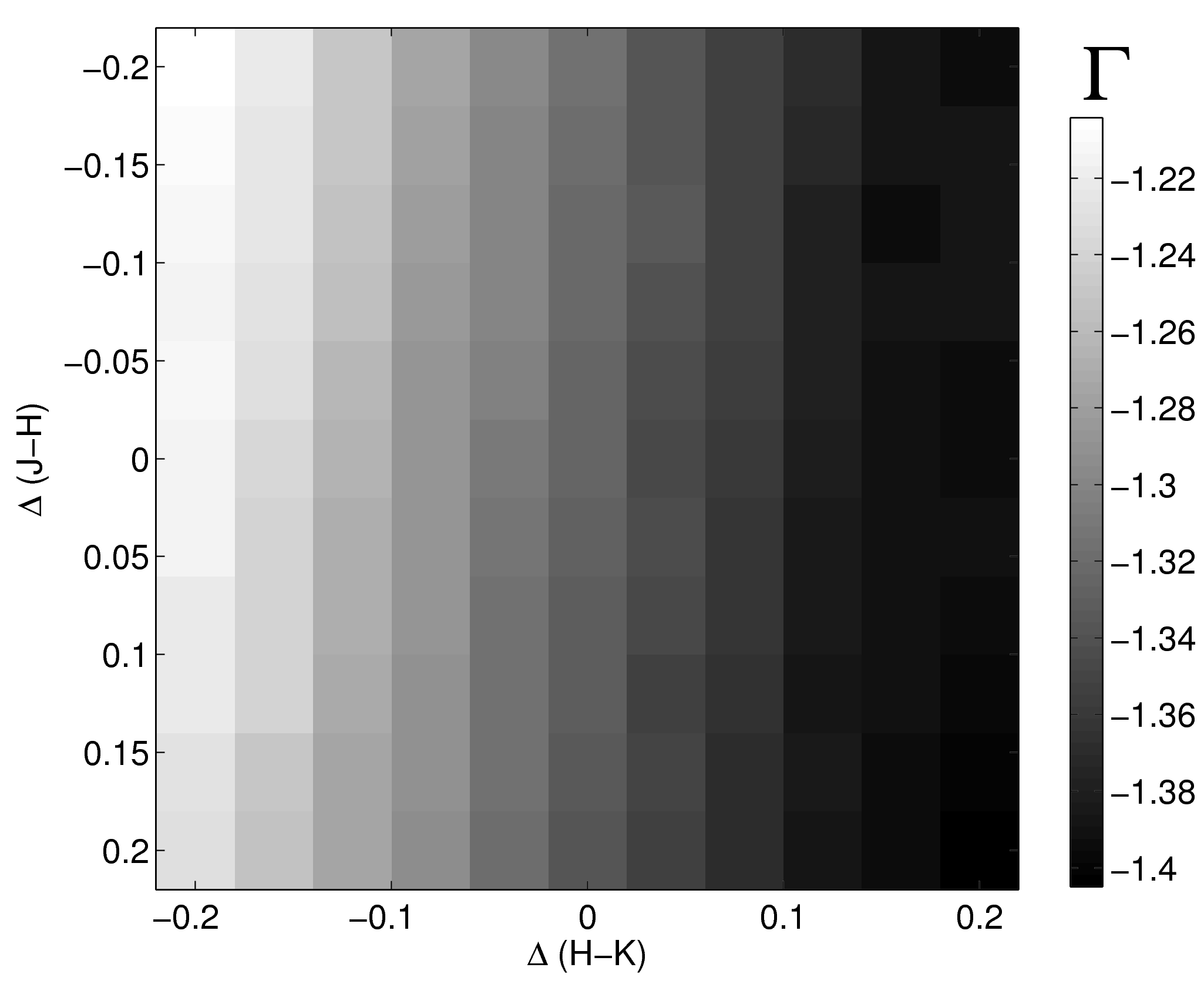}
\end{center}
\caption{Results of the CMS simulations. $\Delta(H-K_{S})$ and $\Delta(J-H)$ stand for variations in color when placing the unreddened theoretical isochrone in the NACO system. Each cell and its color represents a value of $\Gamma$ obtained after a linear fit to $10^{4}$ artificial stars.}
\label{sim1}
\end{figure}

With this procedure, a CMS resembling real observations can be simulated. It is clear that if we recover physical quantities for our set of $10^{4}$ stars using the Bayesian approach, we would recover an IMF with the input slope, i.e. a Salpeter IMF for masses $\geq 10M_{\odot}$ (small variations can occur because of scatter due to photometric errors). This case represents the central cell of Fig.~\ref{sim1}, with $\Delta(H-K_{S})=\Delta(J-H)=0$.

The rest of the map in Fig.~\ref{sim1} is built by the same procedure described above, but slightly displacing the data in the CMS before recovering physical parameters. This is equivalent to keeping the data points fixed and shifting the unreddened isochrone by $\Delta(H-K_{S})$ and $\Delta(J-H)$, which is the easiest way of thinking about the zero point uncertainty effect in the photometric transformation equations. The results can be expressed by a second order polynomial that fits the recovered slopes $\Gamma$ in a least squares sense:

\begin{eqnarray*}
  \Gamma & =  & - 1.3269 - 0.4639\Delta(H-K_{S}) \\ 
 & & - 0.0474\Delta(J-H) + 0.5315\Delta(H-K_{S})^{2} \\
 & & +  0.1092\Delta(H-K_{S})\Delta(J-H) - 0.0105\Delta(J-H)^{2}
\end{eqnarray*}

\end{appendix}

\begin{appendix}
\section{Reddening Parameters in Standard Bandpasses}

We argued in the main text that an optimal solution to the problem of transforming near-IR (but also optical) broad-band photometry of highly reddened early type stars to a standard photometric system (such as the Johnson, 2MASS, or HST/NICMOS systems) is  to work in the \textit{natural} photometric system defined by the instrument.  This approach has two main disadvantages:  it renders comparison with previous photometry difficult, and it makes necessary to compute synthetic colors of stars in the \textit{natural} photometric system.  In addition, as we showed in the main text, the reddening law of Rieke \& Lebofsky (\cite{rieke}, hereafter RL85), that is the common choice used to correct broad-band near-IR (NIR) photometry for the effects of interstellar extinction (and which has been used in all previous studies of the Arches cluster) cannot be used when one observes with filters that deviate from the standards of Johnson. RL85 used the Johnson bandpasses and obtained  photometry for 7 stars, 5 of which are heavily obscured due to their location near the Galactic center (GC). From the photometry and spectral type of these GC stars, RL85 derived a ratio of total to selective extinction of $R_{V}=3.09$. 

This value  $R_{V}$ is not strongly correlated with the shape of the extinction law at NIR and IR wavelengths, where the extinction law is generally considered to be an spatially invariant described by a power law. Thus, RL85 determine their reddening law based on the work of Schultz \& Wiemer (\cite{schultzwiemer}). i.e., assuming  $\frac{E_{V-K}}{E_{B-V}} = 2.744$. Combining the assumptions on $\frac{E_{V-K}}{E_{B-V}}$, $R_{V}$, and also assuming normal unreddened JHKLM colors for the spectral types of stars near the GC, it is rather straightforward to compute the color excesses at NIR bands and to derive an average reddening law for wavelengths redder than $J$. For the $UBVRI$ bands, the extinction law is taken from Nandy et. al (\cite{nandy}) and Schultz \& Wiemer (\cite{schultzwiemer}).

The assumption of  $\frac{E_{V-K}}{E_{B-V}} = 2.744$,  which was determined for relatively unreddened stars ($E(B-V) \leq 1$;  Schultz \& Wiemer \cite{schultzwiemer}), is not valid for highly reddened stars such as those in the GC where $E(B-V) \sim 10$.  Before the work of RL85 it was well known from empirical observations and numerical simulations that bandwidth effects must be considered for highly reddened stars (Olson \cite{olson}, FitzGerald \cite{fitzgerald}).

At visual wavelengths, due to the wide bandpass of the $UBV$ filters, reddening shifts the effective wavelength towards the red, so the reddening line has a curvature term that depends on the intrinsic colors of the stars (Schmidt-Kaler, \cite{schmidt}),
\begin{displaymath}
 \frac{E_{U-B}}{E_{B-V}} = \left\{ \begin{array}{rcl}
0.65-0.05 (U-B)_{0} +0.05 E_{B-V} & \mbox{for} & (U-B)_{0} < 0 \\ 
0.64+0.26 (B-V)_{0} +0.05 E_{B-V} & \mbox{for} & (B-V)_{0} > 0 
\end{array}\right.
\end{displaymath}
The ratio of total-to-selective absorption $R_{V}$ also depends on the amount of reddening and on the spectral energy distribution of the star as (Schmidt-Kaler, \cite{schmidt}),
\begin{displaymath}
 R_{V}=\frac{A_{V}}{E_{B-V}} = 3.30+0.28 (B-V)_{0} + 0.04 E_{B-V}
\end{displaymath}

\begin{figure}[!ht]
\begin{center}
\includegraphics[width=9cm]{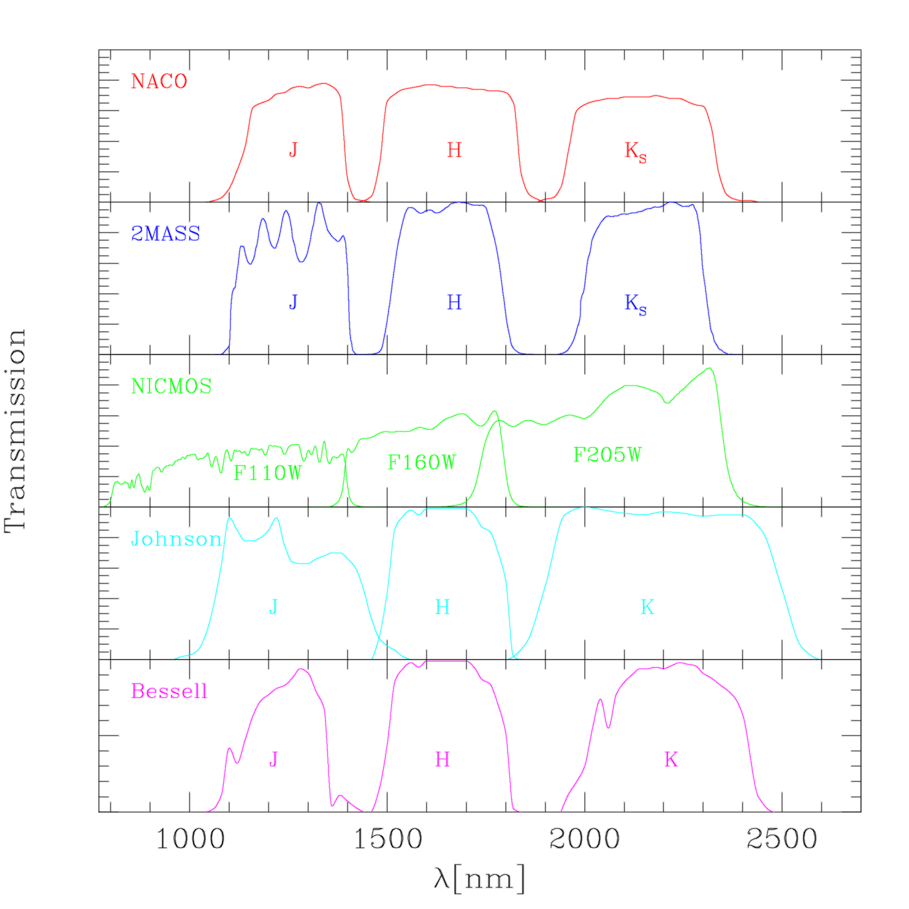}
\end{center} 
\caption{Passbands of the different photometric systems used to derive reddening parameters. From top to bottom, NACO, 2MASS, NICMOS/HST, Johnson (Arizona) and the homogenized Bessel \& Brett system. Transmission curves have been collected from Moro \& Munari (\cite{moro}, and references therein).}
\label{passbands}
\end{figure}

Because the spectral distributions of normal stars in the NIR tend to be flat and featureless, and since the extinction decreases with wavelength, these dependencies tend to be ignored in NIR studies. The significant tilt of our data points relative to the standard reddening line shown in the left panel of Fig.~\ref{twocolor} led us to investigate the issue of how the reddening path depends on the amount of extinction on the basis of numerical simulations.

We used the extinction law parameterization of Fitzpatrick (\cite{fitzpatrick2}) to quantify how extinction varies across a given filter. We used the average Galactic extinction curve, defined by the choice of $R_{V}=3.1$. This curve represents a significant improvement over previous work (Fitzpatrick, \cite{fitzpatrick}) as it is based on a large body of infrared data from 2MASS. The synthetic photometry is performed using the spectral energy distributions from the Kurucz (\cite{kurucz}) ATLAS9 model atmospheres for solar metallicity, microturbulent velocity of $\xi=2\;km\;s^{-1}$, and mixing length parameter $\alpha=1.25$. Different standard bandpasses were used to convolve the stellar spectra as shown in Fig.~\ref{passbands}. The parameters of these photometric systems have been collected from Moro \& Munari (\cite{moro}, and references therein).

\begin{figure}[!ht]
\begin{center}
\includegraphics[width=9cm]{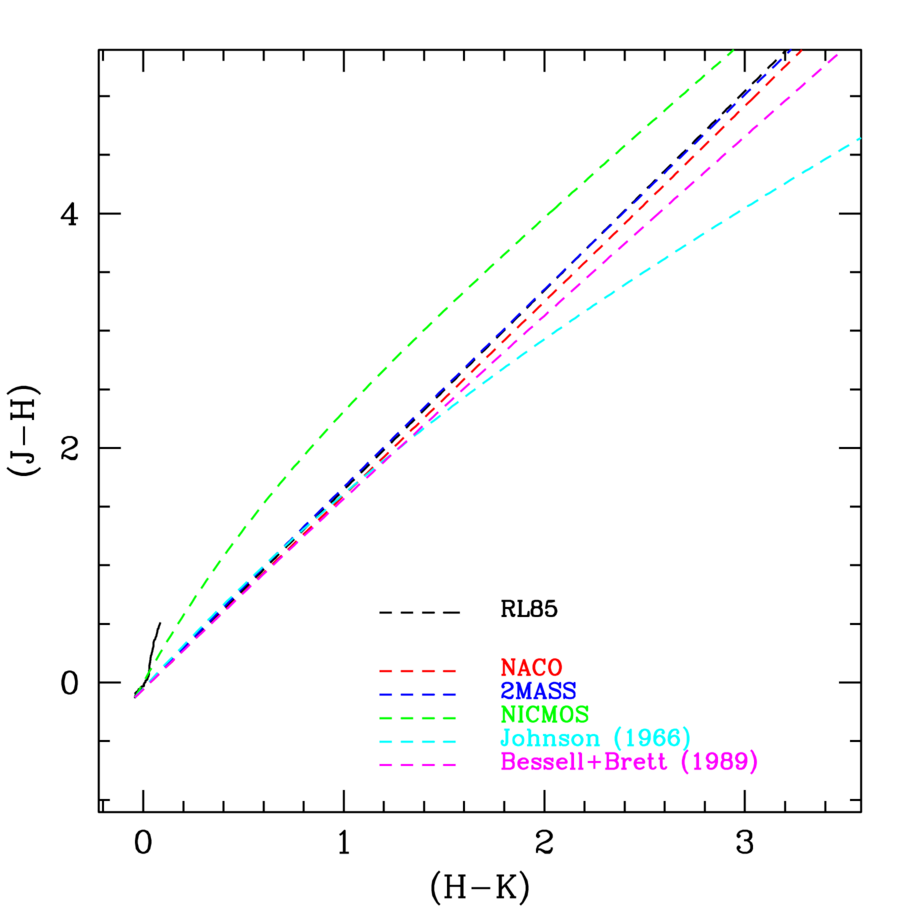}
\end{center}
\caption{Reddening lines obtained with from simulations using the filter bandpasses and Kurucz model atmospheres. Note the evident curvature when the standard Johnson (Arizona) filters are used to derive the extinction law (see text for details). The reddening lines for all photometric systems have been shifted so that they start from the NACO unreddened colors of an $O4V$ star.}
\label{twocolor2}
\end{figure}

Reddening paths for the five filter sets shown in Fig.\ref{passbands} are presented in Fig.\ref{twocolor2}. They are derived by reddening the spectral energy distribution of an $O4V$ star in the range  $0 < E(B-V) < 20$. The reddening parameters remain essentially constant for stars of different spectral types ($O$ to $M$) and luminosities (supergiants to dwarfs) in our simulation. A curvature term can be notably observed in the reddening lines for broad $J$ and $K$ (i.e. Johnson) filters, as expected. Although both colors, $(J-H)$ and $(H-K)$ show a departure from a linear behavior, the effect is more pronounced in the ordinate. The larger shift of $\lambda_{eff, J}$ with respect to $\lambda_{eff, K}$ can be explained because, although $K$ is a wider filter, extinction increases rapidly towards bluer wavelengths making the extinction variation across $J$ the dominant factor over filter width.

The results are in qualitatively agreement with those of Kim et al. (\cite{kim1pasp}, \cite{kim2pasp}). They have used the Padova stellar evolutionary models of Girardi et al. (\cite{girardi02}) to redden isochrones up to 6 magnitudes in the $K$ band. They also find that the width of the filters is important because sets of similar filters can behave as if they follow power laws with different exponents with respect to RL85 (where the transmission functions are assumed to be Dirac deltas). When a longer baseline ($J-K$) is introduced in Kim et al. (\cite{kim2pasp}), nonlinear effects in the reddening path of the wider filters, i.e. HST NICMOS F110W and F205W, also arise.

Our simulations result in $R_{K}=\frac{A_{K}}{E(H-K)}=1.61$ and $S=\frac{E(J-H)}{E(H-K)}=1.66$ after applying a linear fit to the NACO reddening path.

\end{appendix}

\begin{appendix}
\section{Effects on the IMF Slope due to Binning}

Ma\'iz Apell\'aniz \& \'Ubeda (\cite{jesus}) have given a detailed explanation on why the choice of bins that are variable in width are the safest choice to avoid biases in the IMF slope determination. We have seen in Section 4.1, however, that in our case for the Arches we are not subject to large deviations, and that systematic errors dominate over this effect. We will focus then in the uncertainties derived from our scheme of uniform-size logarithmic mass bins to derive the IMF slope $\Gamma$. Thus we have considered different widths and starting points.

Fig.~\ref{resumen3} illustrate our procedure, in which nine set of points are plotted for each panel. The first three ones, from left to right, represent a bin size of $\Delta log(M/M_{\odot})=0.1$. They are followed by three set of points computed with $\Delta log(M/M_{\odot})=0.15$. The rightmost points were derived with the wider bin, i.e. $\Delta log(M/M_{\odot})=0.2$. For each of these bin sizes, we distinguished three starting points, spaced between them by $\frac{1}{3} \Delta log(M/M_{\odot})$. It is noted that all linear fits to the data only included stars above $10\;M_{\odot}$.

For a particular age and metallicity of the models, in this case $2.5\;Myr$ and $Z=0.02$, we consider the average slope value as our best estimate for $\Gamma$. The dispersion, our estimate of uncertainty in the IMF slope due to a particular binning, is $0.09$. 

\begin{figure}[!ht]
\begin{center}
\includegraphics[width=9cm]{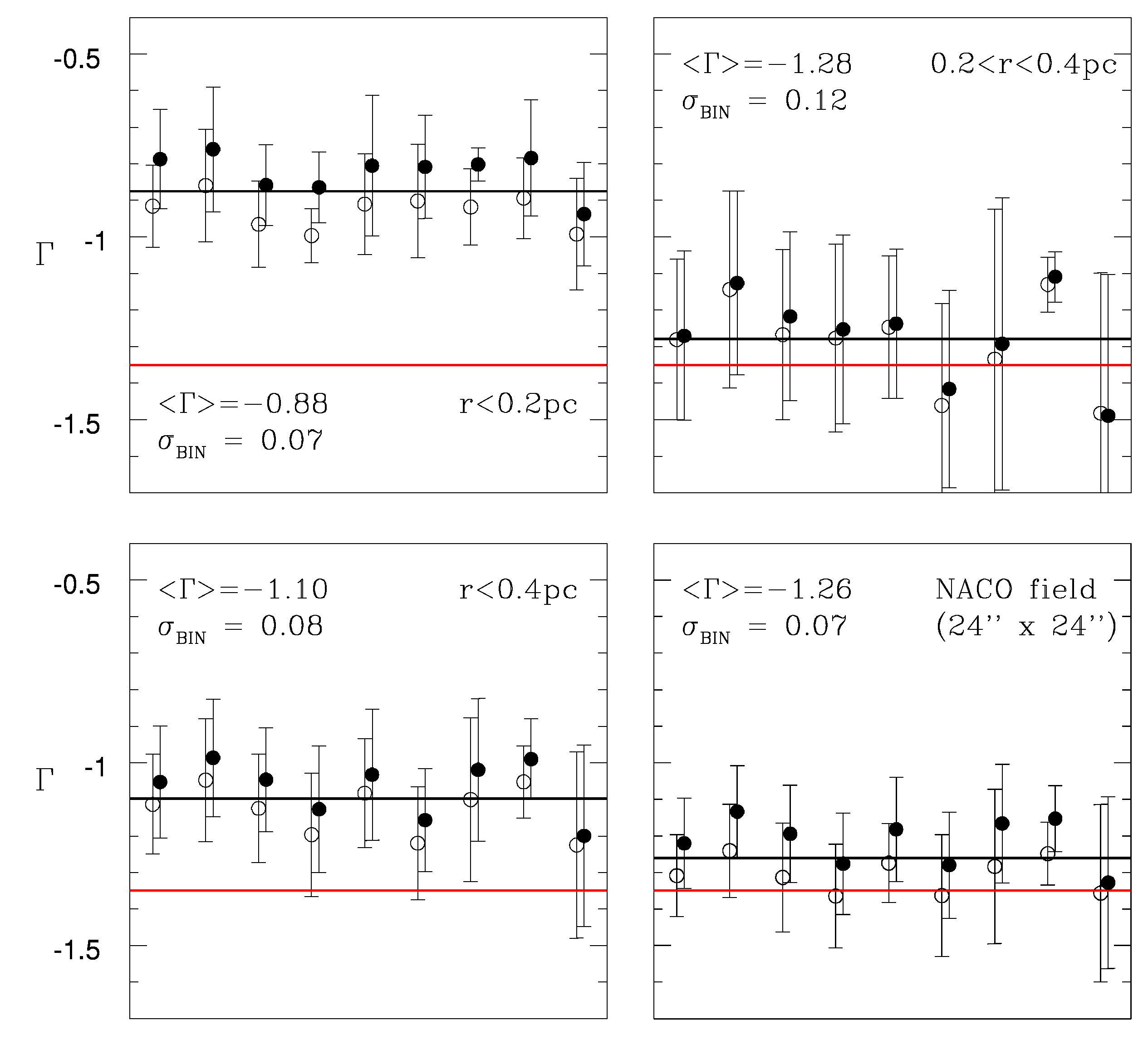}
\end{center}
\caption{Each panel shows the IMF slopes obtained with different binnings at a given distance from the cluster center. Solid (open) circles correspond to IMF determinations that take into account stars within the $50\%$ ($25\%$) completeness limit. Error bars represent formal fitting uncertainties. The average slope value, cited in each panel along with the uncertainty due to binning, is represented by a black line. The red line represents the Salpeter value, $\Gamma=-1.35$, as a reference.}
\label{resumen}
\end{figure}

\end{appendix}

\begin{appendix}
\section{Spatial Distribution of Reddening in the Arches Field}

\begin{figure}[!ht]
\begin{center}
\includegraphics[width=9cm]{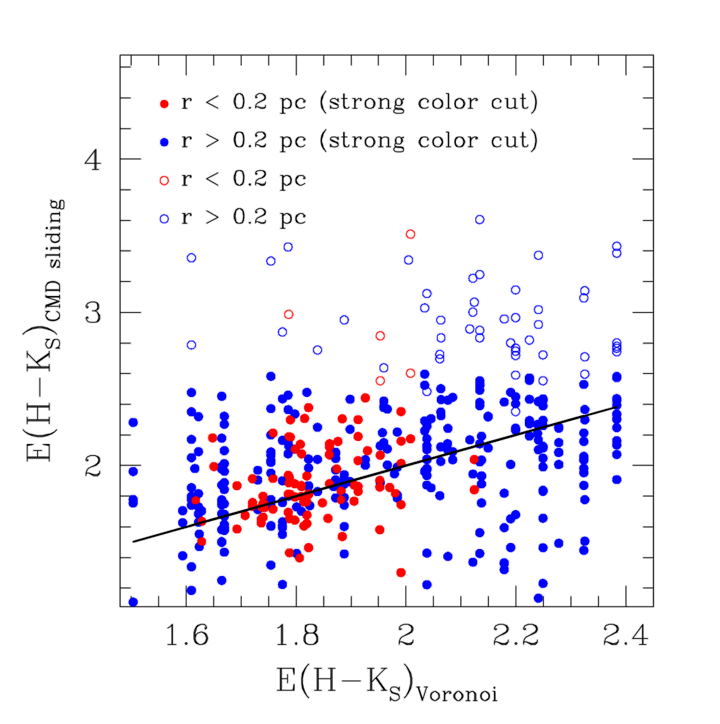}
\end{center}
\caption{Comparison between the color excesses derived from the Voronoi and CMD Sliding methods for stars with $HK_{S}$ photometry. The color distinguishes stars inside (red) and outside (blue) the core of the cluster. Solid circles represent stars within the strong color cut; open circles show the (red) objects rejected by the strong color cut criterion. The straight line correspond to X=Y.}
\label{excesses_app}
\end{figure}

The Bayesian and CMD Sliding methods give independent estimates for the physical parameters and color excesses. We tried a third approach to explore the extinction variations within the NACO field of view. This is shown in Fig~\ref{voron_redd_maps}(a), where the color of each cell corresponds to the $E(H-K_{S})$ value of a star with three-band photometry and within the strong color cut. The set of stars that fulfill this two requirements are called ``owners'' of Voronoi cells. We can then assign to each star with only $HK_{S}$ data the reddening of the ``owner'' of the cell where they fall in. Fig.~\ref{excesses_app} shows the comparison between the color excesses obtained by the CMD Sliding and Voronoi methods.

\end{appendix}

If extinction would vary smoothly with radius one could expect a better match between the two methods, at least for stars located at $r<0.2pc$, due to the fine sampling achieved in the center of the cluster. The poor correlation exposed in this Figure indicates that this is not the case. Reddening can vary randomly from star to star, making the choice of describing extinction by average values impracticable.

At greater distances from the center, the extinction is again extremely patchy and the size of Voronoi cells is large: the scatter can be as big as one magnitude! This remarks the need for individual dereddening in the Arches field, and strengthens the choice of the CMD Sliding over the Voronoi approach.

\end{document}